\documentclass[reprint,superscriptaddress,preprintnumbers,amsmath,amssymb,aps,prx,raggedbottom]{revtex4-2}

\usepackage{bibunits}
\usepackage{graphicx}
\graphicspath{{./Fig/}{./supp_figs/}}
\usepackage{multirow}
\usepackage{bm}
\usepackage[colorlinks=true,linkcolor=red,urlcolor=blue,citecolor=blue]{hyperref}
\usepackage{braket}
\usepackage{xspace}
\usepackage{xcolor}
\usepackage[separate-uncertainty=true,multi-part-units=single]{siunitx}
\usepackage{booktabs}
\usepackage{textcomp}

\newcommand{\vac}{\ket{\mathcal{V}}}

\makeatletter
\def\maketitle{\@author@finish\title@column\titleblock@produce\suppressfloats[t]}
\makeatother

\begin{document}

\title{Analog quantum simulation of paraparticle statistics}

\def\RLEaffil{Research Laboratory of Electronics, Massachusetts Institute of Technology, Cambridge, MA 02139, USA}
\def\LLaffil{Lincoln Laboratory, Massachusetts Institute of Technology, Lexington, MA 02421, USA}
\def\Physaffil{Department of Physics, Massachusetts Institute of Technology, Cambridge, MA 02139, USA}
\def\EECSaffil{Department of Electrical Engineering and Computer Science, Massachusetts Institute of Technology, Cambridge, MA 02139, USA}

\author{Gyunghun~Kim}
\email{gyunghun@mit.edu}
\affiliation{\RLEaffil}
\affiliation{\EECSaffil}

\author{Cora~N.~Barrett}
\affiliation{\RLEaffil}
\affiliation{\Physaffil}

\author{Sarah~E.~Muschinske}
\altaffiliation[Present address: ]{Google Research, Mountain View, CA, USA}
\affiliation{\RLEaffil}
\affiliation{\EECSaffil}

\author{Junyoung~An}
\altaffiliation[Present address: ]{Amazon Web Services (Center for Quantum Computing), Pasadena, CA, USA}
\affiliation{\RLEaffil}
\affiliation{\EECSaffil}

\author{Helin~Zhang}
\affiliation{\RLEaffil}

\author{Ilan~T.~Rosen}
\altaffiliation[Present address: ]{IBM Quantum, IBM Research Cambridge, Cambridge, MA, USA}
\affiliation{\RLEaffil}

\author{Rabindra~Das}
\affiliation{\LLaffil}
\author{Michael~A.~Gingras}
\affiliation{\LLaffil}
\author{David~K.~Kim}
\affiliation{\LLaffil}
\author{Bethany~M.~Niedzielski}
\affiliation{\LLaffil}
\author{Meghan~Schuldt}
\affiliation{\LLaffil}
\author{Jonilyn~L.~Yoder}
\affiliation{\LLaffil}
\author{Mollie~E.~Schwartz}
\affiliation{\LLaffil}
\author{Kyle~Serniak}
\affiliation{\RLEaffil}
\affiliation{\LLaffil}
\author{Jeffrey~A.~Grover}
\affiliation{\RLEaffil}

\author{Max~Hays}
\email{maxhays@mit.edu}
\affiliation{\RLEaffil}

\author{William~D.~Oliver}
\email{william.oliver@mit.edu}
\affiliation{\RLEaffil}
\affiliation{\EECSaffil}
\affiliation{\Physaffil}

\date{\today}

\begin{bibunit}[apsrev4-2]

\begin{abstract}
Bosonic and fermionic quantum statistics describe all known elementary and emergent particles in three dimensions, although many attempts have been made to discover other valid quantum statistical frameworks. One recent formulation proposes ``$R$-paraparticles,'' indistinguishable particles with nonlocal internal degrees of freedom governed by generalized exchange rules~\cite{wangParticleExchangeStatistics2025}. Here, we emulate a species of paraparticle that exhibits a generalized Pauli exclusion principle using a superconducting qubit lattice. We confirm its unique quantum statistics by measuring a $180^\circ$ phase shift under paraparticle exchange, observing antibunching independent of internal degrees of freedom under two-particle interference at a beamsplitter, and finding a thermal occupation inconsistent with both Bose--Einstein and Fermi--Dirac statistics. These results demonstrate that parastatistics are observable on quantum simulation platforms, further motivating their use to explore theorized new phases of matter.
\end{abstract}

\maketitle

\let\oldaddcontentsline\addcontentsline
\renewcommand{\addcontentsline}[3]{}

\section{Introduction}

All known elementary particles obey either fermionic or bosonic statistics. Identical bosons preferentially occupy the same quantum state due to final-state stimulation, whereas identical fermions are forbidden from occupying the same state due to the Pauli exclusion principle~\cite{Pathria2011}. These contrasting statistics govern fundamental physical phenomena ranging from quantum coherent states of light to the fundamental structure of atoms. Particle statistics also plays a central role in condensed matter physics, where collective excitations or composite particles may acquire effective quantum statistics distinct from those of their microscopic constituents~\cite{chenWhenSuperconductivityCrosses2024,greiterFractionalStatistics2024}.

Many theoretical efforts have explored the possibility of particle statistics beyond fermions and bosons~\cite{Green1953Parastatistics,LeinaasMyrheim1977Identical,goldinRepresentationsLocalCurrent1981,Wilczek1982FractionalSpin,wuGeneralTheoryQuantum1984,arovasFractionalStatisticsQuantum1984,Greenberg1990InfiniteStats,greenbergParticlesSmallViolations1991,haldaneFractionalStatisticsArbitrary1991a,wuStatisticalDistributionGeneralized1994}. One prominent example is the anyon~\cite{LeinaasMyrheim1977Identical,goldinRepresentationsLocalCurrent1981,Wilczek1982FractionalSpin,wuGeneralTheoryQuantum1984,arovasFractionalStatisticsQuantum1984,Bartolomei2020AnyonCollisions,nakamuraDirectObservationAnyonic2020}, whose exchange statistics form representations of the braid group and are therefore restricted to two spatial dimensions. Several higher-dimensional generalizations of quantum statistics have also been proposed~\cite{Green1953Parastatistics,Greenberg1990InfiniteStats,greenbergParticlesSmallViolations1991,haldaneFractionalStatisticsArbitrary1991a,wuStatisticalDistributionGeneralized1994}. However, further studies on Green parastatistics~\cite{greenbergSelectionRulesParafields1965,druhlParastatistics1970} and Doplicher--Haag--Roberts (DHR) superselection theory~\cite{DHR1971LocalObsStatsI,DHR1974LocalObsStatsII,doplicherWhyThereField1990} established that a broad class of such generalizations can be reduced to bosons or fermions with local internal degrees of freedom.

A recently developed framework introduced a new class of particles---designated $R$-paraparticles---with generalized exchange statistics~\cite{wangParticleExchangeStatistics2025}. The key ingredient is a nonlocal degree of freedom that is hidden (cannot be accessed via local measurements) and transforms non-trivially under particle exchange according to the $R$ tensor as introduced in Refs.~\cite{wangParticleExchangeStatistics2025,wangRparastatisticsFoundation2026}. We hereafter refer to this internal degree of freedom as ``color'' to easily associate it with our figure data, although it has also been referred to as flavor~\cite{salinelBosonization$R$paraparticleLuttinger2026,schurichtParastatisticsInteractingPeriodic2026,thakurStatisticalMechanicsParaparticles2025,liuParaparticlesIntrinsicallyExhibit2026} or simply internal states~\cite{wangParticleExchangeStatistics2025,wangRparastatisticsFoundation2026}. Whereas exchanging bosons or fermions multiplies the many-body wavefunction by a single scalar factor, exchanging paraparticles applies a matrix transformation in the color space. Unlike previous formulations of parastatistics~\cite{Green1953Parastatistics,greenbergSelectionRulesParafields1965,druhlParastatistics1970}, this framework cannot be reduced to ordinary bosons or fermions with local internal labels.

\begin{figure*}[!t]
\includegraphics{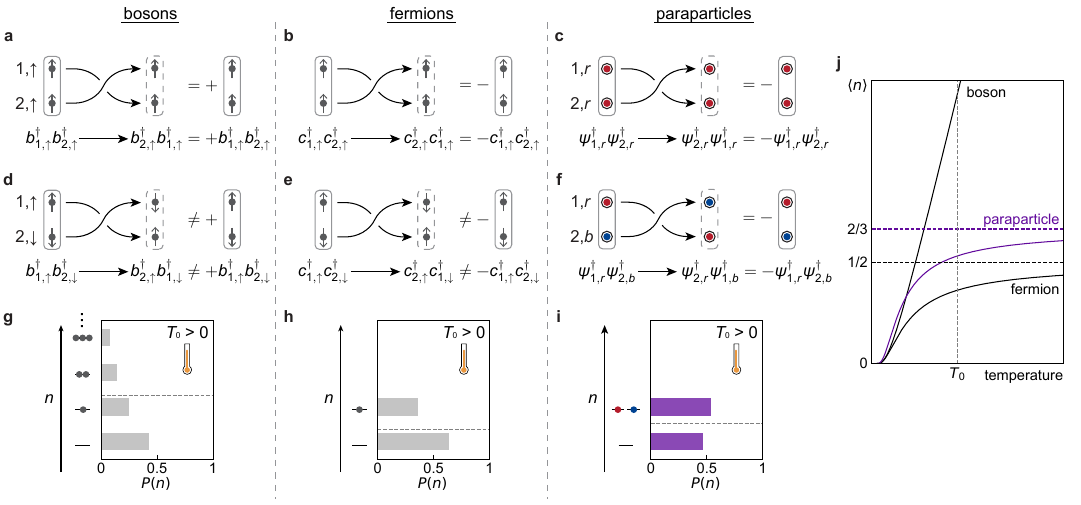}
\caption{\textbf{Quantum statistics of bosons, fermions, and paraparticles.}
\textbf{a-c,} Exchange statistics of identical quantum particles with the same internal state. Exchanging two bosons results in the same state with a factor of $+1$, which promotes bosonic final-state enhancement. Exchanging fermions (\textbf{b}) and paraparticles (\textbf{c}) results in the same state with a factor of $-1$, enforcing an exclusion principle for both types of particles.
\textbf{d-f,} Exchange statistics of identical quantum particles with different internal states. Exchanging bosons (\textbf{d}) or fermions (\textbf{e}) with different spins produces a physically distinct state, as the spin is a local observable that labels and uniquely distinguishes the individual particles. In contrast, the color of an $m=2$ paraparticle is a nonlocal attribute. Exchanging $m=2$ paraparticles (\textbf{f}) again results in the same state with a factor of $-1$, as with indistinguishable fermions and paraparticles of the same color. Consequently, m=2 paraparticles are subject to a generalized exclusion principle that is color-agnostic.
\textbf{g-i,} Thermal occupation statistics for bosons (\textbf{g}), fermions (\textbf{h}), and $m=2$ paraparticles (\textbf{i}). The bar plots show calculated occupation probabilities of each population state after thermalization at temperature $T_0$, while the dotted lines indicate the corresponding average occupations. Due to the generalized exclusion principle, the resulting thermal distribution differs from the Bose--Einstein and Fermi--Dirac distributions.
\textbf{j,} Thermal occupations of $m=2$ paraparticles, bosons, and fermions as a function of temperature.}
\label{fig: concept}
\end{figure*}

We consider here a species of paraparticles with an internal color structure of dimension $m$ and generalized antisymmetric exchange statistics~\cite{wangParticleExchangeStatistics2025}, originally proposed in Refs.~\cite{mishraAlgebraFermionsNew1991,mishraGeneralizedFockSpaces2000}. Paraparticles, like all quantum particles, manifest their quantum statistics through their behavior under particle exchange, as illustrated in Fig.~\ref{fig: concept}. Paraparticles with $m=1$ have a single color only (e.g., color-red, Fig.~\ref{fig: concept}c) and are identical to ordinary single-spin fermions (e.g., spin-up, Fig.~\ref{fig: concept}b), which under exchange result in the same state with a factor of $-1$ (cf. $+1$ for single-spin bosons in Fig.~\ref{fig: concept}a).

For $m\geq2$, however, the paraparticles acquire an internal color structure that differs fundamentally from internal degrees of freedom like spin. Spin is a local observable attached to each particle, whereas the paraparticle color---although represented here by a colored dot inside a circle---is a nonlocal (hidden) property of the quantum state that is globally encoded across the many-body wavefunction. Exchanging two $m=2$ paraparticles produces the same state with a factor of $-1$, independent of the particle colors (red and blue) or their permutations. This color-agnostic exchange rule is distinct from ordinary spinful Bose or Fermi statistics (Figs.~\ref{fig: concept}d-f).

The $m=2$ exchange statistics enforces a generalized exclusion principle across color sectors: two paraparticles with the same (Fig.~\ref{fig: concept}c) or different (Fig.~\ref{fig: concept}f) color states cannot occupy the same mode. In contrast, while two fermions with the same spin state (Fig.~\ref{fig: concept}b) cannot occupy the same mode due to the Pauli exclusion principle, two fermions with different spin states (Fig.~\ref{fig: concept}e) do not have this restriction. As a result, the $m=2$ paraparticle system exhibits nontrivial thermal occupation statistics distinct from those of both Bose--Einstein and Fermi--Dirac statistics (Figs.~\ref{fig: concept}g-j).

Here, we use a superconducting qubit lattice to emulate $m=2$ paraparticles. Our implementation uses interacting pairs of transmon qubits as the elementary paraparticle sites. Each pair supports three relevant states: vacuum, red, and blue. In a system of multiple particles, the doubly occupied state of each pair (red and blue) is energetically excluded by a large $ZZ$ interaction.

The interacting pairs of qubits are mapped to a 1D chain of non-interacting (free) paraparticles via a generalization of the Jordan--Wigner (JW) transformation~\cite{JordanWigner1928,wangParticleExchangeStatistics2025}. This mapping allows us to emulate paraparticle dynamics using pairs of qubits, with each qubit representing a nonlocal color of the paraparticle.

Using subsets of a strongly coupled $4\times4$ transmon lattice, we experimentally probe the paraparticle statistics on 1D chains of up to length four through three complementary experiments: direct exchange of two free paraparticles, two-particle interference at a beamsplitter, and thermalization of a single paraparticle site coupled to an emulated bath.

\section{Qubit implementation and paraparticle mapping}

We use the language of second quantization to create and annihilate paraparticles from the vacuum state $\vac$. We create a single paraparticle at site $i$ with color $\alpha \in \{r, b\}$  using the operator $\psi^\dagger_{i, \alpha}$, and we annihilate it with the operator $\psi_{i, \alpha}$. For example, a single red paraparticle at site $i$ is represented by $\psi^\dagger_{i, r}\vac$ and a blue paraparticle by $\psi^\dagger_{i, b}\vac$.

As with fermions and bosons, paraparticle statistics becomes relevant when moving beyond single-particle states to many-body states, as governed by the following commutation relations~\cite{wangParticleExchangeStatistics2025}:
\begin{equation}
\label{eq:commutation_relations}
\begin{aligned}
    \psi^\dagger_{i,\alpha} \psi^\dagger_{j,\beta} &= -\psi^\dagger_{j,\alpha}\psi^\dagger_{i,\beta}, \\
    \psi _{i,\alpha} \psi _{j,\beta} &= -\psi _{j,\alpha}\psi _{i,\beta}, \\
    \psi _{i,\alpha} \psi^\dagger_{j,\beta} &= \delta_{\alpha \beta}\left(\delta_{ij}-\sum_\gamma\psi^\dagger_{j,\gamma}\psi _{i,\gamma}\right),
\end{aligned}
\end{equation}
where the summation is over all possible colors $\gamma$.

The commutation relations are reminiscent of fermionic anticommutation relations, but intertwine the color sectors, imbuing the paraparticles with interesting properties. First, $\psi^\dagger_{i,\alpha} \psi^\dagger_{i,\beta} = 0$ enforces the generalized exclusion principle across color: two paraparticles cannot occupy the same site $i$ even for different colors $\alpha\neq \beta$ (cf. two fermions with different spins may occupy the same site). Second, the equality $\psi^\dagger_{i,\alpha} \psi^\dagger_{j,\beta}\vac = -\psi^\dagger_{j,\alpha}\psi^\dagger_{i,\beta}\vac$ indicates that color is nonlocal: colors $\alpha$ and $\beta$ cannot be interpreted as uniquely attached to site $i$ or $j$. Instead, the ordered colors $(\alpha, \beta)$ define a global color sector to which the state belongs. The operational significance of these properties will be revealed in the exchange, interference, and thermalization experiments herein.

\begin{figure}
\includegraphics{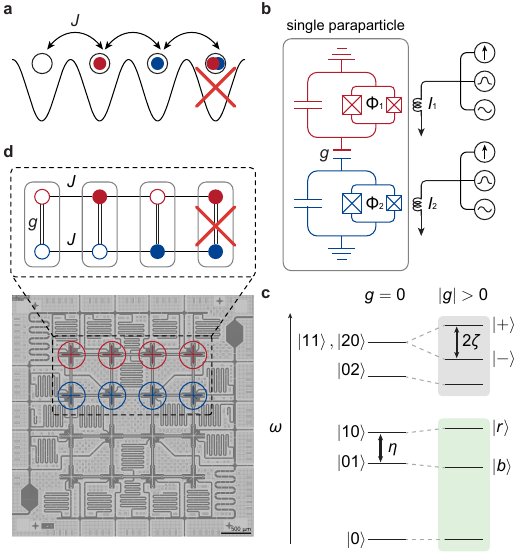}
\caption{\textbf{Realizing a 1D chain of paraparticles on a transmon qubit array.}
\textbf{a,} A 1D chain of paraparticles. Each site can host vacuum (empty circle), or a red or blue paraparticle (colored dot inside a circle), but not both colors simultaneously. The sites are coupled with hopping strength $J$.
\textbf{b,} Each paraparticle site comprises two tunable transmon qubits capacitively coupled with strength $g$, where the color (red or blue) is encoded by whether the top or bottom qubit is excited.
\textbf{c,} Energy-level diagram of a paraparticle site. Detuning the pair of qubits such that $\ket{11}$ and $\ket{20}$ are resonant maximizes the $ZZ$ interaction $\zeta$ between the qubits. In the rotating frame, where the dressed states are on resonance, the zero- and single-photon states shaded in green form the degenerate paraparticle Hilbert space $\{\ket{0}, \ket{r}, \ket{b}\}$, while the transitions to the states in gray become off-resonant and energetically unfavorable.
\textbf{d,} Qubit-ladder implementation of the paraparticle chain depicted in \textbf{a} and its placement on a 16-qubit array of transmons.}
\label{fig: implementation}
\end{figure}

Analogous to the mapping between a fermionic chain and a spin-$\tfrac{1}{2}$ (two-level system) chain under the JW transformation, a 1D chain of paraparticles (Fig.~\ref{fig: implementation}a) can be directly mapped to a chain of three-level systems via a generalized JW transformation. The Hilbert space of the $i$th three-level system is spanned by the ground state $\ket{0}_i$ and two excited states $\ket{r}_i$ and $\ket{b}_i$, with corresponding spin (transition) creation operators $\sigma^{+}_{i, r}=\ket{r}_i\bra{0}_i$ and $\sigma^{+}_{i,b}=\ket{b}_i\bra{0}_i$. In this representation, the paraparticle vacuum is mapped as $\vac \leftrightarrow \ket{0}=\otimes_i \ket{0}_i$, and the single-paraparticle states are mapped as $\ket{r}_i \leftrightarrow \sigma^{+}_{i,r}\ket{0}_i$ and $\ket{b}_i \leftrightarrow \sigma^{+}_{i,b}\ket{0}_i$. The many-body states can be obtained by directly interchanging paraparticle creation operators for the three-level system creation operators, provided that operators are expressed in ascending order by site (as in the conventional JW transformation, see supplementary materials). For example, we have $\psi_{i,\alpha}^\dagger \psi_{j,\beta}^\dagger \vac \leftrightarrow \sigma_{i,\alpha}^+ \sigma_{j,\beta}^+ \ket{0}$ for any $\alpha, \beta\in \{r,b\}$ if and only if $i\leq j$. This additional ordering requirement is necessary because the paraparticle creation operators, which generally do not commute, are mapped to creation operators $\sigma^+_{i,\alpha}$, which do commute across sites.

We realize an effective three-level system (single paraparticle site) using two flux-tunable transmon qubits that are capacitively coupled with strength $g$ (Fig.~\ref{fig: implementation}b). Within a two-level system model for the transmons, the two-transmon system spans the state space $\ket{00}$, $\ket{01}$, $\ket{10}$, and $\ket{11}$. The vacuum state represents the paraparticle vacuum of that site $(\ket{0}\equiv\ket{00})$, an excitation on the upper qubit represents color-red ($\ket{r}\equiv\ket{10}$), and an excitation on the lower qubit represents color-blue ($\ket{b}\equiv\ket{01}$). We suppress population transfer to the doubly occupied $\ket{11}$ state by leveraging the two-excitation states of the transmons ($\ket{02}$ and $\ket{20}$) to maximize the $ZZ$ interaction between the two qubits and thereby detune the $\ket{11}$ state from its bare value~\cite{Fors2024ZZComprehensive, Zhao2021SuppressStaticZZ, zhao2020highcontrast}. In Fig.~\ref{fig: implementation}c, the left column depicts the bare frequencies of the two-qubit system for $g=0$, and the right column depicts the dressed frequencies for $|g|>0$. Detuning the bare qubit frequencies by their nominally identical anharmonicities $\eta$ renders the bare states $\ket{11}$ and $\ket{20}$ frequency-degenerate (see supplementary material). This maximally hybridizes these states into $\ket{\pm} = \tfrac{1}{\sqrt{2}}\ket{11} \pm \tfrac{1}{\sqrt{2}}\ket{20}$ with average level splitting $2\zeta=2\sqrt{2}g\approx 2\pi\times40~\text{MHz}$, where $\zeta = \vert\omega_{\ket{\pm}}-\omega_{\ket{r}}-\omega_{\ket{b}}+\omega_{\ket{0}}\vert$ is the resulting $ZZ$ interaction strength. Therefore, in the rotating frame in which $\ket{r}$, $\ket{b}$, and $\ket{0}$ are on resonance, the two-excitation manifold is detuned by $\zeta$. We are left with an effective three-level system that represents a paraparticle site and can be occupied by a red paraparticle, a blue paraparticle, or neither.

We probe the exchange and interference physics of paraparticles using a one-dimensional chain of tunnel-coupled paraparticle sites, as shown in Fig.~\ref{fig: implementation}a. Under the generalized JW transformation, the paraparticle Hamiltonian maps to the three-level-system Hamiltonian
\begin{equation}\label{eq:tight-binding Hamiltonian}
    \begin{aligned}
        H &= \sum_{i} \sum_{\alpha\in\{r,b\}} J_i(\psi^\dagger_{i,\alpha} \psi _{i+1,\alpha} + \psi^\dagger_{i+1,\alpha} \psi _{i,\alpha}) \\
        &\leftrightarrow \sum_{i} \sum_{\alpha\in\{r,b\}} J_i(\sigma^+_{i,\alpha} \sigma^-_{i+1,\alpha} + \sigma^+_{i+1,\alpha} \sigma^-_{i,\alpha}),
    \end{aligned}
\end{equation}
where $J_i\lesssim2\pi\times3~\text{MHz}$ is the hopping amplitude between neighboring sites $i$ and $i+1$ ($\hbar=1$), which is realized via parametric couplings~\cite{strandFirstorderSidebandTransitions2013a,rosenSyntheticMagneticVector2024}. Importantly, the Hamiltonian is color agnostic ($J_i$ is color independent). To emulate this Hamiltonian, we arrange qubit pairs representing paraparticle sites in a laddered configuration (Fig.~\ref{fig: implementation}d). We emphasize that although the 2D ladder of qubits is strongly interacting, it directly maps to the \textit{non-interacting} 1D paraparticle chain via the generalized JW transformation~\cite{wangParticleExchangeStatistics2025}. This direct correspondence enables analog quantum simulation of the paraparticle Hamiltonian evolution and makes the state preparation and measurement straightforward: paraparticles with nonlocal color are created by applying $\pi$-pulses to selected individual qubits, and their populations are measured with local qubit readout.

Throughout this work, we employ different graphical symbols when depicting these two representations, as illustrated in Figs.~\ref{fig: implementation}a and \ref{fig: implementation}d. A solid-colored bullseye on an otherwise white target with a black border (whether a circle, square, or triangle) denotes a paraparticle excitation ($\psi_{i,\alpha}^\dagger$), with the color a global property of the state. A solid-colored shape without a black border, on the other hand, denotes an excitation of the three-level system local to the site ($\sigma_{i,\alpha}^+$).

\section{Exchange statistics}
\label{sec: exchange}

\begin{figure*}
\includegraphics{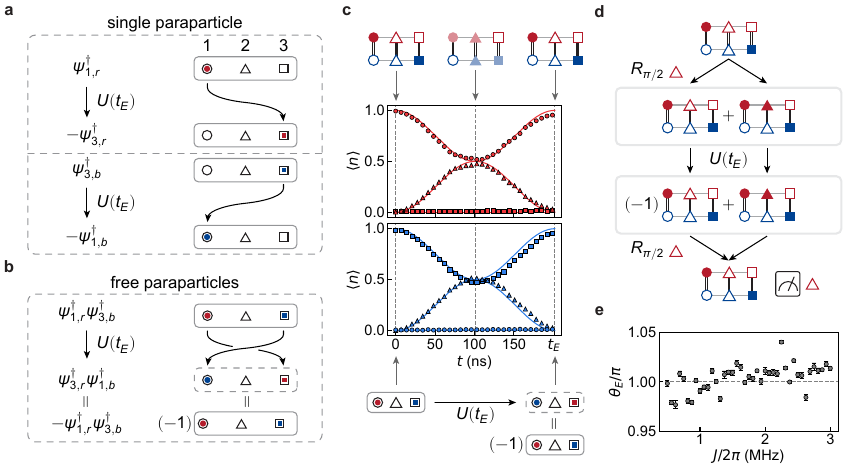}
\caption{\textbf{Exchange of paraparticles.}
\textbf{a,} Single-paraparticle evolution across a three-site chain. Regardless of the color, the unitary evolution realized by activating the coupling $J$ between sites for a time $t_E$ transports a particle from site $1$ to $3$ and vice versa.
\textbf{b,} Two-paraparticle exchange across a three-site chain. Starting at each end of the chain, the two non-interacting free paraparticles traverse each other under the same unitary evolution as in the single-particle case. The final state after the exchange is identical to the initial state up to a factor $-1$ (phase $\pi$).
\textbf{c,} Experimental qubit populations for red and blue qubits across sites 1 (circles), 2 (triangles) and 3 (squares) as a function of time. The exchange concludes at $t_E=\pi/(\sqrt{2}J)\approx202$~ns, when the qubit populations return to their initial values. The markers represent measurements, and the solid lines show theoretical predictions for an ideal chain. Bottom: corresponding paraparticle picture.
\textbf{d,} Interferometric sequence for detecting the exchange phase. We prepare a superposition of two-particle and three-particle states by applying a $\pi/2$ pulse on qubit $(2,r)$. When the coupling between the sites is activated, only the two-particle state (left arm) undergoes the exchange evolution $U(t_E)$ and accumulates a $\pi$ phase shift, while the Pauli blockaded (fully occupied) three-particle state (right arm) remains frozen. We apply a second $\pi/2$ pulse to extract the accumulated phase from the population of the $(2,r)$ qubit (red triangle).
\textbf{e,} Experimentally measured exchange phase $\theta_E=(1.00\pm0.01)\pi$, where the uncertainty is the standard deviation across $J$ values.}
\label{fig: exchange}
\end{figure*}

We first probe the defining exchange phase of paraparticles. We implement the exchange in a three-site chain of paraparticles that follows the Hamiltonian in Eq.~\eqref{eq:tight-binding Hamiltonian} with a uniform coupling $J_i=J$ for all sites $i$. Because the paraparticles are free (non-interacting) under this quadratic Hamiltonian, we can solve the many-body dynamics by tracking the evolution of the creation operators. After a time $t_E=\pi/(\sqrt{2}J)$, the operators evolve as
\begin{equation}\label{eq:exchange operator evolution}
\begin{aligned}
    \psi^\dagger_{1,\alpha} &\to -\psi^\dagger_{3,\alpha},\\
    \psi^\dagger_{2,\alpha} &\to -\psi^\dagger_{2,\alpha},\\
    \psi^\dagger_{3,\alpha} &\to -\psi^\dagger_{1,\alpha},
\end{aligned}
\end{equation}
for each color individually (Fig.~\ref{fig: exchange}a, see supplementary materials for derivation).

For two paraparticles initially occupying sites 1 and 3 of the chain, this free evolution (Fig.~\ref{fig: exchange}b) results in exchange of the particles (see also supplementary materials):
\begin{equation}
\begin{aligned}
    \psi^\dagger_{1,r}\psi^\dagger_{3,b}\vac &\to (-\psi^\dagger_{3,r})(-\psi^\dagger_{1,b})\vac \\
    &=+\psi^\dagger_{3,r}\psi^\dagger_{1,b}\vac \\ & =-\psi^\dagger_{1,r}\psi^\dagger_{3,b}\vac,
\end{aligned}
\end{equation}
where the last equality follows from the commutation relations in Eq.~\eqref{eq:commutation_relations} upon enforcing an established ordering of the operators by increasing site index. Due to the nonlocal character of color, the ordering $(r,b)$ is preserved as a global property of the state. With the established operator ordering, it can be seen that the physical consequence of the exchange operation is a factor $\exp(i \pi)=-1$ (i.e., a $\pi$ phase) acquired by the two-particle state that is color-agnostic (Fig.~\ref{fig: exchange}b) and distinct from spinful fermions (Fig.~\ref{fig: concept}e,f).

We test this prediction in two steps. First, we verify that the qubit lattice implements the desired exchange evolution by measuring the population dynamics of a three-site chain (six-qubit ladder). At $t=0$, we prepare the initial state $\psi^\dagger_{1,r}\psi^\dagger_{3,b}\vac \leftrightarrow \sigma^+_{1,r}\sigma^+_{3,b}\ket{0}$ by applying a $\pi$-pulse to the red qubit at site 1 and the blue qubit at site 3. We then parametrically activate inter-site hopping (Fig.~\ref{fig: implementation}d) with strength $J/2\pi=1.75$~MHz for a variable time $t$ (see supplementary materials for similar results using different $J$ values). Finally, we read out the qubit populations (Fig.~\ref{fig: exchange}c), which yield both the paraparticle site populations and the nonlocal color through the generalized JW transformation.

The observed dynamics can be interpreted in the equivalent qubit and paraparticle representations. In the qubit representation, the initially localized excitations at the two ends of the chain each attempt to flow to the middle site due to the inter-site couplings $J$. However, because of the strong $ZZ$ interaction, a double excitation is effectively forbidden there. Consequently, the red and blue qubit excitations each \textit{reflect} back to their initial positions at $t_E\approx202$~ns (Fig.~\ref{fig: exchange}c). Additionally, it is the $ZZ$ interaction that maintains the color ordering $(r, b)$ throughout the evolution. In the paraparticle representation, on the other hand, the free particles traverse each other. In this picture, the color and its ordering are preserved as global properties of the state and not attached to the individual particles, which do change places. However, these population dynamics alone cannot probe the global exchange phase.

As the second step, we use a Ramsey interferometric sequence to transform the global exchange phase into a directly measurable relative phase between two coherent pathways. We prepare the system in a superposition of two states, only one of which follows the exchange evolution and accumulates the phase $\pi$, then measure the relative phase between the two states. The experimental sequence is shown in Fig.~\ref{fig: exchange}d. We start with the same state, $\psi^\dagger_{1,r}\psi^\dagger_{3,b}\vac$, and prepare a superposition of two-particle (left arm) and three-particle (right arm) states by applying a $\pi/2$ pulse to the $(2,r)$ qubit. We then perform the exchange operation by activating the coupling between the sites for a time $t_E$, during which the two-particle state undergoes the exchange operation $U(t_E)$, while the three-particle state remains frozen as each site is already occupied. We measure the phase accumulated on the two-particle state by applying a second $\pi/2$-pulse on the same $(2,r)$ qubit and reading out its population. To account for the additional dynamical phase, we sweep the phase of the second $\pi/2$ pulse and compare the result with that obtained without the exchange operation (see supplementary materials).

In Fig.~\ref{fig: exchange}e, we plot the measured exchange phase as a function of the parametrically activated coupling strength $J$, applied for a sufficient time period for full exchange of the paraparticles. The two-particle state acquires a phase $\theta_E=(1.00\pm0.01)\pi$, where the uncertainty is the standard deviation across $J$ values (see supplementary materials for different initializations and $J$ values). In the paraparticle picture, the $\pi$-phase originates from the exchange parastatistics, defined by the commutation relations in Eq.~\eqref{eq:commutation_relations}; in the qubit picture, it is a geometric phase that accumulates during the complete cycle of starting and returning to the same initial state~\cite{aharonovPhaseChangeCyclic1987} (see supplementary materials).

\section{Two-Paraparticle Interference}

\begin{figure*}
\includegraphics{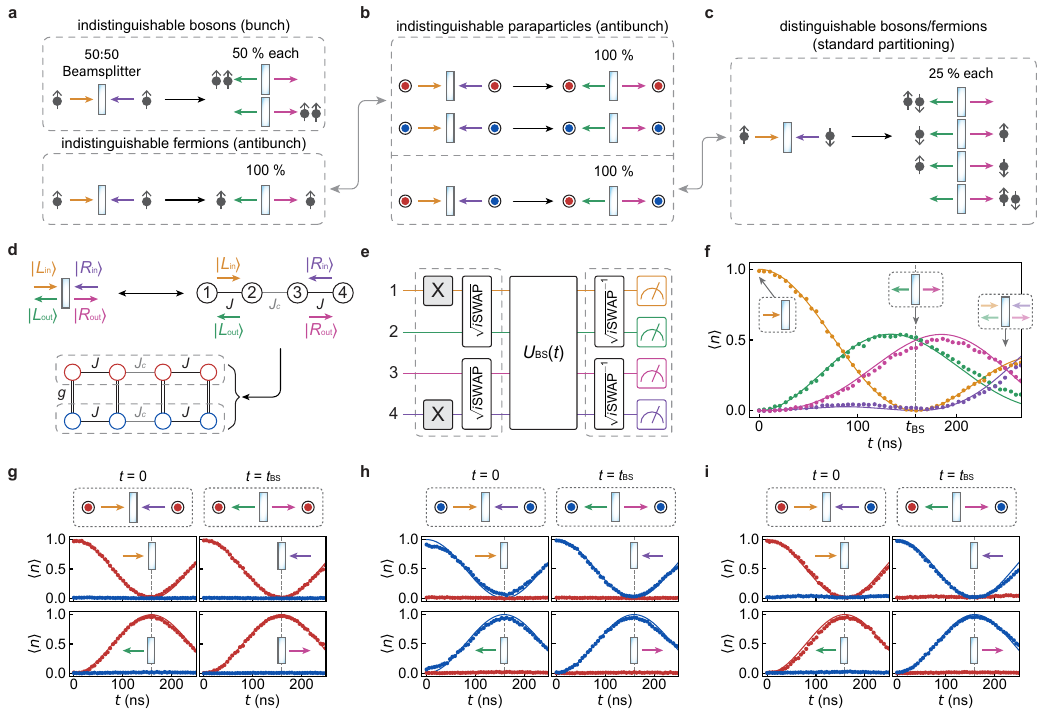}
\caption{\textbf{Two-paraparticle interference.}
\textbf{a,} Two-particle interference of indistinguishable bosons and fermions at a beamsplitter. Due to quantum statistics and the resulting quantum interference, bosons bunch at the outputs whereas fermions antibunch.
\textbf{b,} Two-particle interference of indistinguishable paraparticles. Two paraparticles of the same color antibunch in a manner similar to two indistinguishable fermions (cf., gray arrow comparing panels \textbf{a} and \textbf{b}). Two paraparticles with different colors remain indistinguishable (due to color nonlocality) and also antibunch.
\textbf{c,} In contrast, fermions (and bosons) with different spins are distinguishable, do not quantum mechanically interfere, and thus scatter independently according to standard partitioning (cf. gray line comparing panels \textbf{b} and \textbf{c}).
\textbf{d,} Paraparticle beamsplitter. The input modes ($L_{\mathrm{in}}$ and $R_{\mathrm{in}}$), the output modes ($L_{\mathrm{out}}$ and $R_{\mathrm{out}}$), and the beamsplitter are realized across 4 paraparticle sites (i.e., 8 qubits). Sites $(1,2)$ define the left ports, and sites $(3,4)$ define right ports, with rightward and leftward propagating states defined on each pair. The ratio of couplings, $J_c/J$, determines the beamsplitting ratio, while $J$ and $J_c$ together determine the evolution time $t_{BS}$.
\textbf{e,} Pulse sequence applied to the four paraparticle sites. An $\textrm{X}$ gate creates a paraparticle at site 1 or site 4 only (single-particle input) or at both sites 1 and 4 (two-particle input) of a color determined by the excited qubit(s). Excitations are converted to inward propagating states using $\sqrt{i\textrm{SWAP}}$ gates, which are then followed by the beamsplitting operation ${U_{\textrm{BS}}(t)}$ (see text). The inverse-$\sqrt{i\textrm{SWAP}}$ gates map the output propagating modes back to qubit excitations, which are measured to determine the beamsplitter input and output populations.
\textbf{f,} Time evolution of particle populations following a single input from the left port (yellow). The incoming mode splits equally into the outgoing modes (green and magenta) at time $t_{\mathrm{BS}}\approx 159$~ns. For times $t>t_{\mathrm{BS}}$, incoming modes (yellow and purple) gradually reappear due to reflection from the ends (sites 1 and 4).
\textbf{g,h,} Two-paraparticle interference (same color). Time evolution of the mode populations starting with two red \textbf{(g)} and two blue (\textbf{h}) paraparticles at the inputs. Markers represent measurements; solid lines are theoretical predictions of an ideal chain.
\textbf{i,} Two-paraparticle interference (different colors). Time evolution of the mode populations starting with a red and blue paraparticle at the left and right inputs, respectively. In all three cases \textbf{(g-i)}, the paraparticles antibunch.}
\label{fig: collision}
\end{figure*}

We next probe the paraparticle quantum statistics through two-particle interference, an established method for probing quantum statistics~\cite{loudon1998fermion,hance2025fermion}, including those of bosons~\cite{Hong1987HOM,Lopes2015AtomicHOM}, fermions~\cite{liu1998QuantumInterferenceElectron,Bocquillon2013ElectronicHOM}, and anyons~\cite{Bartolomei2020AnyonCollisions,nakamuraDirectObservationAnyonic2020}. The canonical version injects a pair of non-interacting particles simultaneously into the two input ports of a 50:50 beamsplitter. For indistinguishable bosons and fermions, the possible outcomes interfere quantum mechanically through the exchange phase, resulting in bunching and antibunching, respectively (Fig.~\ref{fig: collision}a). In contrast, when the particles are distinguishable, for example by their local internal spin states, the particles instead independently undergo single-particle standard partitioning (Fig.~\ref{fig: collision}c).

Our paraparticles should antibunch at the beamsplitter output, regardless of color, as the color is a nonlocal attribute that cannot uniquely label the paraparticles (Fig.~\ref{fig: collision}b). This is a manifestation of the generalized exclusion principle. We use a two-port beamsplitter hosting two incoming modes from the left and right ($L_{\mathrm{in}}$ and $R_{\mathrm{in}}$) and two outgoing modes ($L_{\mathrm{out}}$ and $R_{\mathrm{out}}$), as illustrated in Fig.~\ref{fig: collision}d. Using the commutation relations in Eq.~\eqref{eq:commutation_relations}, a two-paraparticle state incident from both ports,  $\psi^\dagger_{L_{\mathrm{in}},\alpha}\psi^\dagger_{R_{\mathrm{in}},\beta}\vac$, evolves into the state, $\psi^\dagger_{L_{\mathrm{out}},\alpha}\psi^\dagger_{R_{\mathrm{out}},\beta}\vac$, with one paraparticle in each outgoing port after the interference (see supplementary material). The two-paraparticle state preserves $(\alpha,\beta)$ as a global property throughout the dynamics.

We measure the two-paraparticle interference using a beamsplitter implemented within a four-site chain of paraparticles. To realize the beamsplitter, we first consider the red and blue qubit chains separately (Fig.~\ref{fig: collision}d, dashed gray boxes). In each four-qubit chain, the first two sites are the left arm (port) of the beamsplitter and the last two sites are the right arm (port). On each arm, we define the incoming and outgoing modes as the eigenstates of the current operator $j_{m+1,m}=-i(\sigma_{m+1}^+\sigma_{m}^- - \sigma_{m}^+\sigma_{m+1}^-)$~\cite{zotos1997TransportConservationLaws}, which describes the qubit population flow from site $m$ to $m+1$. The eigenstate with a positive (negative) eigenvalue corresponds to a rightwards- (leftwards-) propagating mode $\ket{\rightarrow}=(\ket{10} - i\ket{01})/\sqrt{2}$ ($\ket{\leftarrow}=(\ket{10} + i\ket{01})/\sqrt{2}$). The rightwards propagating mode on the left arm is $L_\mathrm{in}$, and we similarly relate the other three modes to $R_\mathrm{in}$, $L_\mathrm{out}$, and $R_\mathrm{out}$. To prepare and read out the input and output modes, we use the $\sqrt{i\mathrm{SWAP}}$ gate and its inverse to map between single-qubit excitations and propagating modes, associating $\mathrm{Q} 1\leftrightarrow L_\mathrm{in}$, $\mathrm{Q} 2\leftrightarrow L_\mathrm{out}$, $\mathrm{Q} 3\leftrightarrow R_\mathrm{out}$, and $\mathrm{Q} 4\leftrightarrow R_\mathrm{in}$, for each color $\alpha\in\{r,b\}$, where $\mathrm{Q}i$ is the qubit at site $i$.

We show the experimental sequence in Fig.~\ref{fig: collision}e. Starting with all qubits in their ground states, we apply an $\textrm{X}$ gate to create a particle at site 1 or site 4 only (single-particle input) or at both sites 1 and 4 (two-particle input). We then convert to inward propagating states using the $\sqrt{i\textrm{SWAP}}$ gates. To perform the beamsplitting operation, we turn on the intra-arm ($J$) and inter-arm ($J_{\textrm{c}}$) couplings, where the ratio $J_c/J$ determines the beamsplitting ratio (here, set to be 50:50). The beamsplitting operation is complete at a time $t_{BS}=\pi/\sqrt{J_c^2+4J^2}$ (see supplementary materials), immediately after which the state is in a superposition of outgoing modes only. Finally, we convert back from propagating modes using inverse-$\sqrt{i\textrm{SWAP}}$ gates and measure the qubits to determine the beamsplitter input and output populations. For example, Fig.~\ref{fig: collision}f shows the experimental data of a 50:50 beamsplitter, where $J/2\pi=1.5$~MHz, $J_c/J\approx 0.637$, and $t_{BS}\approx 1.5/J\approx159$~ns (see supplementary materials for extended results of 50:50 beamsplitting with different $J$ and $J_c$ values).

To make a paraparticle beamsplitter, as was done before, we couple the red and blue qubits at each site with a strength $g$ to create a strong $ZZ$ interaction (Fig.~\ref{fig: collision}d, lower middle). This four-rung ladder again realizes a non-interacting paraparticle chain governed by the Hamiltonian described in Eq.~\eqref{eq:tight-binding Hamiltonian}. With the same $J_c$ and $J$, each incoming paraparticle undergoes 50:50 beamsplitting into the outgoing modes at $t=t_{BS}$.

We present the experimental results for two-paraparticle interference for the same color particles, $(r, r)$ in Fig.~\ref{fig: collision}g and $(b, b)$ in Fig.~\ref{fig: collision}h, and for different colors $(r, b)$ in Fig.~\ref{fig: collision}i. As expected for paraparticles, we observe antibunching at $t=t_{BS}$ for all cases.

The measured antibunching again has two equivalent interpretations. In the interacting qubit representation, the incoming modes reflect back to their initial states due to the strong $ZZ$ interaction. In the paraparticle representation, antibunching arises from the quantum interference between two indistinguishable paraparticles, with the generalized exclusion principle enforcing antibunching for all cases, including the case of different colors.

\section{Thermalization}

\begin{figure*}
\includegraphics{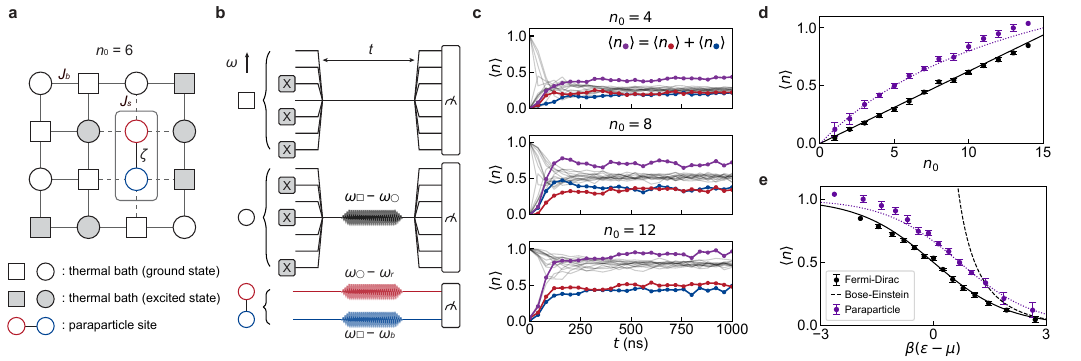}
\caption{\textbf{Thermalization of a paraparticle in a bath of qubits.}
\textbf{a,} A paraparticle site (red and blue qubits) with 14 bath qubits embedded in a 16-qubit lattice. The bath qubits are divided into two groups of 7 qubits (gray circles and squares). We initiate the experiment by populating $n_0$ bath qubits (colored gray). The coupling strengths between bath-bath and paraparticle-bath qubits are $J_b$ and $J_s$, respectively.
\textbf{b,} Experimental sequence. All qubits start detuned from one another to enable initializing the $n_0$ bath excitations. Once initialized, all square and circle bath qubits are brought onto resonance at frequencies $\omega_{\Box}$ and $\omega_{\bigcirc}$, respectively, which remain far detuned from the red and blue qubits at frequencies $\omega_r$ and $\omega_b$. The bath-bath couplings $J_b$ are activated by parametrically driving the bath circle qubits at the difference frequency $\omega_{\Box} - \omega_{\bigcirc}$. Similarly, the red and blue paraparticle qubits are driven at the difference frequencies $\omega_{\bigcirc} - \omega_{r}$ and $\omega_{\Box} - \omega_{b}$, respectively, to activate the $J_s$ couplings. The system evolves for a time $t$ before all bath qubits are returned to their initial detuned frequencies and read out.
\textbf{c,} Time evolution of qubit populations with $n_0=4$, $8$, and $12$. The red and blue traces show the populations of the corresponding paraparticle qubits, while the gray traces show the populations of the bath qubits. The bath and paraparticle qubits saturate to different thermal values.
\textbf{d,} Average population in the paraparticle and fermionic sites after $t=1$~\textmu s as a function of $n_0$. For each $n_0$, we initiate the system by exciting $10$ randomly chosen combinations of $n_0$ bath qubits (one combination for $n_0=14$). The error bars show the standard deviation across 10 realizations.
\textbf{e,} The same data plotted against $\beta(\epsilon-\mu)$. The measured $m=2$ paraparticle statistics show a clear deviation from Bose--Einstein and Fermi--Dirac statistics, but match paraparticle theory.}
\label{fig: thermalization}
\end{figure*}

Lastly, we emulate the impact of paraparticle quantum statistics on the population of paraparticle states in thermal equilibrium. As illustrated in Figs.~\ref{fig: concept}g-j, paraparticles possess distinct thermal statistics due to the generalized exclusion principle, which determines the number of possible microstates at a site. A single $m$-color paraparticle site has a partition function $Z=1+mx$, where $x=e^{-\beta(\epsilon-\mu)}$ is the Boltzmann factor with mode energy $\epsilon$, chemical potential $\mu$, and inverse temperature $\beta=1/k_BT$~\cite{wangParticleExchangeStatistics2025}. For the $m=2$ paraparticle emulated here, this counting gives an average site occupation $\braket{n} = 2x/(1+2x)$, distinct from both Bose--Einstein and Fermi--Dirac statistics (see supplementary materials).

To experimentally probe the occupation statistics, we thermalize a paraparticle site by embedding it in a 16-qubit lattice (Fig.~\ref{fig: thermalization}a). Two qubits (red and blue) are reserved to emulate the paraparticle, and the remaining $N=14$ qubits constitute the bath, providing an energetically dense set of states to thermalize the paraparticle. We further divide the bath into two sets of 7 qubits, alternating circles and squares, representing a bath for the red and blue qubits, respectively (Fig.~\ref{fig: thermalization}a). The red and blue paraparticle qubits are nearest-neighbor coupled to their neighboring bath qubits with strength $J_s$ (dashed line), and the bath qubits interact with one another across sets with strength $J_b$ (solid line). To ensure the two-excitation states of the paraparticle qubit pair remain inaccessible to bath excitations, the width of the bath spectrum, proportional to the bath-bath and bath-site couplings, $J_{b}$ and $J_{s}$, must remain smaller than the $ZZ$ splitting $2\zeta$.

We accumulate thermal statistics by measuring the red and blue qubit populations for different configurations of $n_0$ bath excitations (gray qubits), with $n_0$ running from 1 (single bath excitation) to 14 (all bath qubits excited). For a given $n_0$ excitations, we choose uniformly from all possible combinations of $n_0$ excited bath qubits. See supplementary materials and Fig.~\ref{fig: thermalization} caption for more details.

The experimental protocol is shown in Fig.~\ref{fig: thermalization}b. We begin with all qubits in their ground states and detuned in frequency with the red and blue qubits fixed at frequencies $\omega_{r}$ and $\omega_{b}$, respectively, to maintain a strong $ZZ$ interaction (see supplementary materials for the frequencies of the bath qubits). In this idling layout, we first populate the chosen combination of $n_0$ bath qubits. Next, we rapidly bring the square and circle bath qubits separately onto resonance at frequencies $\omega_{\Box}$ and $\omega_{\bigcirc}$, respectively, and activate the couplings by parametrically driving the bath circle qubits ($J_b/2\pi=1$~MHz) and the two paraparticle qubits ($J_s/2\pi=1$~MHz) at the appropriate difference frequencies (see Fig.~\ref{fig: thermalization}b). After a thermalization time $t=1$~\textmu s, we stop driving the interactions, return the qubits to their idling layout, and read out their populations, post-selecting on shots where the total number of excitations $n_0$ across all 16 qubits (to account for paraparticle qubit excitations from the bath qubits) is conserved. Examples of this thermalization evolution are shown in Fig.~\ref{fig: thermalization}c for $n_0=4$, $8$, and $12$ (see caption for description). The resulting average paraparticle population, defined as the sum of red and blue qubit populations, equilibrates to a value that depends on the number of excitations in the bath.

In this quantum simulation, the 16-qubit lattice is ideally a closed system and does not itself interact with an external bath. Nonetheless, the eigenstate thermalization hypothesis (ETH)~\cite{rigol2008ThermalizationItsMechanism, kaufmanETH2016, deutsch2018EigenstateThermalizationHypothesis} predicts the thermalization of local observables, such as the average occupation of bath and paraparticle qubits, in such closed systems. The average occupations shown in Fig.~\ref{fig: thermalization}c are consistent with this hypothesis, saturating to thermal values that depend on the total number of bath excitations $n_0$. In practice, the lattice is not an ideally closed system; the post-selection to $n_0$ excitations we employ helps account for depolarization with other external degrees of freedom.

We calculate the Boltzmann factor $x$ assuming the ETH, which predicts thermalization to a microcanonical ensemble defined by the total energy $E$ of the initial state. The system can then be approximated as a canonical ensemble with inverse temperature $\beta_e$ giving the same energy $E=\mathrm{Tr }(He^{-\beta_e H})/\mathrm{Tr }(e^{-\beta_e H})$, where $H$ is the total system Hamiltonian. The product states, which are the initial states of this experiment, carry zero energy under the effective hopping Hamiltonian, and thus give a canonical ensemble with $\beta_e=0$. At this infinite temperature, every state in the accessible fixed-$n_0$ sector becomes equally probable. Therefore, the effective Boltzmann factor $x=e^{-\beta(\epsilon-\mu)}$ that produces the partition function is given by $x = \Omega_{B}(1)/\Omega_{B}(0)$, where $\Omega_{B}(1)$ ($\Omega_{B}(0)$) is the Hilbert space dimension of the bath when the paraparticle site is occupied (unoccupied). Since $\Omega_{B}(k) = \binom{N}{n_0-k}$, where $N$ is the total number of bath qubits and $n_0$ is the total number of excitations in the system, the Boltzmann factor becomes $x=n_0/(N-n_0+1)$.

Using this approach, the saturated population of paraparticles (purple) is plotted versus $n_0$ (Fig.~\ref{fig: thermalization}d) and the corresponding Boltzmann factor (Fig.~\ref{fig: thermalization}e). For comparison, we also measure and plot the experimental results for fermions (black) and calculated results for bosons (dashed line). For the fermionic thermalization, we apply the same protocol to an alternative frequency layout that removes the $ZZ$ interaction between the paraparticle qubit pair and measures the mean occupation of those qubits (see supplementary materials).

The distinct thermal statistics of paraparticles can again be interpreted equivalently in both the qubit and paraparticle representations. In the qubit representation, the strong $ZZ$ interaction between the paraparticle qubit pair restricts the Hilbert space within which the system can thermalize. Similarly, in the paraparticle representation, the same saturated occupation is obtained by counting microstates in a system with a two-color paraparticle and a finite number of bath modes.

\section{Conclusion}

We have experimentally probed the quantum statistics of paraparticles, a recently proposed class of particles whose internal degree of freedom, color, is a global attribute that cannot locally distinguish individual particles. We focused on a fermion-like species that obeys a generalized (color-agnostic) exclusion principle. Using the generalized JW transformation~\cite{wangParticleExchangeStatistics2025}, we mapped a 1D chain of non-interacting paraparticles onto a ladder of qubits, with each paraparticle lattice site encoded by a pair of qubits coupled by a strong $ZZ$ interaction. Within this setup, we introduced and performed three experiments to investigate the paraparticle quantum statistics: the exchange of two free paraparticles, two-paraparticle interference at a 50:50 beamsplitter, and thermalization. Together, these measurements demonstrate that a strongly interacting qubit lattice can emulate the characteristic behavior of non-interacting paraparticles in a 1D chain.

A natural extension of this work is to introduce interactions between paraparticles. Nearest-neighbor paraparticle interactions in one dimension can be realized with the present qubit implementation, where additional $ZZ$ couplings between adjacent qubit pairs could realize a paraparticle Luttinger liquid~\cite{luttingerExactlySolubleModel1963,haldaneLuttingerLiquidTheory1981} within a fixed color sector. Such a fixed color sector, as used in the exchange and interference experiments of this work, may also interplay with other global structures, including boundary conditions~\cite{schurichtParastatisticsInteractingPeriodic2026} and the topology of the Hamiltonian, under which the color structure could duplicate protected modes, thereby increasing the available capacity of protected information. Going beyond a fixed sector, more exotic, potentially nonlocal interactions may give rise to color waves~\cite{salinelBosonization$R$paraparticleLuttinger2026}. The configurable connectivity and interactions of quantum simulators provide a natural setting for exploring these directions.

Furthermore, other species of paraparticles, particularly in higher dimensions, remain interesting but challenging to realize. For example, a type of paraparticle emerging in a 2D lattice is constructed within a constrained sector of a Hamiltonian involving eight-body interactions~\cite{wangParticleExchangeStatistics2025}. Repeated qubit-assisted measurements, a distinctive capability of quantum simulators, may offer a way to engineer the required sector by reducing the complexity of the interactions that must be implemented. Even for the system demonstrated in this work, measurements using an auxiliary qubit could detect and discard states with two or more photons on a qubit-pair site, analogous to stabilizing the code space in quantum error correction~\cite{terhalQuantumErrorCorrection2015}.

\section*{Acknowledgments}

The authors gratefully acknowledge R\'{e}ouven Assouly for his help operating the cryogenic system, Amir H. Karamlou for his initial contributions to the infrastructure, Christopher O’Connell for packaging and imaging the device, Zhiyuan Wang, Kaden Hazzard, and Terry P. Orlando for helpful discussions. Codex and Claude were used for the development of measurement and analysis routines, formatting plots, and proofreading the manuscript. The authors take full responsibility for the correctness of the reported results. This material is based upon work supported by the U.S. Department of Energy, Office of Science, National Quantum Information Science Research Centers, Quantum Systems Accelerator (Award No. DE-SCL0000121). Additional support is acknowledged from the Army Research Office Grant W911NF-23-1-0045 and Air Force Contract No. FA8702-15-D-0001. GK acknowledges support from Korea Foundation for Advanced Studies (KFAS). CNB acknowledges support from the National Science Foundation under award DMR-2141064. The views and conclusions contained herein are those of the authors and should not be interpreted as necessarily representing the official policies or endorsements, either expressed or implied, of the U.S. Government.

\section*{Author Contributions}

GK, MH, and CNB designed the experiment. GK conducted the experiment and analyzed the data with MH and CNB. CNB performed numerical simulations. GK developed the experimental infrastructure with assistance from JA, HZ, and ITR. SEM designed the device with layout assistance from MS. RD, MAG, DKK, and BMN fabricated the device with coordination from JLY, MES, and KS. JAG, MH, and WDO supervised the project, and all authors discussed the results. GK, CNB, JAG, MH, and WDO wrote the manuscript with input from all authors.

\putbib[ref]

\clearpage

\setcounter{equation}{0}
\setcounter{figure}{0}
\setcounter{table}{0}
\setcounter{section}{0}
\setcounter{page}{1}

\renewcommand{\theequation}{S\arabic{equation}}
\renewcommand{\thefigure}{S\arabic{figure}}
\renewcommand{\thetable}{S\arabic{table}}
\renewcommand{\thesection}{S\arabic{section}}
\renewcommand{\thepage}{S\arabic{page}}

\renewcommand{\theHequation}{S\arabic{equation}}
\renewcommand{\theHfigure}{S\arabic{figure}}
\renewcommand{\theHtable}{S\arabic{table}}
\renewcommand{\theHsection}{S\arabic{section}}

\title{Supplementary Material for ``Analog quantum simulation of paraparticle statistics''}

\let\addcontentsline\oldaddcontentsline

\clearpage

\maketitle

\onecolumngrid

\vspace{-1.5cm}
\tableofcontents

\clearpage

\begin{bibunit}[apsrev4-2]

\section{Experimental setup}

\subsection{Device parameters}

We use a $4\times4$ array of capacitively coupled transmons. Each transmon is flux-tunable via an asymmetric SQUID. Across the 16 qubits, the maximum and minimum qubit frequencies span the ranges $f_q^\mathrm{max} \in [3.783, 4.035]~\mathrm{GHz}$ and $f_q^\mathrm{min}\in [2.149, 2.462]~\mathrm{GHz}$. The anharmonicities are $\eta/2\pi\in [212, 230]~\mathrm{MHz}$ at the upper sweet spots. The resonator frequencies are $f_r\in [5.976, 6.350]~\mathrm{GHz}$, with the corresponding qubits at their upper sweet spots. The qubits are divided between two feedlines for multiplexed readout: QB1-QB6, QB9, and QB10 are connected to the first feedline, whereas QB7, QB8, and QB11-QB16 are connected to the second. Table~\ref{tab:sample_parameters} lists the individual device parameters.

\begin{table*}[ht]
\centering
{\renewcommand{\arraystretch}{1.5}
\begin{tabular}{l*{16}{c}}
\toprule
Parameter & QB1 & QB2 & QB3 & QB4 & QB5 & QB6 & QB7 & QB8 & QB9 & QB10 & QB11 & QB12 & QB13 & QB14 & QB15 & QB16 \\
\midrule
$f_{{q}}^{\mathrm{max}}$ (GHz) & $4.012$ & $4.035$ & $3.884$ & $3.993$ & $3.973$ & $3.922$ & $3.823$ & $3.842$ & $3.882$ & $3.959$ & $3.800$ & $3.783$ & $3.893$ & $3.898$ & $3.933$ & $3.901$ \\
$f_{{q}}^{\mathrm{min}}$ (GHz) & $2.277$ & $2.308$ & $2.342$ & $2.411$ & $2.301$ & $2.392$ & $2.300$ & $2.271$ & $2.349$ & $2.381$ & $2.149$ & $2.237$ & $2.348$ & $2.233$ & $2.286$ & $2.462$ \\
$\eta/2\pi$ (MHz) & $230$ & $227$ & $225$ & $222$ & $229$ & $221$ & $217$ & $215$ & $228$ & $220$ & $217$ & $214$ & $226$ & $220$ & $212$ & $215$ \\
$f_r$ (GHz) & $6.102$ & $6.224$ & $6.083$ & $6.350$ & $6.263$ & $6.144$ & $6.099$ & $6.029$ & $6.175$ & $6.268$ & $6.008$ & $6.072$ & $6.344$ & $6.002$ & $6.052$ & $5.976$ \\
\bottomrule
\end{tabular}
}
\caption{Qubit parameters. Maximum and minimum qubit transition frequencies, $f_q^\mathrm{max}$ and $f_q^\mathrm{min}$, anharmonicity $\eta/2\pi$, and resonator frequency $f_r$. The anharmonicities and resonator frequencies are measured with the qubits at their upper sweet spots.}
\label{tab:sample_parameters}
\end{table*}

This device features strong nearest-neighbor couplings, which are required to obtain a sufficiently large blockade against double occupancy. We characterize the capacitive interaction by fitting avoided crossings measured at several bias frequencies to values derived by diagonalizing the two-transmon Hamiltonian~\cite{kochChargeinsensitiveQubitDesign2007}:
\begin{equation}\label{eq:circuit-level two transmons}
    H_{\mathrm{NN}}=\sum_{i=1}^{2}\left[ 4E_{C,i}\hat n_i^2 -E_{J,i}(\Phi_i)\cos\left(\hat\varphi_i+\varphi_{0,i}(\Phi_i)\right) \right] +J_C\hat n_1\hat n_2,
\end{equation}
where $\hat n_i$ and $\hat\varphi_i$ are the charge and phase operators, $\Phi_{i}$ is the net external flux, $\Phi_0=h/(2e)$ is the flux quantum, $E_{J,i}(\Phi_i)$ and $\varphi_{0,i}(\Phi_i)$ are the flux-dependent effective Josephson energy and phase offset of the SQUID, respectively, and $J_C$ is the charge-charge coupling strength of the qubit pair.

\begin{figure*}[ht]
\centering
\includegraphics{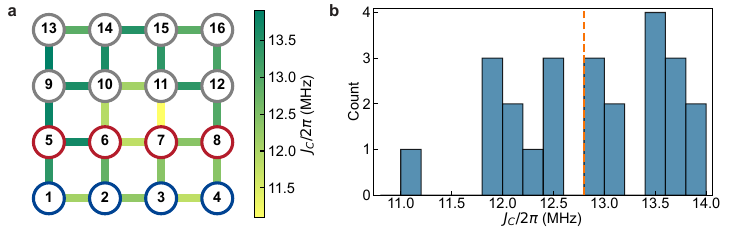}
\caption{\textbf{a,} Capacitive coupling strengths $J_C/2\pi$ between nearest-neighbor qubits. QB1-QB8 are used to form the paraparticle sites in the exchange and interference experiments.
\textbf{b,} Histogram of these coupling strengths. The dashed line indicates the average.}
\label{fig: supp. bare couplings}
\end{figure*}

Figure~\ref{fig: supp. bare couplings} shows the measured $J_C/2\pi$ values and their histogram.

\subsection{Electronics}\label{sec: electronics}

\begin{figure*}[ht]
\centering
\includegraphics[width=0.85\textwidth]{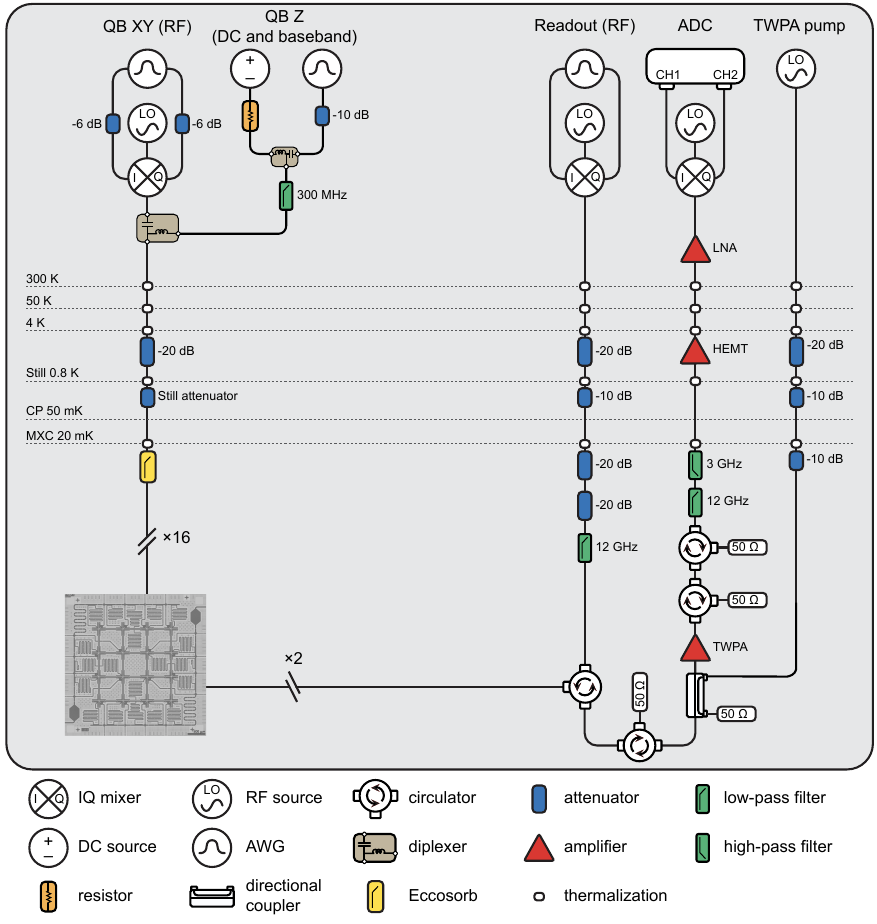}
\caption{The wiring diagram.}
\label{fig: supp. wiring diagram}
\end{figure*}

The wiring diagram is shown in Fig.~\ref{fig: supp. wiring diagram}. Three qubit control signals in different frequency ranges---DC, baseband, and RF---are generated separately and sent down through a common flux line to a qubit. The DC bias is generated by a QDAC-I, all baseband and I/Q signals for RF pulses by Keysight M3202A arbitrary waveform generators (AWGs), and all RF signals by Rohde \& Schwarz SGS100A local oscillators. The readout signals are downconverted and digitized by Keysight M3102A digitizers. Inside the dilution refrigerator, each control line includes a \qty{20}{\dB} attenuator below the 4 K stage and either \qty{6}{\dB} or \qty{10}{\dB} of attenuation below the still plate, separately chosen per line to maximize noise suppression while avoiding overheating the system. A low-cutoff Eccosorb filter (Quantum Microwave QMC-CRYOIRF-001) is placed at the mixing-chamber stage to further suppress high-frequency noise. The values of the room temperature resistors and still attenuators are provided in Table~\ref{tab:resistor_and_atten}.

\begin{table*}[ht]
\centering
{\renewcommand{\arraystretch}{1.5}
\begin{tabular}{l*{16}{c}}
\toprule
Parameter & QB1 & QB2 & QB3 & QB4 & QB5 & QB6 & QB7 & QB8 & QB9 & QB10 & QB11 & QB12 & QB13 & QB14 & QB15 & QB16 \\
\midrule
RT resistor ($\Omega$) & $500$ & $500$ & $500$ & $500$ & $500$ & $500$ & $500$ & $1000$ & $500$ & $500$ & $1000$ & $1000$ & $500$ & $500$ & $500$ & $500$ \\
Still attenuation (dB) & $10$ & $6$ & $6$ & $6$ & $6$ & $10$ & $10$ & $10$ & $6$ & $10$ & $10$ & $6$ & $6$ & $10$ & $6$ & $10$ \\
\bottomrule
\end{tabular}
}
\caption{The room-temperature resistor and still attenuator values on each qubit-control line.}
\label{tab:resistor_and_atten}
\end{table*}

\section{System calibration and control}

\begin{figure*}[ht]
\centering
\includegraphics[width=0.8\textwidth]{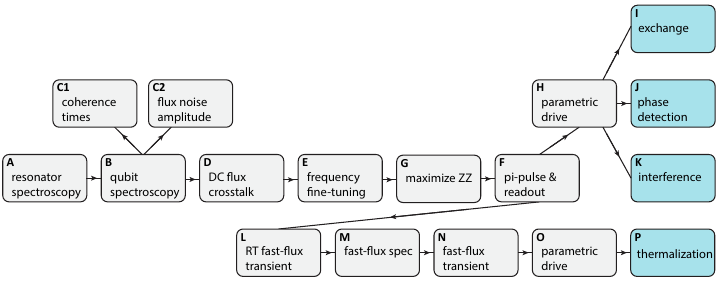}
\caption{The overall calibration process. Gray boxes are the preliminary calibration steps explained throughout this section. Blue boxes are the paraparticle experiments introduced in the main text and discussed in more detail in the following sections.}
\label{fig: supp. calibration}
\end{figure*}

The overall calibration routine is shown in Fig.~\ref{fig: supp. calibration}. In this section, we describe common calibration routines for the paraparticle experiments in this work. We employ iterative methods for DC-bias and parametric-drive calibrations to achieve both accuracy and efficiency. We refer the reader to Refs.~\cite{krantzQuantumEngineersGuide2019,naghilooIntroductionExperimentalQuantum2019,Karamlou2024EntanglementHCBH} for procedures not described here in detail, including single-qubit gate calibration, readout calibration, and noise characterization.

\subsection{Spectroscopy and DC flux crosstalk}

\begin{figure*}[ht]
\centering
\includegraphics{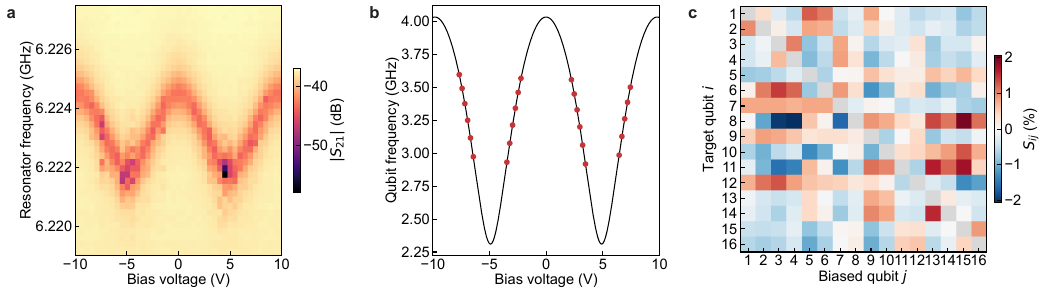}
\caption{\textbf{a,} Resonator spectroscopy. We probe the resonator frequency as a function of qubit bias voltage.
\textbf{b,} Qubit spectroscopy. We probe the qubit frequency as a function of qubit bias voltage. The line shows a fit to the model in Eq.~\eqref{eq:transmon model}.
\textbf{c,} DC crosstalk sensitivity matrix. The diagonal elements, $S_{ii}=1$, are filled with gray.}
\label{fig: supp. spec and dc xtalk}
\end{figure*}

We read out the qubits through two common feedlines supporting eight qubits each. Therefore, the scattering parameter of each feedline shows eight resonator dips, distinguished by the resonance frequencies. To pair up each resonator feature with the corresponding qubit, we first measure the resonator spectrum while biasing one of the qubits (Fig.~\ref{fig: supp. spec and dc xtalk}a), and find the resonator whose frequency responds to the bias. This step gives us a lookup table of the resonator frequency versus bias voltage of each qubit.

Next, we perform a two-tone spectroscopy to probe the qubit frequency as a function of bias voltage. While keeping the other qubits' biases at zero, we sweep the voltage $V_i$ and fit the qubit frequency $f_{q,i}$ to the following model~\cite{kochChargeinsensitiveQubitDesign2007}:
\begin{equation}\label{eq:transmon model}
\begin{aligned}
f_{q,i} &= \left(f^\mathrm{max}_{q,i} + \frac{E_{C,i}}{h}\right) \left[d_i^2 + (1-d_i^2)\cos^2\left(\pi\frac{\Phi_i}{\Phi_0}\right)\right]^{1/4} - \frac{E_{C,i}}{h}, \\
\frac{\Phi_i}{\Phi_0} &= \frac{V_i}{V^{\Phi_0}_i}+\frac{\Phi^\mathrm{offset}_i}{\Phi_0},
\end{aligned}
\end{equation}
where $\Phi_i$ is the external flux applied to the qubit. We fit the spectroscopy data to the frequency curve (Fig.~\ref{fig: supp. spec and dc xtalk}b) with five free parameters: the upper sweet spot frequency $f^\mathrm{max}_{q,i}$, charging energy $E_{C,i}$, junction asymmetry $d_i$, voltage per flux quantum $V^{\Phi_0}_i$, and flux offset $\Phi^\mathrm{offset}_i$. Because all sixteen qubits have similar upper sweet spot frequencies, we measure the shoulder of the curve to avoid frequency collision at the peak. The exact sweet spots can be characterized by biasing all the other qubits away from their sweet spots, which is done in a later step when we perform flux-transient calibration.

After the two-tone spectroscopy of individual qubits, we characterize the DC crosstalk of flux biases. The current injected through the flux line of a qubit returns through distributed paths in the shared ground plane and produces unintended magnetic flux in other SQUID loops. Due to this mechanism, the flux crosstalk is generally asymmetric, and its magnitude need not vary monotonically with physical separation.

We use a linear model to relate the applied voltages to the fluxes threading the SQUID loops~\cite{barrett2023LearningBasedCalibrationFlux}. Denoting the normalized response of qubit $i$ to flux line $j$ by $S_{ij}$, the full model is
\begin{equation}\label{eq:dc xtalk}
\frac{\Phi_i}{\Phi_0}=\sum_j \frac{S_{ij} V_j}{V^{\Phi_0}_i} + \frac{\Phi^\mathrm{offset}_i}{\Phi_0},
\end{equation}
where $j$ runs over all flux lines, and $S_{ii}=1$ by definition of $V_{i}^{\Phi_0}$ in the second line of Eq.~\eqref{eq:transmon model}. We measure $S_{ij}$ by probing the sensitivity of qubit $i$ to the bias voltage $V_{j}$ directly, parking the probed qubit far from either sweet spot to leverage flux sensitivity. The measured DC crosstalk matrix is shown in Fig.~\ref{fig: supp. spec and dc xtalk}c.

\subsection{Frequency fine-tuning}\label{sec:frequency fine-tuning}

\begin{figure*}[ht!]
\centering
\includegraphics{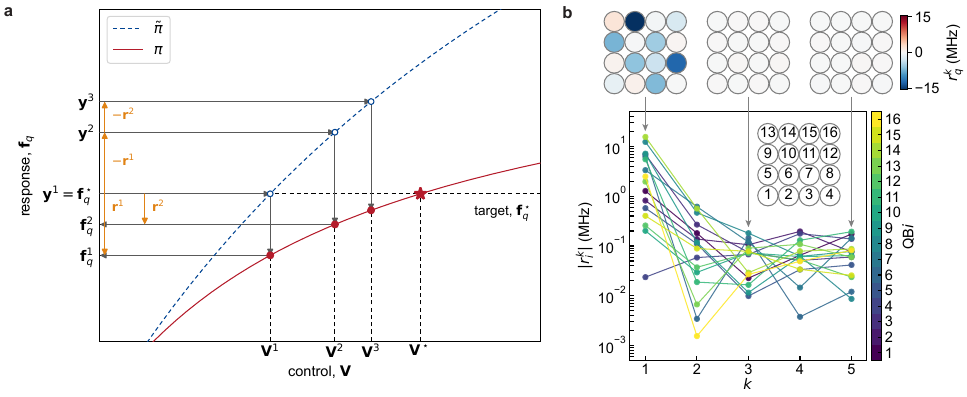}
\caption{\textbf{a,} Iterative process for fine-tuning qubit frequency. Let $\pi$ be the true relation between bias voltage $\mathbf{V}$ and frequency $\mathbf{f}_q$ of all qubits in vector form. The individual transmon model with the DC crosstalk compensation provided in Eqs.~\eqref{eq:transmon model} and \eqref{eq:dc xtalk} gives an approximation, denoted by $\tilde \pi$. The objective is to find voltage $\mathbf{V}^\star$ that gives the target frequency $\mathbf{f}_q^\star$ in the real system: $\mathbf{f}_q^\star = \pi(\mathbf{V}^\star)$. The iteration starts from the initial guess, by applying $\mathbf{V}^1=\tilde\pi^{-1}(\mathbf{f}_q^\star)$ to the system. Using the measured frequency $\mathbf{f}^1_q$, we adjust the voltage to $\mathbf{V}^2=\tilde\pi^{-1}(\mathbf{f}_q^\star-\mathbf{r}^1)$, where $\mathbf{r}^1=\mathbf{f}^1_q-\mathbf{f}^\star_q$ is the residual. We probe the frequencies at $\mathbf{V}^2$, and further adjust the voltage following Eqs.~\eqref{eq:freq iteration 1}-\eqref{eq:freq iteration 3}. As the red dots show, given that $\pi$ and $\tilde\pi$ are close enough, the iteration converges to the answer $\mathbf{V}^\star$ that gives the target frequency layout $\mathbf{f}_q^\star$ when applied to the real system.
\textbf{b,} The residual frequency over iteration. The plot exemplifies how the residual is corrected in this iterative process. The presented data is from one of the thermalization experiments, where we place the qubit at the fermionic state preparation and readout frequency layout shown in Fig.~\ref{fig: supp. thermalization frequency layout}a. The maximum residual decays from $>\qty{10}{\MHz}$ to around $\qty{100}{\kHz}$ within three rounds. The measurement uncertainty dominates the residuals in subsequent rounds.}
\label{fig: supp. iterative frequency}
\end{figure*}

Individual two-tone spectroscopy measurements and the DC crosstalk characterization yield a map from bias voltages to qubit frequencies. Given a set of desired qubit frequencies, we can invert the map to find the appropriate bias voltages. However, the inverted map is not accurate enough to set the qubits to the target frequencies with a precision better than $\qty{1}{MHz}$, which is the strength of the effective coupling between the qubits in the main experiments. This inaccuracy is partly attributed to the strong capacitive coupling between neighboring qubits, which hybridizes the qubit modes. For example, if $\Delta\approx2\pi\times\qty{500}{\MHz}$ is the detuning between a qubit pair with a coupling strength $g\approx2\pi\times\qty{15}{\MHz}$, the dispersive shift is $g^2/\Delta\approx2\pi\times\qty{0.45}{MHz}$.

Instead of adding extra complexity to the model by including couplings, we use an imperfect model and correct the error iteratively to find the accurate bias voltages for a target frequency layout. The method is illustrated in Fig.~\ref{fig: supp. iterative frequency}a. Let us denote by $\tilde\pi$ the map from the voltages $\mathbf{V}=(V_1,\ldots,V_{16})$ to the qubit frequencies $\mathbf{f}_q=(f_{q,1},\ldots,f_{q,16})$ obtained from the DC crosstalk characterization in Eqs.~\eqref{eq:transmon model} and \eqref{eq:dc xtalk}. First, we invert the model $\tilde \pi$ to get voltages $\mathbf{V}^1=\tilde\pi^{-1}(\mathbf{f}_q^\star)$ for a target frequency layout $\mathbf{f}_q^\star$. Next, we apply $\mathbf{V}^1$ and probe the qubit frequencies $\mathbf{f}^1_q$. After calculating the residual $\mathbf{r}_{1}=\mathbf{f}_q^1 - \mathbf{f}_q^\star$, we adjust the target by the same amount but in the opposite direction of the residual to obtain the next voltage $\mathbf{V}^2=\tilde\pi^{-1}(\mathbf{f}_q^\star - \mathbf{r}^1)$. The iteration proceeds by applying $\mathbf{V}^2$ again and probing the qubit frequencies $\mathbf{f}_q^2$. The new residuals $\mathbf{r}^2=\mathbf{f}_q^2 - \mathbf{f}_q^\star$ are used to further adjust the target to obtain the voltage $\mathbf{V}^3=\tilde\pi^{-1}(\mathbf{f}_q^\star - \mathbf{r}^1 - \mathbf{r}^2)$ for the next trial. These steps are repeated until the probed frequencies converge to the target $\mathbf{f}_q^\star$ within the measurement uncertainty.

The iteration gives the target frequency layout when it converges. Denoting the true relation between the bias voltages and the frequency layout by $\pi:\mathbf{V}\mapsto\mathbf{f}_q$, the iteration can be written as
\begin{align}
    \text{Estimate:}\quad & \mathbf{V}^k     = \tilde{\pi}^{-1}(\mathbf{y}^k),\label{eq:freq iteration 1} \\
    \text{Measure:}\quad  & \mathbf{r}^k     = \pi(\mathbf{V}^k) - \mathbf{f}_q^\star, \label{eq:freq iteration 2}\\
    \text{Adjust:}\quad   & \mathbf{y}^{k+1} = \mathbf{y}^{k} - \mathbf{r}^k,\label{eq:freq iteration 3}
\end{align}
where we define $\mathbf{y}^1 = \mathbf{f}_q^\star$. By merging the three equations, we obtain a fixed-point iteration $T$:
\begin{equation}
    \mathbf{V}^{k+1} = \tilde\pi^{-1}\left[\mathbf{f}_q^\star + \tilde \pi(\mathbf{V}^k) - \pi(\mathbf{V}^k)\right] \equiv T(\mathbf{V}^{k}).
\end{equation}
Here, the fixed-point solution $\mathbf{V}^\star$ satisfying $T(\mathbf{V}^\star)=\mathbf{V}^\star$ also satisfies $\pi(\mathbf{V}^\star)=\mathbf{f}^\star_q$, which means that these bias voltages realize the target frequency layout when applied to the system.

Figure \ref{fig: supp. iterative frequency}b shows the measured residual $\mathbf{r}^k$ versus iteration. The process converges within three steps, with the initial error of up to $\approx\qty{10}{\MHz}$ suppressed to $\lesssim\qty{100}{\kHz}$. After the third round, measurement uncertainty from, e.g., two-tone spectroscopy dominates the residual.

\subsection{Maximizing the intra-pair $ZZ$ interaction}\label{sec: maximizing ZZ}

\begin{figure*}[h!]
\centering
\includegraphics[width=\textwidth]{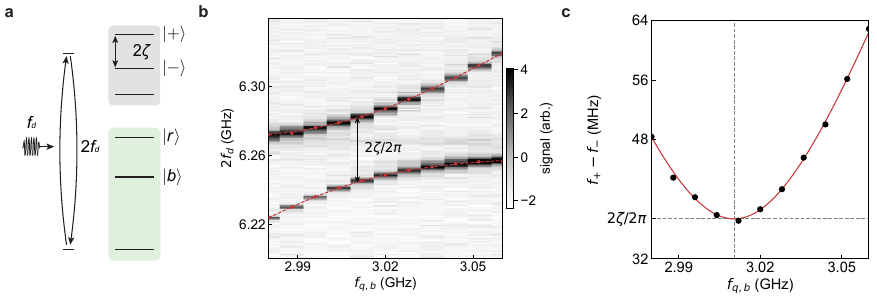}
\caption{\textbf{a,} Spectroscopy of two-photon transitions. To perform spectroscopy of the two-photon dressed states $\ket{+}$ and $\ket{-}$, we drive the system with frequency $f_d$, which is half of the state frequency that we want to probe. The avoided crossing between $\ket{+}$ and $\ket{-}$ equals twice the $ZZ$ interaction, $2\zeta$, when the bare $\ket{11}$ and $\ket{20}$ are maximally hybridized.
\textbf{b,} Spectroscopy of the states $\ket{+}$ and $\ket{-}$, sweeping one of the qubit frequencies ($f_{q,b}$) while keeping the other qubit frequency ($f_{q,r}$) constant.
\textbf{c,} Frequency difference between the dressed states $\ket{+}$ and $\ket{-}$. The data is fitted to a hyperbola to extract the optimal $f_{q,b}$ that maximizes the $ZZ$ interaction.}
\label{fig: supp. intra pair ZZ}
\end{figure*}

A key element of the paraparticle site implementation in this work is the large intra-pair blockade that originates from the $\ket{11}$--$\ket{20}$ avoided crossing. The $ZZ$ interaction $\zeta$, which is maximized when the states are fully hybridized to
\begin{equation}
    \ket{\pm} = \frac{1}{\sqrt{2}}\left(\ket{11} \pm \ket{20}\right),
\end{equation}
where $\ket{11}$ and $\ket{20}$ denote bare states, must be much larger than the effective inter-site coupling $J$ to isolate the paraparticle subspace at each site (the pair of qubits).

As discussed in the main text, the maximum $\zeta$ is achieved when the qubits are detuned by the anharmonicity of the higher-frequency qubit. To show this, let us consider the paraparticle pair qubits at frequencies $\omega_r=2\pi f_{q,r}$ and $\omega_b = 2\pi f_{q,b}$ and assume $\omega_r > \omega_b$. We denote their transverse coupling by $g$, and their anharmonicities by $\eta_r$ and $\eta_b$. The effective Hamiltonian describing the qubit pair then becomes
\begin{equation}\label{eq:qubit pair Hamiltonian}
    H_\mathrm{pair} = \sum_{\alpha\in\{r,b\}} \left(\omega_\alpha a_\alpha^\dagger a_\alpha - \frac{\eta_\alpha}{2}a_\alpha^\dagger a_\alpha^\dagger a_\alpha a_\alpha\right)+ g(a_r^\dagger a_b + a_b^\dagger a_r),
\end{equation}
where $a_\alpha^\dagger$ adds an excitation to qubit $\alpha$.

Here, we can write the $ZZ$ interaction as
\begin{equation}
    \zeta = \min_{s\in\{+,-\}} \left\lvert \omega_{\ket{s}} - \omega_{\ket{\tilde{10}}} -\omega_{\ket{\tilde{01}}} + \omega_{\ket{00}}\right\rvert,
\end{equation}
where $\ket{\tilde{10}}$ and $\ket{\tilde{01}}$ are single-photon dressed states. When $\omega_r-\omega_b=\eta_r$ is satisfied, the bare $\ket{11}$ and $\ket{20}$ are on resonance, and thus fully hybridized. The maximal $ZZ$ interaction becomes: $\zeta_{\max}\sim\sqrt{2}g=\bra{20}g(a_r^\dagger a_b + a_b^\dagger a_r)\ket{11}$.

In the experiment, we probe the spectrum of dressed states $\ket{+}$ and $\ket{-}$ as a function of $\omega_b$ while keeping $\omega_r$ constant to find the optimal value of $\omega_b$ that maximizes $\zeta$. We probe the two-photon states using a two-photon process, in which the drive frequency is half of the state frequency (Fig.~\ref{fig: supp. intra pair ZZ}a). The avoided crossing between the two branches is clearly shown in Fig.~\ref{fig: supp. intra pair ZZ}b, and fitted to a hyperbola in Fig.~\ref{fig: supp. intra pair ZZ}c. The qubit frequency layout and the maximal $ZZ$ interactions used in the exchange and interference experiments are shown below.
\begin{table*}[ht]
\centering
{\renewcommand{\arraystretch}{1.5}
\begin{tabular*}{0.35\textwidth}{@{\extracolsep{\fill}}l*{3}{c}}
\toprule
Parameter & Pair 1 & Pair 2 & Pair 3 \\
\midrule
$f_{q,r}$ (GHz) & 3.2000 & 3.9000 & 3.2500 \\
$f_{q,b}$ (GHz) & 2.9547 & 3.6672 & 3.0104 \\
$\zeta_{\max}/2\pi$ (MHz) & 17.1 & 22.4 & 18.7 \\
\bottomrule
\end{tabular*}
}
\caption{Frequency layout and maximal $ZZ$ interaction used in the exchange and phase detection experiments.}
\end{table*}
\begin{table*}[ht]
\centering
{\renewcommand{\arraystretch}{1.5}
\begin{tabular*}{0.45\textwidth}{@{\extracolsep{\fill}}l*{4}{c}}
\toprule
Parameter & Pair 1 & Pair 2 & Pair 3 & Pair 4 \\
\midrule
$f_{q,r}$ (GHz) & 3.2700 & 3.8000 & 3.1500 & 3.6800 \\
$f_{q,b}$ (GHz) & 3.0248 & 3.5662 & 2.9088 & 3.4501 \\
$\zeta_{\max}/2\pi$ (MHz) & 17.3 & 21.7 & 18.2 & 19.1 \\
\bottomrule
\end{tabular*}
}
\caption{Frequency layout and maximal $ZZ$ interaction used in the interference experiment.}
\end{table*}

\clearpage

\subsection{Baseband transient calibration}

\begin{figure*}[ht!]
\centering
\includegraphics[width=\textwidth]{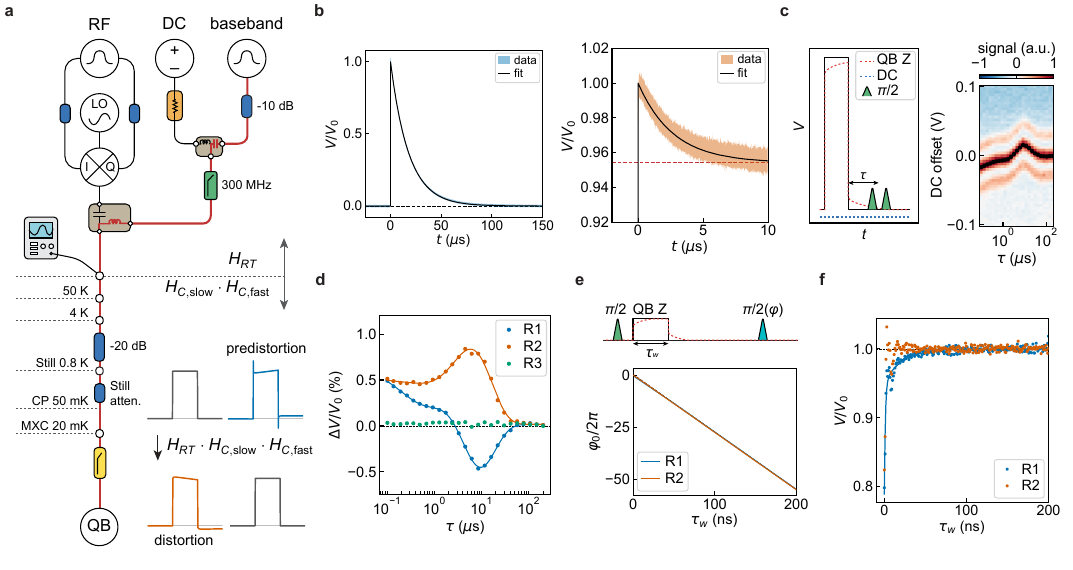}
\caption{\textbf{a,} The wiring diagram for the qubit control signal. The red path shows the baseband path that we characterize and compensate for the pulse distortion. The baseband signal passes through two bias-tees (fast and slow signal paths, respectively), attenuator, and the filter outside the dilution refrigerator. Inside the refrigerator, there are two attenuators and an Eccosorb IR filter on the path. These electrical components and other features of the control line can distort the baseband pulse profile depicted as an orange line in the bottom-right inset. The signal can be ``predistorted'' (blue line) to compensate for the distortion, by applying the inverse transfer function of the signal path to the ideal waveform. We model the signal path as three cascaded transfer functions: the room temperature response $H_\mathrm{RT}$, and slow- and fast-cryogenic responses $H_\mathrm{C,slow}$ and $H_\mathrm{C,fast}$. $H_\mathrm{RT}$ is characterized with an oscilloscope by directly measuring the room-temperature path before the signal enters the dilution refrigerator. $H_{\mathrm{C,slow}}$ and $H_{\mathrm{C,fast}}$ are characterized by probing the qubit, as explained below.
\textbf{b,} Characterization of the room-temperature response $H_{\mathrm{RT}}$. The plots show the step response and fit to an exponential decay for the two consecutive rounds of measurements.
\textbf{c,} Characterization of the slow-cryogenic response $H_{\mathrm{C,slow}}$. The system response is measured by probing the residual response of a rectangular pulse. With a qubit placed at a flux-sensitive bias point, a rectangular pulse is played with a small DC bias on top of it. The two consecutive $\pi/2$ pulses following the rectangular pulse form a ``Ramsey probe'', which is calibrated to populate the qubit when there is no baseband or DC bias offset. By sweeping the small DC bias with a fixed $\tau$, we can measure the residual at $\tau$ by finding the DC offset that cancels out the residual response and results in the most population after the Ramsey probe.
\textbf{d,} Slow-cryogenic responses. Each round is measured after applying compensations from the previous rounds. The flat signal in the third round indicates that the predistortion from the first two rounds cancels out most of the signal distortion.
\textbf{e,} The cryoscope method characterizing the fast-cryogenic response $H_{\mathrm{C,fast}}$~\cite{rolTimedomainCharacterizationCorrection2020a,hellingsCalibratingMagneticFlux2025a}. A Ramsey sequence encompassing a baseband pulse is applied to a qubit at its upper sweet spot. We measure the accumulated phase by sweeping the pulse width $\tau_w$ with the minimal step size that the AWG allows ($\qty{1}{ns}$). The plot shows the accumulated phase as a function of $\tau_w$.
\textbf{f,} Fast-cryogenic responses. The voltage response is extracted from the phase accumulation by differentiating the signal shown in \textbf{e}. The signal from the first round is fitted to a sum of two exponential decays with an offset and used to characterize $H_\mathrm{{C,fast}}$.}
\label{fig: supp. cryoscope}
\end{figure*}

In this section, we discuss the characterization of baseband signal distortion and how we compensate for it. As shown in Fig.~\ref{fig: supp. cryoscope}a and explained in Section \ref{sec: electronics}, each qubit in this work is controlled through an individual flux line carrying the DC, baseband, and RF signals generated from different sources, which are combined outside the dilution refrigerator (i.e., the room temperature). The signal therefore passes through a variety of electronic components on its path to the qubit, which distort the signal from its original profile. The baseband square pulse suffers the most from this distortion, because it spans across a wide range of low frequencies unlike DC bias or RF pulses. To mitigate the distortion, we characterize the transfer function of the signal path and invert it to ``predistort" the waveform~\cite{hofheinzSynthesizingArbitraryQuantum2009,guoUniversalScalableCharacterization2024,an$ZZ$freeTwotransmon$CZ$2026,rolTimedomainCharacterizationCorrection2020a,hellingsCalibratingMagneticFlux2025a}, assuming the system is linear and time-invariant. The system response of the baseband signal path (red in Fig.~\ref{fig: supp. cryoscope}a) is divided into three components---room-temperature, slow-cryogenic, and fast-cryogenic parts---to be individually characterized.

The first part, the response of the room-temperature electronics, is characterized using an oscilloscope. To combine with the DC signal, the baseband pulse is routed to the high-frequency port of the first bias-tee on the path (Fig.~\ref{fig: supp. cryoscope}a), which has a zero DC gain. Assuming the system of combined electronics is linear and time-invariant, we model the response function as a cascade of two single-pole systems:
\begin{align}
    H_\mathrm{RT}(s) &= H_{\mathrm{RT},1}(s)H_{\mathrm{RT},2}(s),\\
    H_{\mathrm{RT},1}(s) &= A_1\frac{s}{s+1/\tau_1},\\
    H_{\mathrm{RT},2}(s) &= C_2+A_2\frac{s}{s+1/\tau_2}.
\end{align}
For a unit-step input, these transfer functions give
\begin{align}
y_1(t) &= A_1 e^{-t/\tau_1},\\
y_2(t) &= C_2+A_2e^{-t/\tau_2},
\end{align}
where $y_i(t)$ is the ratio of measured voltage $V(t)$ to the step height $V_0$. The coefficients are extracted by fitting the step response to these exponential decays. We first characterize $H_{\mathrm{RT},1}$, and proceed to $H_{\mathrm{RT},2}$ after compensating for $H_{\mathrm{RT},1}$, as illustrated in the plots of Fig.~\ref{fig: supp. cryoscope}b.

Next, we characterize the slow-cryogenic response on $\qty{100}{\ns}$ to $\qty{1}{\ms}$ timescales using DC bias and Ramsey probe~\cite{hofheinzSynthesizingArbitraryQuantum2009,guoUniversalScalableCharacterization2024,an$ZZ$freeTwotransmon$CZ$2026}. The measurement sequence is shown in Fig.~\ref{fig: supp. cryoscope}c. Placing a qubit at a flux-sensitive frequency, a rectangular pulse of a fixed width is applied. At the same time, a small DC bias is applied as shown by the blue dotted line in Fig.~\ref{fig: supp. cryoscope}c. Then, we probe the qubit frequency using a short Ramsey sequence with a consecutive pair of $\pi/2$ pulses. The calibrated Ramsey sequence is sensitive to qubit frequency, i.e., the final population quickly falls off from $1$ when the qubit frequency is detuned, and is therefore used as a probe in this sequence. Using the probe, we can find the DC bias that maximizes the final population, which means that the residual baseband signal is canceled out. The measured fringe pattern is shown in Fig.~\ref{fig: supp. cryoscope}c, the second panel. The transfer function is constructed from the response by fitting the curve to a sum of four exponential functions, assuming no complex pole. To ensure stability of the inverse transfer function, we force the fit to have time constants separated by at least a factor of $3$. We have empirically found that adding the second round fitting with a three-exponential sum helps to further compensate the residual. The measured response after each round R1, R2, and the final validation R3, is plotted in Fig.~\ref{fig: supp. cryoscope}d.

Finally, we apply the cryoscope method~\cite{rolTimedomainCharacterizationCorrection2020a,hellingsCalibratingMagneticFlux2025a} to characterize fast transients down to approximately $\qty{100}{\ns}$. The pulse sequence is illustrated in Fig.~\ref{fig: supp. cryoscope}e. First, the target qubit is parked at its upper sweet spot while the other qubits are detuned by more than $\qty{500}{MHz}$. Next, a square pulse of width $\tau_w$ is applied between two $\pi/2$ pulses separated by a fixed delay. The accumulated phase is measured while increasing $\tau_w$ in steps of $\qty{1}{ns}$ corresponding to the AWG sampling period. We obtain the contribution of each time slice to the phase by calculating the gradient of accumulated phase, which is directly convertible to the time-dependent voltage seen by the qubit. A key feature of this method is that the phase accumulation due to the transient after the pulse is negligible, because the idling point is the upper sweet spot. The results are plotted in Fig.~\ref{fig: supp. cryoscope}f. We find that a sum of two exponential functions with no finite-impulse response (FIR) is sufficient to fit and mitigate the residual fast-timescale transient for the thermalization experiment.

\clearpage

\subsection{Parametric drive calibration}
\label{sec: parametric drive calibration}

\subsubsection{Parametric coupling}

The neighboring transmons in the chip are capacitively coupled as described in Eq.~\eqref{eq:circuit-level two transmons}. To implement tunable inter-site couplings essential to the experiments presented in the main text, we parametrically couple neighboring qubit pairs by detuning them and applying a time-dependent flux at the detuning frequency~\cite{strandFirstorderSidebandTransitions2013a,rosenSyntheticMagneticVector2024}. In this section, we discuss the working principle and the iterative calibration method of the parametric coupling used in this work.

\begin{figure*}[ht]
\centering
\includegraphics[width=0.7\textwidth]{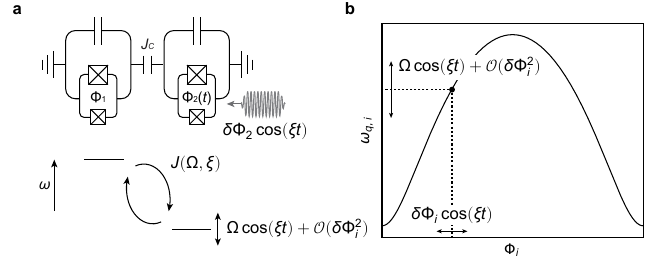}
\caption{\textbf{a,} Two capacitively coupled transmons. The center capacitor gives the charge-charge coupling term, $J_C\hat n_1 \hat n_2$, as discussed in Eq.~\eqref{eq:circuit-level two transmons}. In the parametric coupling scheme, one of the qubits is driven by modulating the flux. We assume in this section that the second-qubit flux is modulated as $\delta \Phi_2 \cos(\xi t)$. The bottom plot shows the transition frequency between $\ket{0}$ and $\ket{1}$ of each qubit. With the qubits initially detuned, the flux drive on the second qubit induces an effective frequency modulation $\Omega\cos(\xi t)$ with a second-order drive-induced frequency shift. Both the modulation amplitude $\Omega$ and frequency $\xi$ determine the effective coupling strength $J(\Omega,\xi)$ at a given detuning between qubit transition frequencies.
\textbf{b,} Origin of the drive-induced frequency shift. The plot shows the qubit transition frequency as a function of flux. Because the qubit frequency depends parametrically on the flux, a harmonic modulation on flux leads to the same harmonic modulation on frequency to the first order with additional higher-order terms including the second-order static frequency shift.}
\label{fig:parametric}
\end{figure*}

Let us consider an effective Hamiltonian of a qubit pair:
\begin{equation}
    H_\mathrm{pair} = \sum_{i=1,2} \left[\omega_i(t) a_i^\dagger a_i - \frac{\eta_i}{2}a_i^\dagger a_i^\dagger a_i a_i\right]+ g(a_1^\dagger a_2 + a_2^\dagger a_1),
\end{equation}
where the qubit transition frequency $\omega_i(t)$ is tunable in time via flux as described in the caption of Fig. \ref{fig:parametric}, and the coupling $g$ is treated as a constant. In this section, we assume that only the second qubit is modulated as
\begin{equation}\label{eq:periodically modulated qubit frequency}
    \begin{aligned}
        \omega_1(t) &= \omega_1, \\
        \omega_2(t) &= \omega_2 + \Omega \cos(\xi t + \phi),
    \end{aligned}
\end{equation}
where $\Omega$ is the modulation amplitude (in frequency units), $\xi$ is the modulation frequency, and $\phi$ is the modulation phase.

In a rotating frame given by transformations
\begin{equation}
    \begin{aligned}
        a_1 &\to e^{-i\omega_1t} a_1, \\
        a_2 &\to e^{-i\left[\omega_2t + \frac{\Omega}{\xi}\sin(\xi t+\phi)\right]} a_2,
    \end{aligned}
\end{equation}
the Hamiltonian becomes
\begin{equation}
    \begin{aligned}
        H_\mathrm{rot} &= -\sum_{i=1,2}\frac{\eta_i}{2} a_i^\dagger a_i^\dagger a_i a_i + g\left[ e^{i(\omega_1-\omega_2)t-i\frac{\Omega}{\xi}\sin(\xi t+\phi)} a_1^\dagger a_2 + \mathrm{h.c.}\right] \\
        &= -\sum_{i=1,2}\frac{\eta_i}{2} a_i^\dagger a_i^\dagger a_i a_i + g\sum_{m=-\infty}^\infty \mathcal J_m\left(\frac{\Omega}{\xi}\right) \left[e^{i(\omega_1-\omega_2-m\xi)t-im\phi} a_1^\dagger a_2 + \mathrm{h.c.} \right].
    \end{aligned}
\end{equation}
The second equality follows from the Jacobi--Anger expansion, where $\mathcal J_m$ denotes the Bessel function of the first kind of integer order $m$. For a modulation frequency matching the qubit detuning, $\xi=\omega_1-\omega_2$, the interaction becomes resonant. Omitting the anharmonicity by restricting each qubit to its two-level subspace and applying the rotating-wave approximation, the Hamiltonian becomes
\begin{equation}\label{eq:effective coupling first Bessel with phase}
    H_{\mathrm{eff}} = J(\Omega,\xi) \left( e^{-i\phi}\sigma_1^+\sigma_2^- + e^{i\phi}\sigma_2^+\sigma_1^- \right).
\end{equation}
Therefore, the effective coupling strength is
\begin{equation}\label{eq:effective coupling first Bessel}
    J(\Omega,\xi) = g\mathcal J_1\left(\frac{\Omega}{\xi}\right).
\end{equation}

In reality, we modulate the qubit transition frequency $\omega_i$ via the flux. Because the relation between the external flux $\Phi_i$ and $\omega_i$ is nonlinear, as described in Eq.~\eqref{eq:transmon model}, the frequency modulation is distorted from a perfect sinusoidal profile. This distortion induces a static shift in $\omega_i$, which is not captured by the model in Eq.~\eqref{eq:periodically modulated qubit frequency}. To see this effect, we Taylor expand $\omega_i(\Phi_i)$ around the DC bias point $\Phi_i^0$:
\begin{equation}
    \omega_i(\Phi_i) = \omega_i(\Phi_i^0) + \left.\frac{\partial\omega_i}{\partial\Phi_i}\right|_{\Phi_i^0} (\Phi_i-\Phi_i^0) + \frac{1}{2} \left.\frac{\partial^2\omega_i}{\partial\Phi_i^2}\right|_{\Phi_i^0} (\Phi_i-\Phi_i^0)^2 + \mathcal O\left((\Phi_i-\Phi_i^0)^3\right).
\end{equation}
Under a sinusoidal flux modulation, $\Phi_i(t) = \Phi_i^0 + \delta\Phi_i\cos(\xi t)$, we obtain the induced static shift in the qubit transition frequency:
\begin{equation}\label{eq: drive-induced frequency shift}
    \Delta \omega_i = \frac{1}{4} \left.\frac{\partial^2\omega_i}{\partial\Phi_i^2}\right|_{\Phi_i^0} \delta\Phi_i^2 + \mathcal O(\delta\Phi_i^4),
\end{equation}
which is second-order in the modulation amplitude.

\subsubsection{Calibration process}

In our system, we find that the drive-induced frequency shift is comparable to, or even several times larger than, the effective coupling strength $J$ used in the experiments. Therefore, the shift in the modulated qubit, $\Delta\omega_2$, is not negligible and must be compensated for by adjusting the modulation frequency as
\begin{equation}
    \xi = \omega_1 - (\omega_2+\Delta\omega_2) = \omega_1-\omega_2-\Delta\omega_2.
\end{equation}
Together with Eq.~\eqref{eq:effective coupling first Bessel}, the $\delta\Phi_2$ dependence of $\Delta\omega_2$ introduces an implicit relation between the modulation amplitude $\Omega^\star(J)$ and frequency $\xi^\star(J)$ required to simultaneously achieve a given coupling strength $J$ and resonance, which does not admit a simple closed-form solution.

\begin{figure*}[ht]
\centering
\includegraphics[width=\textwidth]{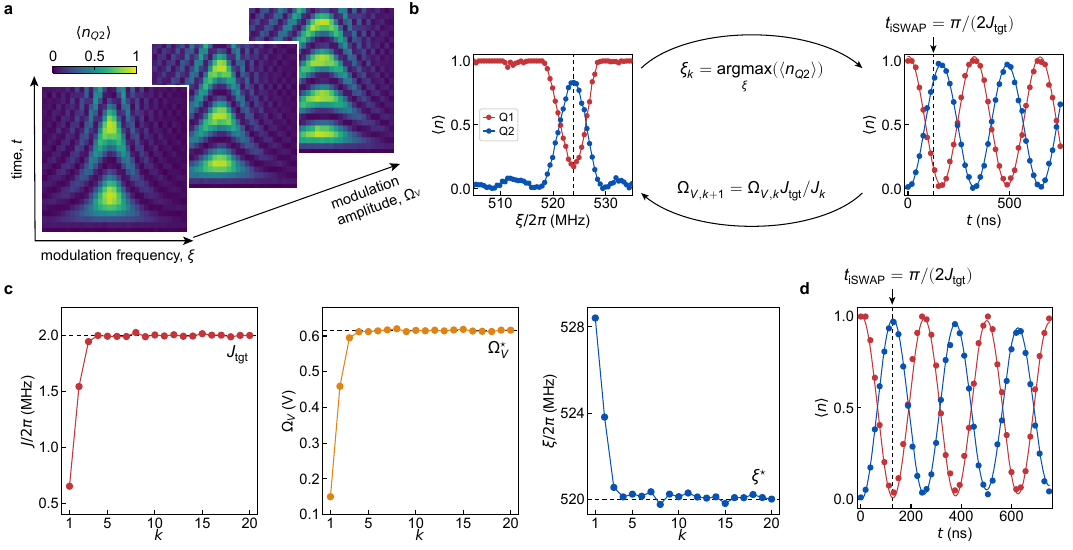}
\caption{\textbf{a,} The Rabi--Chevron plot of the parametric interaction between states $\ket{10}$ and $\ket{01}$ of a pair of qubits. The first qubit is initially populated ($\braket{n_{Q1}}=1$), and the second qubit is modulated to activate the parametric interaction. The population fully swaps between the qubits only when the modulation frequency, $\xi$, brings the two states into resonance in the simultaneous rotating frame of each qubit. The modulation amplitude in voltage units, $\Omega_V$, mainly determines the effective coupling strength, as shown by the three instances with different $\Omega_V$. The center of the pattern shifts depending on $\Omega_V$ due to the drive-induced shift in qubit frequency, which needs to be compensated by adjusting $\xi$. Therefore, without an iterative method, targeting a given coupling strength requires a 3D sweep over the time, modulation-frequency, and modulation-amplitude axes.
\textbf{b,} The iterative method. $J_\mathrm{tgt}$ is the target effective coupling strength. Two parameters, the modulation frequency $\xi$ and time $t$, are alternately swept to find the optimal modulation frequency and amplitude that give a full population swap with the target coupling strength (rate). Starting with $\xi$ set to the difference between the bare qubit frequencies, the $\xi$ sweep finds the value that maximizes the population transfer from the first to the second qubit after a fixed time $t_{i\mathrm{SWAP}}=\pi/(2J_\mathrm{tgt})$. The selected $\xi$ is applied for the time sweep, where the coupling strength is measured by fitting the swap rate between the qubits to sinusoidal curves. The measured coupling rate is then used to linearly update the drive strength for the next round. The iteration then returns to the $\xi$ sweep and is repeated until the population transfer approaches unity with the target coupling strength.
\textbf{c,} Convergence of the target coupling rate, modulation amplitude, and frequency. Here, the initial values are largely off from the optimal values: the initial modulation amplitude is four times smaller than its optimal value, and the initial modulation frequency is $\qty{8}{\MHz}$ away from its optimal value, which is four times larger than the target coupling strength $J_\mathrm{tgt}/2\pi=\qty{2}{\MHz}$. Despite the bad initial guess, the iteration converges to the target within five rounds and fluctuates within $\pm2~\%$ of the target coupling strength in the following rounds.
\textbf{d,} Dynamics with the calibrated parametric drive between the two qubits. A full population transfer occurs at the target $i\mathrm{SWAP}$ duration: $t_{i\mathrm{SWAP}}=\qty{125}{\ns}$.}
\label{fig: supp. param drive tuneup}
\end{figure*}

To determine the pair $(\Omega^\star,\xi^\star)$ for a target coupling $J_\mathrm{tgt}$, one can perform Rabi--Chevron sweeps over interaction time $t$ and modulation frequency $\xi$ for different values of $\Omega$, as illustrated in Fig.~\ref{fig: supp. param drive tuneup}. At fixed $\Omega$, each sweep yields the resonant modulation frequency $\xi^\star$ and the realized coupling $J$ extracted from the swap period. Repeating this procedure for several amplitudes maps out $\Omega^\star(J)$ and $\xi^\star(J)$, allowing their values at $J=J_\mathrm{tgt}$ to be obtained by interpolation.

However, this method is essentially a three-dimensional sweep over $\xi$, $t$, and $\Omega$, and its accuracy is limited by the interpolation. To target a chosen $J_\mathrm{tgt}$ more precisely and efficiently, we develop an iterative method that converges to the pair $(\Omega^\star(J_\mathrm{tgt}),\,\xi^\star(J_\mathrm{tgt}))$ by alternating between sweeps of $\xi$ and $t$: the $\xi$ sweep at fixed amplitude locates the resonance, and the following time sweep measures how far the realized coupling is from the target.

In the $\xi$ sweep of the $k$-th iteration, as shown in Fig.~\ref{fig: supp. param drive tuneup}, we fix the modulation amplitude at $\Omega_k$ and the interaction time at the $i$SWAP duration of $J_\mathrm{tgt}$,
\begin{equation}
    t_{i\mathrm{SWAP}}=\frac{\pi}{2J_\mathrm{tgt}},
\end{equation}
and take $\xi_k$ that gives the maximal population transfer. The seed value $\Omega_1$ can be arbitrary as long as it gives a peak in the sweep.

In the following time sweep, we use the updated $(\Omega_k,\xi_k)$ values and obtain the effective coupling $J_k$ from the hopping rate. Because $\xi_k$ is not yet optimal, the contrast of the population exchange might not reach unity. This is automatically fixed as the iteration proceeds. We then linearly rescale the modulation amplitude as
\begin{equation}
    \Omega_{k+1}=\Omega_k\frac{J_\mathrm{tgt}}{J_k},
\end{equation}
and return to the $\xi$ sweep, alternating between the two sweeps until the coupling strength converges to the target with a full population transfer.

Experimentally, the iteration converges within five rounds as shown in Fig.~\ref{fig: supp. param drive tuneup}. As a fixed-point iteration, the method follows the same principles as the iterative frequency calibration of Section~\ref{sec:frequency fine-tuning}. Convergence is therefore not guaranteed in general, but is fast whenever the update contracts in the parameter space, which is the case for the relatively small couplings targeted here ($J/g=\mathcal{J}_1(\Omega/\xi)\lesssim 0.2$), where the Bessel function of the first kind $\mathcal{J}_1$ is nearly linear.

We use this method for all parametric calibrations in this work. This method is crucial to achieve a symmetric coupling strength between the paraparticle colors in the qubit-ladder implementation, and precise calibration of the beamsplitter coupling in the interference experiment.

\clearpage

\section{Generalized Jordan--Wigner transformation}

Paraparticle creation operators acting at different positions in space in general do not commute, a property shared with ordinary fermions. When a pair of operators do not commute, the action of one operator can alter the expectation value of the other. Therefore, these noncommuting creation and annihilation operators cannot be regarded as local operators confined to a single site, despite the intuitive notion of ``creating/annihilating a particle at a location.'' Indeed, Ref.~\cite{vidalQuantumOperationsInformation2021a} shows that it is possible to perform a faster-than-light communication if one assumes local access to fermionic creation and annihilation operators.

However, the generalized Jordan--Wigner (JW) transformation of Refs.~\cite{wangParticleExchangeStatistics2025,wangRparastatisticsFoundation2026} provides a mapping between a 1D chain of any paraparticle species and a chain of sites with local creation and annihilation operators. This is crucial because fundamental excitations that can be prepared and read out in quantum simulators are local. Therefore, the generalized JW transformation provides the foundation for quantum simulation of paraparticles. In this section, we describe the generalized JW transformation for the paraparticle of this work and derive Eq.~\eqref{eq:tight-binding Hamiltonian} of the main text and the state mapping relation.

As described in Eq.~\eqref{eq:commutation_relations}, the paraparticle studied in this work features $m$ colors and a hard single-occupancy constraint at each site. The corresponding chain consists of local sites, each of which is equipped with an $(m+1)$-dimensional Hilbert space spanned by vacuum $\ket{0}_i$ and states $\{\ket{\alpha}_i\}_{\alpha=1}^{m}$ corresponding to the $m$ colors, where $i$ is the site index. The local creation and annihilation operators $\sigma_{i,\alpha}^{\pm}$ are defined by
\begin{equation}
\begin{aligned}
    \sigma_{i,\alpha}^+ &= \ket{\alpha}_i\bra{0}_i, \\
    \sigma_{i,\alpha}^- &= \ket{0}_i\bra{\alpha}_i.
\end{aligned}
\end{equation}
We can observe that these operators obey
\begin{equation}
\begin{aligned}
    \sigma_{i,\alpha}^-\sigma_{i,\beta}^+ &= \delta_{\alpha\beta}\left(\mathbb{I}_i-\sum_{\gamma=1}^{m}\sigma_{i,\gamma}^+\sigma_{i,\gamma}^-\right),\\
    \sigma_{i,\alpha}^+\sigma_{i,\beta}^+ &= 0,\\
    \sigma_{i,\alpha}^-\sigma_{i,\beta}^- &= 0,
\end{aligned}
\end{equation}
and that operators on distinct sites commute:
\begin{equation}
    [\sigma_{i,\alpha}^{s}, \sigma_{j,\beta}^{t}]=0, \qquad \text{for } i\neq j \text{ and } s,t\in\{+,-\}.
\end{equation}

The generalized JW transformation obtains the paraparticle operators satisfying the commutation relations in Eq.~\eqref{eq:commutation_relations}, by attaching a matrix-product-operator (MPO) string to each local operator:
\begin{equation}\label{eq:generalized JW transformation}
\begin{aligned}
    \psi_{i,\alpha}^\dagger &= \sum_{\gamma_2,\ldots,\gamma_i=1}^{m} \left(\prod_{k=1}^{i-1}T^+_{k,\gamma_k\gamma_{k+1}}\right) \sigma^+_{i,\gamma_i},\\
    \psi_{i,\alpha} &= \sum_{\gamma_2,\ldots,\gamma_i=1}^{m} \left(\prod_{k=1}^{i-1}T^-_{k,\gamma_k\gamma_{k+1}}\right) \sigma^-_{i,\gamma_i},
\end{aligned}
\end{equation}
where $\gamma_1=\alpha$, and the string extends from left (site 1) to right (site $i$). Here, the local MPO tensor is defined by
\begin{equation}
\begin{aligned}\label{eq:MPO definition}
    T^{\pm}_{i,\alpha\beta}\ket{0}_i &= \delta_{\alpha\beta}\ket{0}_i,\\
    T^{+}_{i,\alpha\beta}\ket{\gamma}_i &= -\delta_{\beta\gamma}\ket{\alpha}_i,\\
    T^{-}_{i,\alpha\beta}\ket{\gamma}_i &= -\delta_{\alpha\gamma}\ket{\beta}_i.
\end{aligned}
\end{equation}

To derive the Hamiltonian in Eq.~\eqref{eq:tight-binding Hamiltonian}, we use the following identity:
\begin{equation}\label{eq:MPO identity}
    \sum_{\alpha=1}^m T^+_{i,\alpha\beta} T^-_{i,\alpha\gamma} = \delta_{\beta\gamma} \mathbb{I}_i.
\end{equation}
Applying this identity, we get
\begin{equation}
    \sum_{\alpha=1}^{m} \psi_{i,\alpha}^{\dagger}\psi_{i+1,\alpha} = \sum_{\beta,\gamma=1}^{m} \sigma_{i,\beta}^{+} T^{-}_{i,\beta\gamma} \sigma_{i+1,\gamma}^{-} = \sum_{\alpha=1}^{m} \sigma_{i,\alpha}^{+}\sigma_{i+1,\alpha}^{-},
\end{equation}
where Eq.~\eqref{eq:MPO identity} contracts the overlapping sites of the two strings in the first equality. Equation~\eqref{eq:tight-binding Hamiltonian} then follows directly from this relation.

The generalized Jordan--Wigner transformation in Eq.~\eqref{eq:generalized JW transformation} also provides a direct basis-state correspondence when creation operators are written in ascending order of site index:
\begin{equation}\label{eq:JW state mapping}
    \psi^\dagger_{i_1,\alpha_1} \psi^\dagger_{i_2,\alpha_2}\cdots \psi^\dagger_{i_n,\alpha_n}\vac = \sigma^+_{i_1,\alpha_1} \sigma^+_{i_2,\alpha_2}\cdots \sigma^+_{i_n,\alpha_n}\ket{0},
\end{equation}
\noindent for ordered indices $i_1<i_2<\cdots<i_n$. We can derive it by applying the creation operators from right to left to the vacuum. From the first equality in Eq.~\eqref{eq:MPO definition}, an MPO string applied to the vacuum has no effect:
\begin{equation}
    \psi^\dagger_{i,\alpha}\vac = \sum_{\gamma_2,\ldots,\gamma_i=1}^{m} \left(\prod_{k=1}^{i-1}T^+_{k,\gamma_k\gamma_{k+1}}\right) \sigma^+_{i,\gamma_i}\ket{0} = \sigma^+_{i,\alpha}\ket{0},
\end{equation}
where $\ket{0}=\bigotimes_i\ket{0}_i$ is the vacuum of the system of local sites. Therefore, we obtain ~\eqref{eq:JW state mapping} by applying the creation operators in order.

\section{Exchange of free paraparticles}\label{sec:exchange of free paraparticles}

In this section, we describe free propagation of paraparticles on a chain. These non-interacting paraparticle dynamics enable exchange and interference experiments of this work. The $R$-statistics of paraparticles are preserved under a unitary transformation~\cite{wangParticleExchangeStatistics2025}: if $\psi^\dagger_{i,\alpha}$ and $\psi_{i,\alpha}$ are paraparticle creation and annihilation operators satisfying the commutation relations, then the transformed operators,
\begin{equation}
    \varphi^\dagger_{k,\alpha} = \sum_i W_{ik}\,\psi^\dagger_{i,\alpha}, \qquad \varphi_{k,\alpha} = \sum_i W^*_{ik}\,\psi_{i,\alpha},
\end{equation}
where $W$ is a unitary, also satisfy the same commutation relation. Therefore, $\varphi^\dagger_{k,\alpha}$ and $\varphi_{k,\alpha}$ describe creation and destruction of paraparticles in a different basis. Such a unitary transformation can be leveraged to diagonalize a quadratic Hamiltonian $H = \sum_{i,j} h_{ij} \sum_\alpha \psi^\dagger_{i,\alpha}\psi_{j,\alpha}$, such as the tight-binding Hamiltonian of paraparticles presented in Eq.~\eqref{eq:tight-binding Hamiltonian} in the main text. We can diagonalize such a Hamiltonian as
\begin{equation}
    H = \sum_{i,j} h_{ij} \sum_\alpha \psi^\dagger_{i,\alpha}\psi_{j,\alpha} = \sum_k \epsilon_k \sum_\alpha \varphi^\dagger_{k,\alpha}\varphi_{k,\alpha} = \sum_k \epsilon_k\, n_k .
\end{equation}
Because the number operators,
\begin{equation}
    n_k = \sum_\alpha \varphi^\dagger_{k,\alpha}\varphi_{k,\alpha},
\end{equation}
mutually commute, the propagator simply becomes
\begin{equation}
    U(t) \equiv e^{-iHt} = \prod_k e^{-i\epsilon_k t\, n_k}.
\end{equation}
Using the relations
\begin{equation}
    e^{-i\epsilon_k t\, n_k}\,\varphi^\dagger_{k,\alpha}\,e^{i\epsilon_k t\, n_k} = e^{-i\epsilon_k t}\,\varphi^\dagger_{k,\alpha}, \qquad \psi^\dagger_{i,\alpha} = \sum_k W^*_{ik}\,\varphi^\dagger_{k,\alpha},
\end{equation}
we obtain the time-evolution of the creation operator:
\begin{equation}
    U(t)\,\psi^\dagger_{i,\alpha}\,U^\dagger(t) = \sum_k W^*_{ik}\, e^{-i\epsilon_k t}\, \varphi^\dagger_{k,\alpha} = \sum_j \Big[\sum_k W_{jk}\, e^{-i\epsilon_k t}\, W^*_{ik}\Big]\psi^\dagger_{j,\alpha} = \sum_j \big[e^{-iht}\big]_{ji}\,\psi^\dagger_{j,\alpha}.
\end{equation}
In this form, we see that the evolution is controlled by the single-particle dynamics $u(t)=e^{-iht}$ under the single-particle Hamiltonian $h$, which is an $L\times L$ matrix for an $L$-site chain.

Applying this to the exchange experiment where we use the three-site Hamiltonian in Eq.~\eqref{eq:tight-binding Hamiltonian},
\begin{equation}
    H = J\sum_\alpha\left(\psi^\dagger_{1,\alpha}\psi_{2,\alpha} + \psi^\dagger_{2,\alpha}\psi_{3,\alpha} + \mathrm{h.c.}\right), \qquad h = J
    \begin{pmatrix}
    0 & 1 & 0\\
    1 & 0 & 1\\
    0 & 1 & 0
    \end{pmatrix},
\end{equation}
we obtain the single-particle propagator
\begin{equation}
    u(t) = e^{-iht} =
    \begin{pmatrix}
        \cos^2(\theta/2) & -\tfrac{i}{\sqrt{2}}\sin\theta & -\sin^2(\theta/2)\\[4pt]
        -\tfrac{i}{\sqrt{2}}\sin\theta & \cos\theta & -\tfrac{i}{\sqrt{2}}\sin\theta\\[4pt]
        -\sin^2(\theta/2) & -\tfrac{i}{\sqrt{2}}\sin\theta & \cos^2(\theta/2)
    \end{pmatrix},
    \qquad \theta=\sqrt{2}Jt.
\end{equation}
At the exchange time $t_E=\pi/(\sqrt{2}J)$, the single-particle propagator becomes
\begin{equation}
u(t_E)=
\begin{pmatrix}
0 & 0 & -1\\
0 & -1 & 0\\
-1 & 0 & 0
\end{pmatrix},
\end{equation}
yielding the operator evolutions in Eq.~\eqref{eq:exchange operator evolution} of the main text:
\begin{equation}
U(t_E)\psi^\dagger_{1,\alpha}U^\dagger(t_E) = -\psi^\dagger_{3,\alpha}, \qquad U(t_E)\psi^\dagger_{2,\alpha}U^\dagger(t_E) = -\psi^\dagger_{2,\alpha}, \qquad U(t_E)\psi^\dagger_{3,\alpha}U^\dagger(t_E) = -\psi^\dagger_{1,\alpha}.
\end{equation}
Thus, the particles on sites 1 and 3 are exchanged, while the internal index $\alpha$ remains unchanged.

Now if we start in the state $\psi^\dagger_{1,\alpha}\,\psi^\dagger_{3,\beta}\vac$, then after completing the exchange, we will have:
\begin{equation}
    \begin{aligned}
        U(t_E)\,\psi^\dagger_{1,\alpha}\,\psi^\dagger_{3,\beta}\vac &= U(t_E)\,\psi^\dagger_{1,\alpha}\,U^\dagger(t_E)\;U(t_E)\,\psi^\dagger_{3,\beta}\,U^\dagger(t_E)\;U(t_E)\vac \\
        &= \big(-\psi^\dagger_{3,\alpha}\big)\big(-\psi^\dagger_{1,\beta}\big)\,U(t_E)\vac \\
        &= \psi^\dagger_{3,\alpha}\,\psi^\dagger_{1,\beta}\vac \\
        &= -\psi^\dagger_{1,\alpha}\,\psi^\dagger_{3,\beta}\vac,
    \end{aligned}
\end{equation}
where again, in the last line, we have used the paraparticle commutation relation. So for the three-site exchange, the dynamical phase is 0 and we do accrue an overall $\pi$-phase due only to the exchange, which we are able to probe in the physical qubit picture.

\clearpage

\section{Extended data on the exchange experiment}

\begin{figure*}[ht!]
\centering
\includegraphics[width=\textwidth]{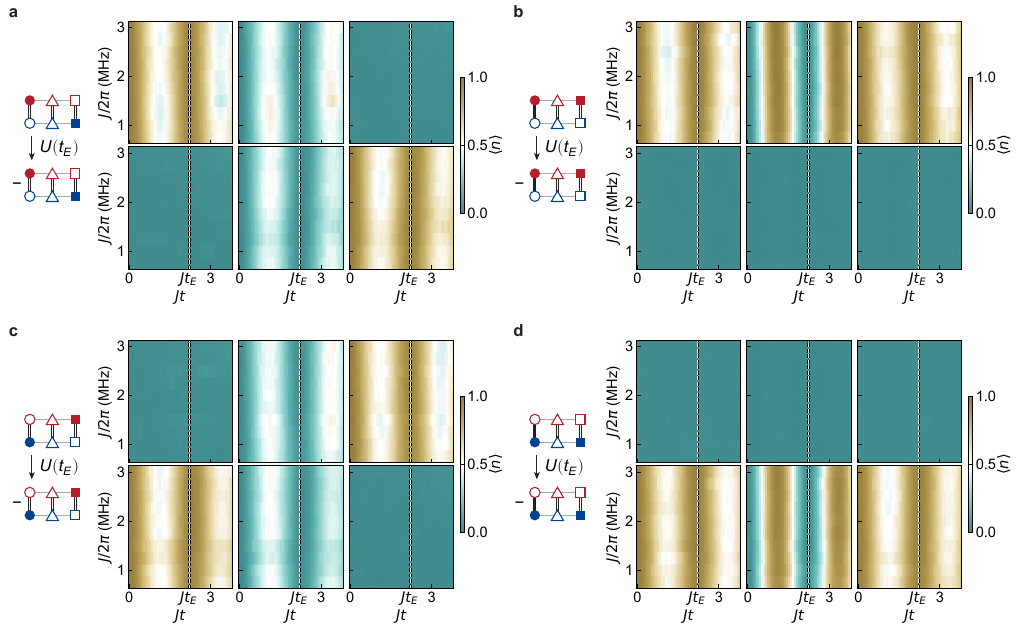}
\caption{\textbf{a-d,} The exchange experiment with different initializations and coupling strengths. The coupling strengths are individually calibrated from $J/2\pi=\qty{0.75}{\MHz}$ to $\qty{3.0}{\MHz}$ and applied to all initializations. For each initial configuration shown, the populations return to their initial values at the same scaled time $Jt_E=\pi/\sqrt{2}\approx2.22$, while the state acquires a global phase of $\pi$, which is not observable from the populations alone. Panels \textbf{a} and \textbf{c} show that the $ZZ$ interaction strongly suppresses the occupation of the initially unoccupied end of each row, consistent with the color-agnostic exclusion principle of paraparticles.}
\label{fig: supp. exchange supp data}
\end{figure*}

\clearpage

\section{Experimental details on the phase detection experiment}

\label{sec:Experimental details on the phase detection experiment}

\subsection{Experimental sequence}

\begin{figure*}[ht]
\centering
\includegraphics[width=\textwidth]{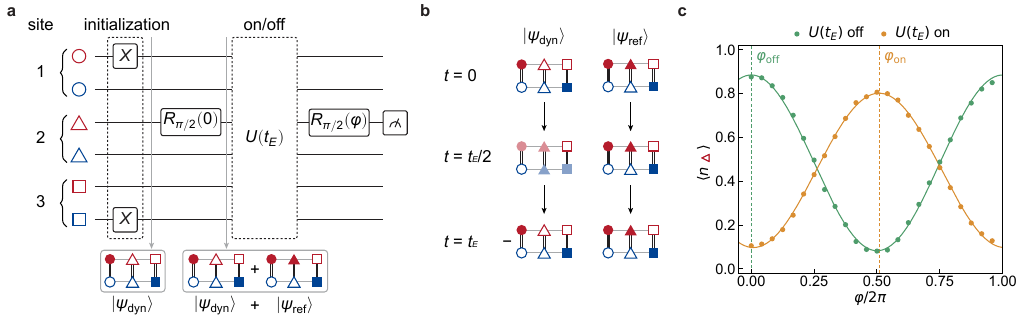}
\caption{\textbf{a,} Experimental sequence for the phase detection experiment. The six-qubit system (three-site paraparticle chain) is initialized with two X gates populating the end sites. The initial state, $\ket{\psi_\mathrm{dyn}}$, is the same state used in the exchange experiment. The following $\pi/2$ pulse is applied to the red-triangle qubit, evolving the system to an equal superposition of the dynamical and reference states. The reference state, $\ket{\psi_\mathrm{ref}}$, is a three-particle state with the red-triangle qubit populated on top of $\ket{\psi_\mathrm{dyn}}$. Next, the parametric drives used in the exchange experiment are turned on, activating all the bonds between the paraparticle sites. The system evolves for the exchange time $t_E$. Finally, a second $\pi/2$ pulse about an axis at angle $\varphi$ from the $y$ axis is applied before reading out the population of the red-triangle qubit, which encodes the exchange phase.
\textbf{b,} The evolution of the dynamical and reference states under $U(t_E)$. The dynamical state undergoes the exchange evolution as discussed in the previous section. However, due to the color-agnostic exclusion principle, the reference state is frozen under the same Hamiltonian because all sites are occupied.
\textbf{c,} Measurement of the $(2,r)$ qubit population as a function of $\varphi$ with the exchange operation on and off. As explained in the text, the offset of the sinusoidal curve, $\varphi_\mathrm{on}$, corresponds to the phase accumulated during the dynamics. To cancel the additional phase arising from the frequency mismatch between the pulse frequency and the transition frequency between the dynamical and reference states, we also measure the offset with no parametric drive ($\varphi_\mathrm{off}$). Finally, the exchange phase is $\theta_E=\varphi_{\mathrm{on}}-\varphi_\mathrm{off}$.}
\label{fig: supp. phase detection exp details}
\end{figure*}

In this section, we explain the experimental sequence of the exchange-phase detection experiment in more detail. The $\pi$ phase accumulated after the exchange of a pair of paraparticles is a global phase and therefore not an observable. However, using Ramsey interferometry, or more generally quantum phase estimation~\cite{NielsenChuangQCQI}, we can observe this exchange phase by preparing the system in a superposition of a reference state and a dynamical state, where only the dynamical state undergoes the exchange evolution.

The experimental sequence is shown in Fig.~\ref{fig: supp. phase detection exp details}a. As in the exchange experiment, we initiate the three-site chain in the state $\ket{\psi_\mathrm{dyn}}$ by exciting two paraparticle sites at the ends (sites $1$ and $3$). Here, we demonstrate the sequence with an example in which we excite the red circle, $(1,r)$, and the blue square, $(3,b)$, following the main text. Extended data in Section~\ref{sec:extended data phase detection} include the results of all four color combinations, $(r,r)$, $(r,b)$, $(b,r)$, and $(b,b)$.

Next, we apply an $R_{\pi/2}(0)$ pulse to the red-triangle qubit, $(2,r)$, preparing it in a superposition of vacuum and populated states. Here, the gate $R_{\pi/2}(\varphi)$ denotes a single-qubit $\pi/2$ rotation around the axis on the equator of the Bloch sphere, where $\varphi$ is the angle between the $y$ axis and the rotation axis. Overall, we prepare the system in a superposition of two- and three-particle states,
\begin{equation}
    \ket{\Psi_0} = \frac{\ket{\psi_\mathrm{dyn}} + \ket{\psi_\mathrm{ref}}}{\sqrt{2}},
\end{equation}
as shown at the bottom of Fig.~\ref{fig: supp. phase detection exp details}a.

After preparing the superposition state, we turn on the parametric drives used in the exchange experiment and evolve the system for the exchange time $t_E$. Because the evolution is linear, the dynamical state $\ket{\psi_\mathrm{dyn}}$ undergoes the exchange operation, $U(t_E)$, described in the main text and in Section~\ref{sec:exchange of free paraparticles} in more detail. On the other hand, as shown in Fig.~\ref{fig: supp. phase detection exp details}b, the reference state has all three sites occupied, a configuration in which the neighboring particles ``block'' each other from moving due to the single-occupancy constraint. Therefore, the three-particle state is frozen, accumulating no exchange phase when the parametric drive is on.

Finally, we apply a second $\pi/2$ pulse with a variable phase $\varphi$, $R_{\pi/2}(\varphi)$, to the red-triangle qubit at $(2,r)$ before readout. The population becomes
\begin{equation}\label{eq:phase to observable}
\langle n_{2,r}\rangle_\varphi = \bra{\Psi_0} U^\dagger(t_E)M(\varphi)U(t_E) \ket{\Psi_0},
\end{equation}
where $M(\varphi)=R_{\pi/2}^\dagger(\varphi)n_{2,r}R_{\pi/2}(\varphi)$. Assuming the following evolution of the state after $t_E$,
\begin{equation}
U(t_E)\ket{\Psi_0} = U(t_E)\frac{\ket{\psi_{\mathrm{dyn}}}+\ket{\psi_{\mathrm{ref}}}}{\sqrt{2}} = \frac{e^{i\theta_E}\ket{\psi_{\mathrm{dyn}}}+\ket{\psi_{\mathrm{ref}}}}{\sqrt{2}},
\end{equation}
we obtain
\begin{equation}
\begin{aligned}
\langle n_{2,r}\rangle_\varphi =\frac12\Big[ &\bra{\psi_{\rm dyn}}M(\varphi)\ket{\psi_{\rm dyn}} +\bra{\psi_{\rm ref}}M(\varphi)\ket{\psi_{\rm ref}}\\
&+e^{-i\theta_E}\bra{\psi_{\rm dyn}}M(\varphi)\ket{\psi_{\rm ref}} +e^{i\theta_E}\bra{\psi_{\rm ref}}M(\varphi)\ket{\psi_{\rm dyn}} \Big] = \frac12\left[1+\cos(\varphi-\theta_E)\right],
\end{aligned}
\end{equation}
after plugging in the relevant matrix elements:
\begin{equation}
\bra{\psi_{\mathrm{dyn}}}M(\varphi)\ket{\psi_{\mathrm{dyn}}} = \bra{\psi_{\mathrm{ref}}}M(\varphi)\ket{\psi_{\mathrm{ref}}} = \frac{1}{2}, \qquad \bra{\psi_{\mathrm{dyn}}}M(\varphi)\ket{\psi_{\mathrm{ref}}} = \frac{1}{2}e^{i\varphi}, \qquad \bra{\psi_{\mathrm{ref}}}M(\varphi)\ket{\psi_{\mathrm{dyn}}} = \frac{1}{2}e^{-i\varphi}.
\end{equation}
To extract the exchange phase $\theta_E$, we measure the average population $\langle n_{2,r}\rangle_\varphi$ as a function of $\varphi$ and fit it to a sinusoidal curve.

However, there are two non-idealities that can enter the measurement and introduce errors in the extraction of $\theta_E$ using this method. First, the $\pi/2$ pulse frequency might not be exactly the same as the frequency difference between $\ket{\psi_\mathrm{dyn}}$ and $\ket{\psi_\mathrm{ref}}$ in the lab frame, due to frequency shifts induced by the $\pi/2$ pulse drive itself. If these frequencies differ, the phase accumulation of the drive over time $t_E$ differs from the relative dynamical phase between the reference and dynamical states. In the rotating frame where all qubits are on resonance, the $\pi/2$ pulse will be effectively detuned in this case, resulting in a phase offset $\varphi_0$, namely
\begin{equation}
\langle n_{2,r}\rangle_\varphi = \frac12\left[1+\cos(\varphi+\varphi_0-\theta_E)\right].
\end{equation}
To compensate for it, we perform two sweeps of $\varphi$, with the parametric drives on and off, illustrated as the dashed box surrounding $U(t_E)$ in Fig.~\ref{fig: supp. phase detection exp details}a. From the phase offset of the sweep with the parametric drive off, we get $\varphi_\mathrm{off} = -\varphi_0$ from the curve fit shown in Fig.~\ref{fig: supp. phase detection exp details}c. We extract $\theta_E$ by comparing it to $\varphi_\mathrm{on}=-\varphi_0 + \theta_E$, which is shown as the orange curve in the same plot.

The second non-ideality is phase decoherence, which reduces the visibility of the $\langle n_{2,r}\rangle_\varphi$ oscillation. Our sequence mitigates this effect by sweeping $\varphi$ and extracting the phase offset from a sinusoidal fit, which is insensitive to the overall oscillation amplitude. In contrast, conventional Ramsey interferometry at a fixed phase is sensitive to dephasing, because a population measurement at a single point of $\varphi$ cannot distinguish between the population decay due to the accumulated phase and the loss of phase information (dephasing).

\clearpage

\section{Extended data on the phase detection experiment}\label{sec:extended data phase detection}

\begin{figure*}[ht]
\centering
\includegraphics{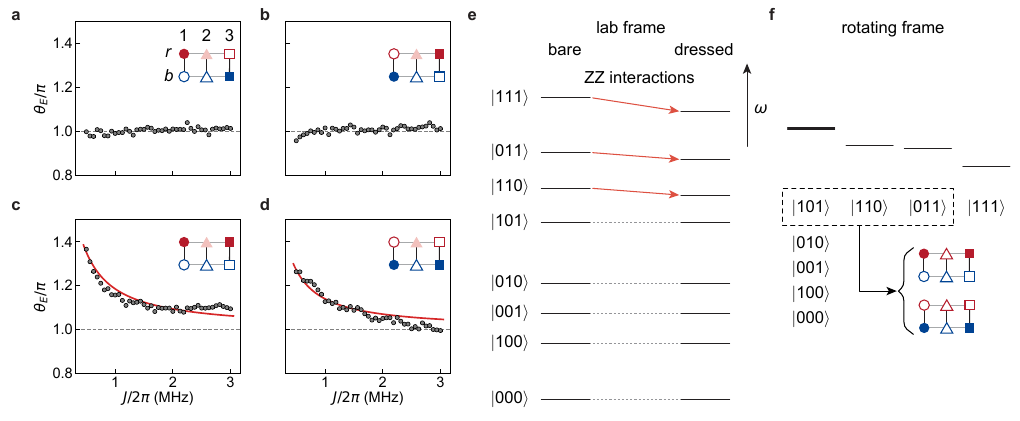}
\caption{\textbf{a-d,} The measured exchange phase for different initial states across coupling strengths. The same qubit at $(2,r)$ is used as an auxiliary qubit, with all four combinations of the end sites initialized. The experiment is performed across different $J/2\pi$ values from $\qty{0.5}{\MHz}$ to $\qty{3}{\MHz}$. The results show a clear difference between same- and different-color initializations at the ends (sites $1$ and $3$). When different-color sites are initialized at sites 1 and 3, respectively, the exchange phase $\theta_E$ remains close to $\pi$, as expected from the ideal exchange dynamics. However, for the same-color cases, the phase deviates from $\pi$ by an amount that grows as $1/J$. We attribute this deviation to the $ZZ$ interaction \textit{between} the sites, as explained in the text and the following panels. The red curves in panels c and d show the theoretical predictions of the $ZZ$-shifted model, calculated using the values in Table~\ref{tab:phase detection ZZ table} with no free parameters.
\textbf{e,} The bare and dressed energy levels of relevant states in the same-color case. The paraparticles remain on the same row for the same-color case. Therefore, we can consider only the three-qubit row, with states from $\ket{000}$ to $\ket{111}$. Although neighboring qubits are detuned by approximately $\qty{650}{\MHz}$, a bare coupling strength of approximately $\qty{15}{\MHz}$ induces a non-negligible inter-site $ZZ$ interaction. Therefore, the dressed two-photon and three-photon states, $\ket{110}$, $\ket{011}$, and $\ket{111}$, are shifted from their bare frequencies by an amount set by the $ZZ$ interaction.
\textbf{f,} The dressed-state energies in the rotating frame. In the simultaneous rotating frame of the individual qubits, the $ZZ$ interaction pulls the affected states down from the degenerate manifold of the other states. Because the parametric drive is calibrated in the single-particle manifold, the drive activates the inter-site couplings leaving the states, $\ket{110}$, $\ket{011}$, and $\ket{111}$, off the manifold. This effect adds diagonal terms to the exchange Hamiltonian, causing deviations from the ideal dynamics.}
\label{fig: supp. phase supp data}
\end{figure*}

Figure~\ref{fig: supp. phase supp data} presents the measured exchange phase with the different initializations as shown in the insets. The qubit at $(2,r)$ is used as an auxiliary control qubit for every initialization that blocks the exchange operation when populated. The exchange phase is read out through the population of the same qubit via the Ramsey interferometry introduced in the main text and Section~\ref{sec:Experimental details on the phase detection experiment}.

The measurements of different-color initializations shown in panels a and b lie around the expected exchange phase $\pi$. However, there are systematic deviations in panels c and d, where same-color qubits are initialized. We attribute this to the $ZZ$ interaction between the same-color qubits. The mechanism by which this $ZZ$ interaction affects our result is closely related to the ``spectator effect'' in the context of two-qubit gates, where the state of a spectator qubit shifts the optimal operating point of the gate and thereby degrades its fidelity.

To understand this effect in our experiment, we simplify the system to a three-qubit chain. This is valid for the situations in panels c and d, because the exchange dynamics occur only within the same-color row. The control qubit thus conditionally blockades the exchange evolution when it is in the $\ket{1}$ state.

The $ZZ$ interaction in a two-qubit transmon pair is dominated by the coupling of the $\ket{11}$ state to the $\ket{20}$ and $\ket{02}$ states. In this three-qubit row, the dressed frequencies of the following three states, $\ket{110}$, $\ket{011}$, and $\ket{111}$, are shifted by the $ZZ$ interaction, as shown in panel e.

The shift becomes prominent in the rotating frame shown in the last panel. Throughout this work, this simultaneously rotating frame of all qubits, where the $\ket{000}$, $\ket{100}$, $\ket{010}$, and $\ket{001}$ states are degenerate, is the frame in which the parametric drives are calibrated and the paraparticle Hamiltonian in Eq.~\eqref{eq:tight-binding Hamiltonian} is described. However, the three shifted states are no longer degenerate with the vacuum in the rotating frame, which affects only the same-color initializations (panels c and d) but not different-color ones (panels a and b). The Hamiltonian involving two- or three-excitation states reads
\begin{equation}
    H =
    \begin{pmatrix}
        0 & J & J & 0 \\
        J & \delta_{110} & 0 & 0 \\
        J & 0 & \delta_{011} & 0 \\
        0 & 0 & 0 & \delta_{111}
    \end{pmatrix}
    \quad \text{in the basis} \quad \left( \ket{101}, \ket{110}, \ket{011}, \ket{111} \right),
\end{equation}
where $\delta_{110}$, $\delta_{011}$, and $\delta_{111}$ are the frequency shifts of the corresponding states due to the $ZZ$ interaction.

\begin{table*}[ht]
\centering
{\renewcommand{\arraystretch}{1.5}
\begin{tabular}{lccc}
\toprule
Row & $\delta_{110}/2\pi$ (MHz) & $\delta_{011}/2\pi$ (MHz) & $\delta_{111}/2\pi$ (MHz) \\
\midrule
Red (panel c)  & $-0.491$ & $-0.575$ & $-1.065$ \\
Blue (panel d) & $-0.413$ & $-0.385$ & $-0.797$ \\
\bottomrule
\end{tabular}
}
\caption{$ZZ$-induced frequency shifts of the doubly and triply excited states of each three-qubit row. The values are calculated from the qubit parameters, coupling strengths, and bias frequencies.}
\label{tab:phase detection ZZ table}
\end{table*}

Table~\ref{tab:phase detection ZZ table} shows the frequency shifts numerically obtained from the model in Eq.~\eqref{eq:circuit-level two transmons} with the qubit parameters and bare coupling rates, at the frequency layout used in this exchange phase experiment.

Using the frequency shifts, we can reconstruct the Hamiltonian and the corresponding unitary evolution $U(t_E)$ for the exchange. We then extract the expected $\theta_E$ by plugging it into Eq.~\eqref{eq:phase to observable} assuming an ideal auxiliary qubit. Finally, we obtain the red curves in panels c and d by fitting the theoretical $\langle n_{2,r}\rangle_\varphi$ as a function of $\varphi$ and reading out the phase offset. The curves match the measurements, identifying the inter-pair $ZZ$ interaction as the dominant cause of the deviation.

\clearpage

\section{Two-paraparticle interference}

\subsection{Color-agnostic antibunching}

A 50:50 beamsplitter is characterized by a unitary operation, which splits an incoming mode into an equal superposition of the two outgoing modes. In this work, we implement a beamsplitter that has the following action:
\begin{equation}
\begin{aligned}
\psi^\dagger_{L_{\mathrm{in}},\alpha} &\rightarrow -\frac{1}{\sqrt{2}}\psi_{R_{\mathrm{out}},\alpha}^\dagger - \frac{1}{\sqrt{2}}\psi_{L_{\mathrm{out}},\alpha}^\dagger \\
\psi_{R_{\mathrm{in}},\alpha}^\dagger &\rightarrow \frac{1}{\sqrt{2}}\psi_{R_{\mathrm{out}},\alpha}^\dagger - \frac{1}{\sqrt{2}}\psi_{L_{\mathrm{out}},\alpha}^\dagger
\end{aligned}
\end{equation}
on creation operators, where $\alpha$ denotes an internal degree of freedom, such as spin for a fermion or boson, or color for a paraparticle. If we have two free incoming particles, the final state becomes:
\begin{equation}
\begin{aligned}
\psi_{L_{\mathrm{in}},\alpha}^\dagger\psi_{R_{\mathrm{in}},\beta}^\dagger\vac &\rightarrow \left( -\frac{1}{\sqrt{2}}\psi_{R_{\mathrm{out}},\alpha}^\dagger -\frac{1}{\sqrt{2}}\psi_{L_{\mathrm{out}},\alpha}^\dagger \right) \left( \frac{1}{\sqrt{2}}\psi_{R_{\mathrm{out}},\beta}^\dagger -\frac{1}{\sqrt{2}}\psi_{L_{\mathrm{out}},\beta}^\dagger \right)\vac \\
&= \left[ -\frac{1}{2} \left( \psi_{R_{\mathrm{out}},\alpha}^\dagger\psi_{R_{\mathrm{out}},\beta}^\dagger - \psi_{L_{\mathrm{out}},\alpha}^\dagger\psi_{L_{\mathrm{out}},\beta}^\dagger \right) + \frac{1}{2} \left( \psi_{R_{\mathrm{out}},\alpha}^\dagger\psi_{L_{\mathrm{out}},\beta}^\dagger - \psi_{L_{\mathrm{out}},\alpha}^\dagger\psi_{R_{\mathrm{out}},\beta}^\dagger \right) \right]\vac.
\end{aligned}
\end{equation}

It is most instructive to compare the behavior of the paraparticles to bosons and fermions. There are two cases, the first of which is when $\alpha=\beta$. Since bosons are symmetric under exchange, only the first two terms survive, and therefore we will see ``bunching'' behavior, where both particles are always measured to be in the same outgoing mode. Since fermions are antisymmetric under exchange, only the last two terms survive, and therefore we will see ``antibunching'' behavior, where both particles are always measured to be in different outgoing modes. If $\alpha\neq\beta$, however, then all terms are unique and therefore we will see all possible combinations. The generalized commutation relations in Eq.~\eqref{eq:commutation_relations} of the main text for paraparticles lead to cancellation of terms, even in the case when $\alpha\neq\beta$. For the paraparticles studied in this work:
\begin{equation}
\psi_{i,\alpha}^\dagger\psi_{j,\beta}^\dagger = -\psi_{j,\alpha}^\dagger\psi_{i,\beta}^\dagger
\end{equation}
The first two terms vanish due to the exclusion principle even when $\alpha\neq\beta$, giving us antibunching behavior. In other words, paraparticles antibunch, even when their colors are different.

\subsection{Four-site beamsplitter}

\begin{figure*}[ht]
\includegraphics{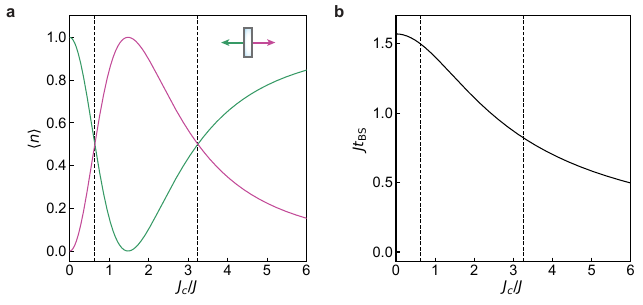}
\caption{\textbf{a,} The mode population after time $t_\mathrm{BS}=\pi/\sqrt{4J^2+J_c^2}$, where $J$ and $J_c$ are the beamsplitter couplings shown in Fig.~\ref{fig: collision}d of the main text. The incoming mode populations become zero simultaneously at $t_\mathrm{BS}$. Two values of $J_c/J$, approximately $0.637$ and $3.247$, give the 50:50 beamsplitting ratio.
\textbf{b,} The normalized beamsplitting time $Jt_\mathrm{BS} = \pi J/\sqrt{4J^2+J_c^2}$ as a function of $J_c/J$. $J_c/J\approx 0.637$, corresponding to the 50:50 beamsplitting ratio, gives $Jt_\mathrm{BS}\approx 1.50$.}
\label{fig: supp. interference beamsplitter}
\end{figure*}

As discussed in the main text, we realize a beamsplitter with a four-site chain, where we label the left two sites as $L$ and the right two sites as $R$. The beamsplitter works on the propagating modes, with the four propagating modes on left ($L$) and right ($R$) arms defined as
\begin{equation}
\begin{aligned}
\ket{L_{\mathrm{in}}} &= \frac{\ket{1000}-i\ket{0100}}{\sqrt{2}}, &\qquad \ket{L_{\mathrm{out}}} &= \frac{\ket{1000}+i\ket{0100}}{\sqrt{2}},\\
\ket{R_{\mathrm{in}}} &= \frac{\ket{0010}+i\ket{0001}}{\sqrt{2}}, &\qquad \ket{R_{\mathrm{out}}} &= \frac{\ket{0010}-i\ket{0001}}{\sqrt{2}}.
\end{aligned}
\end{equation}

Within the $L$ or $R$ arms, the two sites are coupled with strength $J$ and the coupling between $L$ and $R$ has strength $J_c$. The Hamiltonian in the single excitation subspace $\{\ket{1000}, \ket{0100}, \ket{0010}, \ket{0001}\}$ reads:
\begin{equation}\label{eq:beamsplitting Hamiltonian}
H_\mathrm{BS} =
\begin{pmatrix}
0 & J & 0 & 0 \\
J & 0 & J_c & 0 \\
0 & J_c & 0 & J \\
0 & 0 & J & 0
\end{pmatrix}.
\end{equation}
From this, we can consider the unitary evolution of the state $\ket{L_{\mathrm{in}}}$ under $H_\mathrm{BS}$ and solve for the time when there is a 50:50 beamsplitter.

We can find the unitary evolution under $H_\mathrm{BS}$ for time $t$. In the case where the initial state is $\ket{L_{\mathrm{in}}}$, the populations as a function of time are:
\begin{equation}
\begin{aligned}
P_{L_{\mathrm{in}}}(t) &= \cos^2\left(\frac{\lambda t}{2}\right) \cos^2\left(\frac{J_c t}{2}\right) \\
P_{L_{\mathrm{out}}}(t) &= \frac{1}{\lambda^2} \sin^2\left(\frac{\lambda t}{2}\right) \left[ -2J\cos\left(\frac{J_c t}{2}\right) + J_c \sin\left(\frac{J_c t}{2}\right) \right]^2 \\
P_{R_{\mathrm{out}}}(t) &= \frac{1}{\lambda^2} \sin^2\left(\frac{\lambda t}{2}\right) \left[ J_c\cos\left(\frac{J_c t}{2}\right) + 2J \sin\left(\frac{J_c t}{2}\right) \right]^2 \\
P_{R_{\mathrm{in}}}(t) &= \cos^2\left(\frac{\lambda t}{2}\right) \sin^2\left(\frac{J_c t}{2}\right),
\end{aligned}
\end{equation}
where $\lambda = \sqrt{4J^2+J_c^2}$. The first instance when $P_{L_{\mathrm{in}}}(t)=0$ is when $t=t_{BS}=\pi/\lambda$, and at that time, the population in $\ket{R_{\mathrm{in}}}$ also vanishes. In other words, all the population has split between $\ket{L_{\mathrm{out}}}$ and $\ket{R_{\mathrm{out}}}$. The caveat is that the split may not be 50:50. It is 50:50 only when:
\begin{equation}
\left[ -2J\cos\left(\frac{J_c \pi}{2\lambda}\right) + J_c \sin\left(\frac{J_c \pi}{2\lambda}\right) \right]^2 = \left[ J_c\cos\left(\frac{J_c \pi}{2\lambda}\right) + 2J \sin\left(\frac{J_c \pi}{2\lambda}\right) \right]^2.
\end{equation}
This has exactly two positive solutions: $J_c/J \approx 0.637, 3.247$. In this work, we realize a beamsplitter with $J_c/J \approx 0.637$.

\clearpage

\section{Phase alignment}

\begin{figure*}[ht]
\centering
\includegraphics[width=\textwidth]{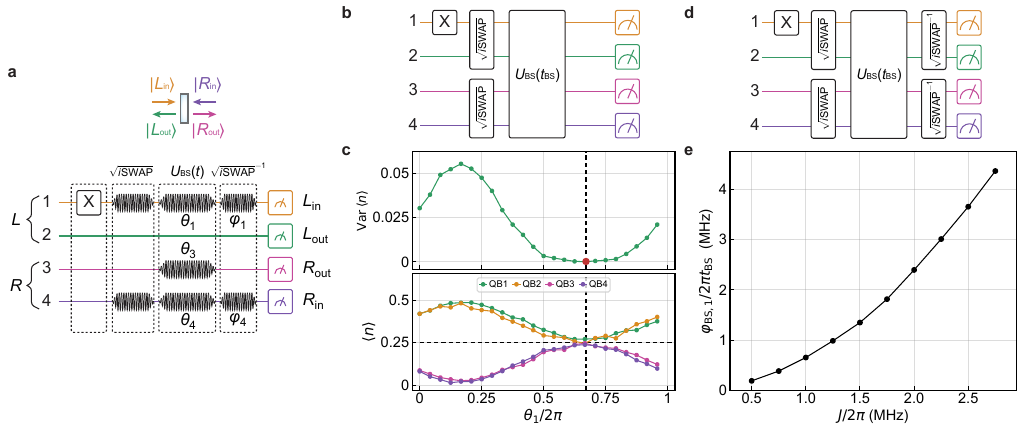}
\caption{\textbf{a,} Pulse sequence of the interference experiment. The beamsplitting operation $U_\mathrm{BS}(t)$ and the $\sqrt{i\mathrm{SWAP}}$ gates are all realized with parametric drives. The phase of each parametric drive is shown: $\theta_i$ for the beamsplitting operation, and $\varphi_i$ for the inverse $\sqrt{i\mathrm{SWAP}}$ layer. The phases of the first $\sqrt{i\mathrm{SWAP}}$ layer can be absorbed into sites 1 and 4. The drive phase on qubit 3, $\theta_3$, can also be absorbed due to gauge freedom, as discussed in the text. Therefore, the four remaining phases should be calibrated for the experiment.
\textbf{b,} Pulse sequence for $\theta_1$ calibration. For $\theta_i$ calibration, we omit the last inverse $\sqrt{i\mathrm{SWAP}}$ layer. The first X gate is moved to site 4 for $\theta_4$ calibration.
\textbf{c,} Calibration of $\theta_1$. Applying the sequence shown in panel \textbf{b}, we measure the qubit populations while sweeping the phase $\theta_1$. When the beamsplitting operation is realized with the correct phase, the two outgoing modes are equally populated, resulting in equal qubit populations of 0.25 each. We choose the $\theta_1$ that minimizes the variance of the qubit populations. $\theta_4$ is calibrated in the same manner with the initial population at site 4.
\textbf{d,} Pulse sequence for $\varphi_1$ calibration. Now with the full sequence, where the beamsplitting operation is sandwiched between two $\sqrt{i\mathrm{SWAP}}$ layers, we sweep $\varphi_1$ to find the value that maximizes the population at site 2.
\textbf{e,} The phase accumulation rate for $\varphi_1$ during the beamsplitting operation. After calibrating $\varphi_1$, we can extract the phase accumulated during the beamsplitting operation, $\varphi_{\mathrm{BS},1}$, by subtracting the $\pi$ phase added to implement the inverse $\sqrt{i\mathrm{SWAP}}$ gate. We attribute the phase accumulation to the different drive-induced frequency shifts during the beamsplitter operation. Dividing $\varphi_{\mathrm{BS},1}$ by $t_\mathrm{BS}$ for different $J$ values, we observe that the accumulation rate scales quadratically and is comparable to the drive-induced frequency shifts discussed in Eq.~\eqref{eq: drive-induced frequency shift}.}
\label{fig: supp. interference tuneup details}
\end{figure*}

The parametric drive can add a phase to the coupling it induces, as shown in Eq.~\eqref{eq:effective coupling first Bessel with phase}. This coupling phase is not observable in the exchange experiment. However, in the interference experiment, it becomes an observable and therefore must be calibrated accordingly.

Let us consider the experimental sequence shown in Fig.~\ref{fig: supp. interference tuneup details}a, which corresponds to the gate representation shown in Fig.~\ref{fig: collision} in the main text. The sequence first prepares the incoming propagating mode on the left arm by turning on the parametric coupling within pairs $(1,2)$ and $(3,4)$:
\begin{equation}
H_{\sqrt{i\mathrm{SWAP}}} = J
\begin{pmatrix}
0 & 1 & 0 & 0\\
1 & 0 & 0 & 0\\
0 & 0 & 0 & 1\\
0 & 0 & 1 & 0
\end{pmatrix},
\qquad \text{in the basis } \{\ket{1000},\ket{0100},\ket{0010},\ket{0001}\}.
\end{equation}
Starting from $\ket{1000}$, a $\sqrt{i\mathrm{SWAP}}$ pulse prepares
\begin{equation}
\ket{L_{\mathrm{in}}} = \frac{\ket{1000}-i\ket{0100}}{\sqrt{2}}.
\end{equation}

Next, we turn on three parametric drives to apply the beamsplitting Hamiltonian. The drive amplitudes and frequencies are calibrated following the method explained in Section~\ref{sec: parametric drive calibration}. Including the coupling phases, the Hamiltonian becomes
\begin{equation}
H_{\mathrm{BS},\boldsymbol{\theta}} =
\begin{pmatrix}
0 & J e^{i\theta_1} & 0 & 0 \\
J e^{-i\theta_1} & 0 & J_c e^{i\theta_3} & 0 \\
0 & J_c e^{-i\theta_3} & 0 & J e^{i\theta_4} \\
0 & 0 & J e^{-i\theta_4} & 0
\end{pmatrix},
\end{equation}
where $\theta_1$, $\theta_3$, and $\theta_4$ are the phases of the parametric drives that activate the $(1,2)$, $(2,3)$, and $(3,4)$ couplings, respectively. Using the gauge freedom to redefine the common phase of the right-arm states, we absorb $\theta_3$. We therefore calibrate the phases $\theta_1$ and $\theta_4$ to implement the phase convention of Eq.~\eqref{eq:beamsplitting Hamiltonian}.

Figures~\ref{fig: supp. interference tuneup details}b and c show the calibration procedure and results. After preparing $\ket{L_{\mathrm{in}}}$, we perform the beamsplitting operation without the second $\sqrt{i\mathrm{SWAP}}^{-1}$ layer, while sweeping $\theta_1$. Because the initial state contains a relative phase between $\ket{1000}$ and $\ket{0100}$ only, the phases $\theta_3$ and $\theta_4$ can be absorbed into phase redefinitions of qubits $3$ and $4$ for this sequence, and the final populations depend only on $\theta_1$. Therefore, we find the value of $\theta_1$ where the final population of every qubit becomes $0.25$, by minimizing the variance of the four populations. The same sequence, except for the initial X gate on qubit $4$ instead of qubit $1$, is used to calibrate $\theta_4$.

Finally, we calibrate the drive phases of the second $\sqrt{i\mathrm{SWAP}}^{-1}$ layer. After the beamsplitting operation $U_{\mathrm{BS}}(t_{\mathrm{BS}})$, the $\sqrt{i\mathrm{SWAP}}^{-1}$ layer should map $\ket{L_{\mathrm{out}}}$ and $\ket{R_{\mathrm{out}}}$ onto qubits $2$ and $3$, respectively. We sweep the phases $\varphi_1$ and $\varphi_4$ of the final $\sqrt{i\mathrm{SWAP}}^{-1}$ pulses to maximize the populations of qubits $2$ and $3$, while minimizing those of qubits $1$ and $4$, as shown in Fig.~\ref{fig: supp. interference tuneup details}d.

Taking the phase of the initial $\sqrt{i\mathrm{SWAP}}$ pulse as zero, the corresponding inverse-pulse phase in the drive frame is
\begin{equation}
\varphi_i=\varphi_{\mathrm{BS},i}+\pi, \qquad i\in\{1,4\},
\end{equation}
where $\varphi_{\mathrm{BS},i}$ is the phase accumulated during the beamsplitting operation for the corresponding arm. Using the calibrated $\varphi_{\mathrm{BS},i}$, we can calculate its accumulation rate, $\dot{\varphi}_{\mathrm{BS},i}=\varphi_{\mathrm{BS},i}/t_{\mathrm{BS}}$, and use it to determine the $\sqrt{i\mathrm{SWAP}}^{-1}$ phase for the interference experiments with $t$ other than $t_{\mathrm{BS}}$, including the data shown in Fig.~\ref{fig: collision}. Figure~\ref{fig: supp. interference tuneup details}e presents $\varphi_{\mathrm{BS},1}/(2\pi t_{\mathrm{BS}})$ as a function of effective coupling $J$, showing an approximately quadratic dependence on $J$, consistent with the drive-induced frequency shift discussed in Section~\ref{sec: parametric drive calibration}. This identifies the origin of the accumulated phase as the drive-induced frequency shift of the qubits during the beamsplitting operation.

\clearpage

\section{Extended data on the interference experiment}

\begin{figure*}[ht]
\centering
\includegraphics[width=\textwidth]{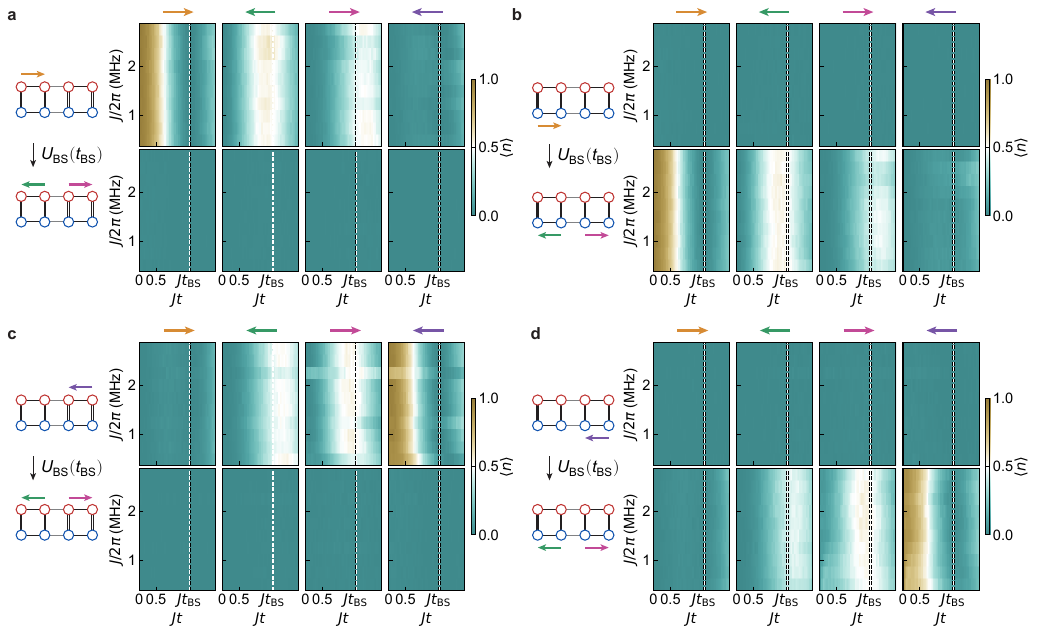}
\caption{Single-paraparticle beamsplitter experiment.
\textbf{a-d,} Each panel shows a different initialization. The incoming mode is prepared following the sequence shown in Fig.~\ref{fig: collision}e of the main text. All possible cases, with red or blue paraparticles incoming from the left or right, are measured. The coupling between the paraparticle sites in each arm, $J$, is swept from $J/2\pi=\qty{0.5}{\MHz}$ to $\qty{2.75}{\MHz}$. The coupling strength at the center, $J_c$, is tuned accordingly to maintain $J_c/J = 0.637$. The $x$ axis shows time on a normalized scale across different $J$ values. Consistent beamsplitter dynamics are observed across $J$: at the beamsplitting time $Jt_\mathrm{BS} = 1.5$, the outgoing modes are split in a 50:50 ratio, while both incoming modes are suppressed to zero.}
\label{fig: supp. interference single particle supp data}
\end{figure*}

\begin{figure*}[ht]
\centering
\includegraphics[width=\textwidth]{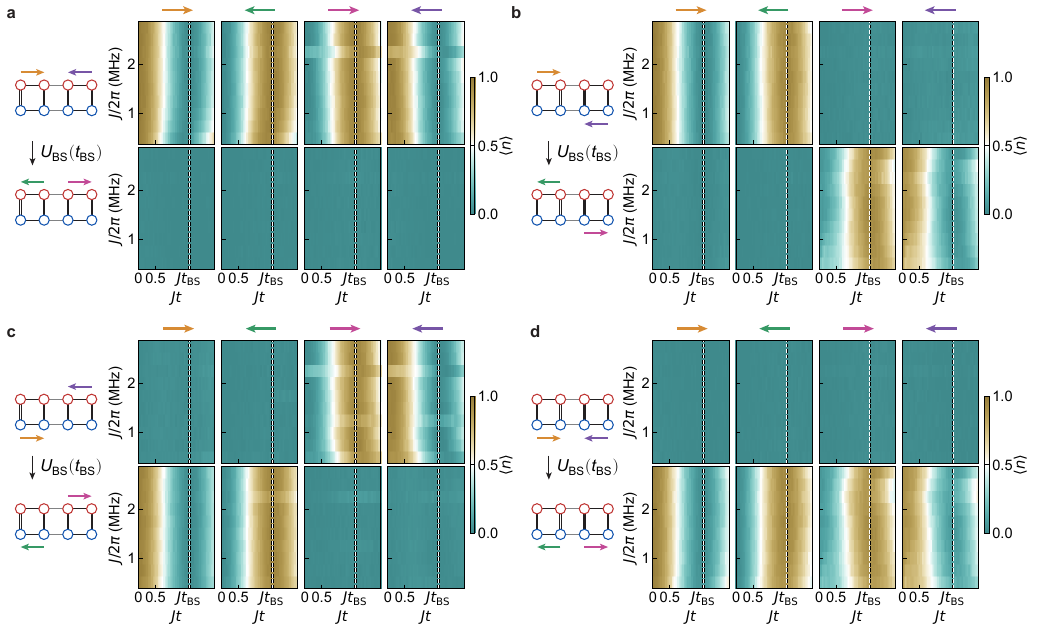}
\caption{Two-paraparticle beamsplitter experiment.
\textbf{a-d,} Each panel shows a different initialization with two incoming paraparticles. All possible cases with different colors and incoming directions are measured. The coupling between the paraparticle sites in each arm, $J$, is swept from $J/2\pi=\qty{0.5}{\MHz}$ to $\qty{2.75}{\MHz}$. The coupling strength at the center, $J_c$, is tuned accordingly to maintain $J_c/J = 0.637$. The $x$ axis shows time on a normalized scale across different $J$ values. The result renders color-agnostic antibunching behavior, as especially shown in panels \textbf{b} and \textbf{c}, where the paraparticles of different colors antibunch.}
\label{fig: supp. interference two particle supp data}
\end{figure*}

\clearpage

\section{Experimental details of the thermalization experiment}
\label{sec: thermalization details}

\begin{figure*}[ht]
\centering
\includegraphics[width=\textwidth]{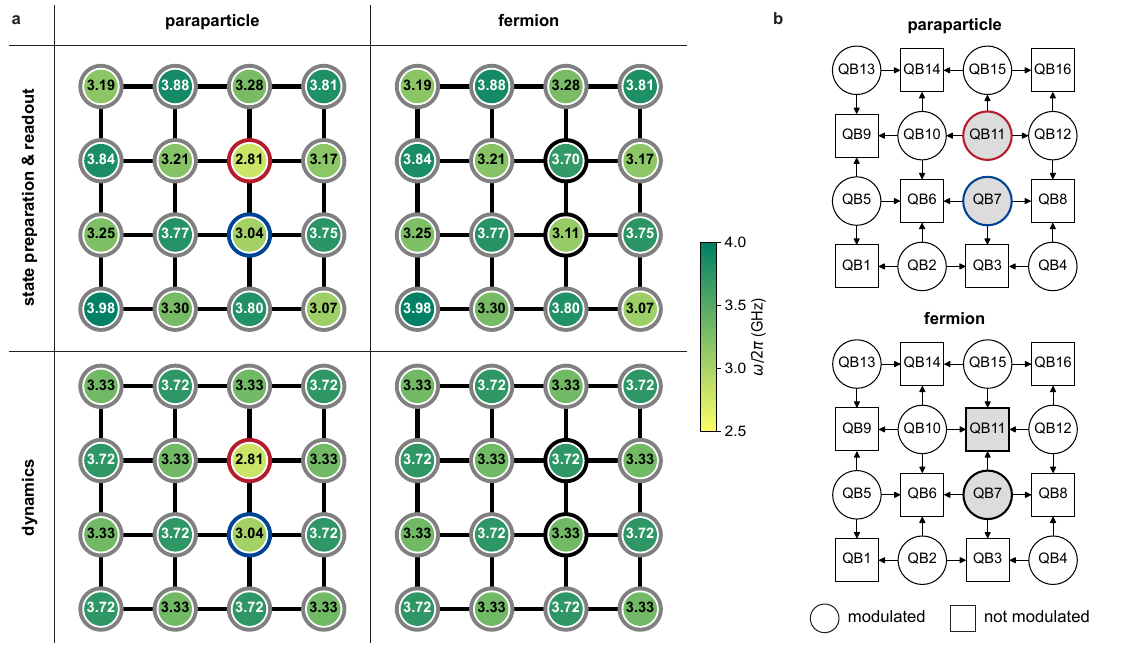}
\caption{\textbf{a,} Frequency layouts used in the thermalization experiment. The top row shows frequency layouts for state preparation and qubit readout. In these layouts, populations are frozen due to detuning between nearest and next-nearest neighboring qubits. Each qubit is assigned a unique frequency to mitigate the RF crosstalk between the qubits when multiple qubits are simultaneously excited. The bottom row shows the frequency layout used for dynamics. The bath qubits are separated into two frequency groups, so that all couplings are activated by modulating only a subset of qubits.
\textbf{b,} Qubit modulation scheme with assignment of paraparticle and fermionic sites. All circle qubits are modulated to parametrically activate the nearest-neighbor couplings. The paraparticle frequency layout maximizes the $ZZ$ interaction between the designated qubits (QB7 and QB11). In the fermionic layout, the qubits for fermionic sites (the same QB7 and QB11) are parametrically coupled with a suppressed $ZZ$ interaction, which allows simultaneous occupation of the pair.}
\label{fig: supp. thermalization frequency layout}
\end{figure*}

As illustrated in Fig.~\ref{fig: thermalization}b of the main text, the thermalization experiment has three stages: initialization, interaction, and readout. For the initialization and readout stages, the qubits are placed in a frequency layout with a checkerboard pattern, which allows independent excitation and readout. These ``state preparation and readout'' layouts for the paraparticle and fermionic thermalization experiments are shown in the first row of Fig.~\ref{fig: supp. thermalization frequency layout}a. Here, in the paraparticle thermalization layout, the red and blue qubits corresponding to each color of the paraparticle site are detuned by the frequency that maximizes the $ZZ$ interaction between the pair, as discussed in Section~\ref{sec: maximizing ZZ}. In the fermionic thermalization layout, on the other hand, we designate the same qubits as fermionic sites but set the qubits to different frequencies that suppress the $ZZ$ interaction so that both qubits can be simultaneously populated. The occupation of this ``fermionic site'' is read out as the average of the two qubit populations. In both experiments, we do not excite these designated qubits in the initialization stage.

In the interaction stage, all qubits are quenched to the ``dynamics'' layout, shown in the second row of Fig.~\ref{fig: supp. thermalization frequency layout}a, through baseband pulses that temporarily shift the qubit frequencies. Then, at the same time, we modulate the qubit frequencies to activate couplings between all neighboring qubits, except for the coupling between the paraparticle qubit pair. As discussed in the main text, not all, but only 9 (8 in fermionic thermalization) circle qubits in Fig.~\ref{fig: thermalization}a are driven. It is the checkerboard pattern that allows the subset of qubits to drive all couplings. For example, the modulation on QB2 simultaneously activates the couplings between pairs (QB2,~QB1), (QB2,~QB3), and (QB2,~QB6). The tail of an arrow in Figure~\ref{fig: supp. thermalization frequency layout}b indicates the qubit that drives each bond.

\begin{figure*}[ht]
\centering
\includegraphics[width=\textwidth]{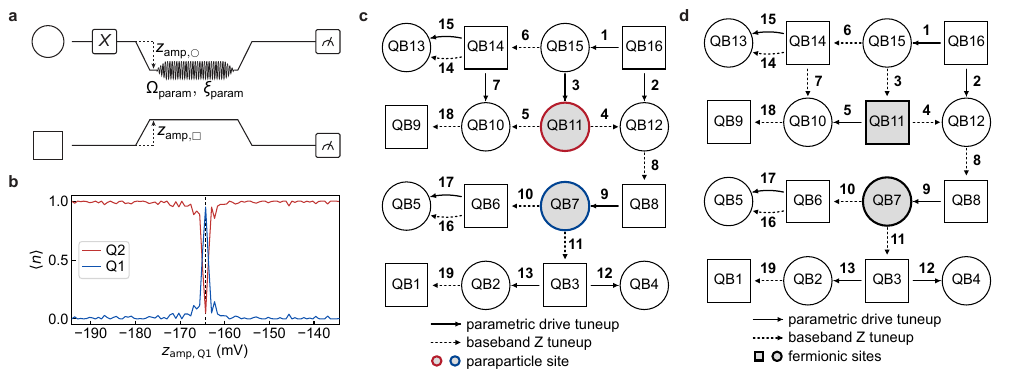}
\caption{\textbf{a,} Coupling between neighboring qubits in the thermalization experiment. After initialization, the qubits are rapidly tuned to the dynamics layout using a baseband square pulse. Only the circle qubits are parametrically modulated to activate the couplings. Four control knobs are shown: the baseband pulse amplitudes for each qubit, $z_{\mathrm{amp},\bigcirc}$ and $z_{\mathrm{amp},\Box}$, modulation amplitude $\Omega_\mathrm{param}$, and modulation frequency $\xi_\mathrm{param}$. The modulation parameters are calibrated following the method discussed in Section~\ref{sec: parametric drive calibration}. The baseband pulse amplitudes are calibrated as shown in panel \textbf{b}.
\textbf{b,} Calibration of the baseband pulse amplitude. The target baseband pulse amplitude is swept at a fixed interaction time $t=\pi/(2J)$, at which the population is expected to fully swap between the qubit pair. The amplitude giving the maximal population transfer is chosen.
\textbf{c,} The calibration sequence for the paraparticle thermalization experiment. The solid (dashed) arrow denotes parametric drive (baseband pulse amplitude) calibration. The numbers on the arrows indicate the order of the calibration steps.
\textbf{d,} The calibration sequence for the fermionic thermalization experiment.}
\label{fig: supp. thermalization tuneup sequence}
\end{figure*}

Figure~\ref{fig: supp. thermalization tuneup sequence} presents the calibration process for the drives used in the interaction stage. The control degrees of freedom that need to be calibrated are shown in Fig.~\ref{fig: supp. thermalization tuneup sequence}a. For a square qubit, which is not modulated, the baseband pulse amplitude is the only knob. For a modulated (circle) qubit, we can tune the modulation amplitude, frequency, and the baseband pulse amplitude. Here, as discussed in Section~\ref{sec: parametric drive calibration}, the modulation amplitude mainly determines the parametric coupling strength, while the baseband pulse amplitude and modulation frequency are fine-tuned to put all qubits on resonance (up to sidebands) when the modulation is on.

The numbers on the arrows in Figs.~\ref{fig: supp. thermalization tuneup sequence}c and d show the order of the calibration for the paraparticle and fermionic thermalization experiments, respectively. In each figure, the arrow at each bond points to the qubit whose parameters are being calibrated, and each arrow is visited sequentially following the number on it. The solid arrow indicates parametric drive calibration following the method introduced in Section~\ref{sec: parametric drive calibration}, and the dashed arrow means calibration of the baseband pulse amplitude. For example, arrow 1 in both panels c and d indicates that QB15's modulation amplitude and frequency are calibrated for the bond between QB15 and QB16. Arrow 6, for another example, means calibration of QB14's baseband pulse amplitude, which is done by sweeping it while keeping the evolution time $\pi/(2J)$ constant, where $J=J_b=J_s$ is the target coupling strength, set to $2\pi\times\qty{1}{MHz}$ throughout the thermalization experiment. The optimal baseband pulse amplitude is chosen to maximize the population transfer between the pairs, which happens when the modulated qubit's sideband is on resonance with the other qubit.

Due to the drive-induced frequency shift and crosstalk between baseband pulses, each pair is calibrated with all modulations and baseband pulses applied. However, it is convenient to isolate each bond to calibrate it individually. To achieve this, we detune all qubits except those forming the target bond by applying baseband pulses that shift their frequencies by $\qty{40}{MHz}$ away from the target frequency. This technique isolates the target bond while preserving the second-order effects, including crosstalk and drive-induced frequency shifts.

\clearpage

\section{Boltzmann factor and thermal populations}

In the main text, we show that the Boltzmann factor in the paraparticle thermalization experiment is given by
\begin{equation}
x=\frac{\Omega_B(1)}{\Omega_B(0)} = \frac{n_0}{N-n_0+1},
\end{equation}
where $\Omega_B(k)=\binom{N}{n_0-k}$ is the Hilbert space dimension of the $N=14$-qubit bath with $n_0-k$ excitations, and $k\in\{0,1\}$ is the occupation of the paraparticle site. $n_0$ is the total number of excitations in the system.

The partition function of the paraparticle site is $Z(x)=1+2x$, and the average population is therefore
\begin{equation}
\braket{n} = x\frac{\partial \log Z(x)}{\partial x} = \frac{2x}{1+2x}.
\end{equation}

In the fermionic thermalization experiment, we have two qubits designated as fermionic sites in the system. In contrast to a paraparticle site, a pair of fermionic sites can host zero, one, or two particles. Therefore, there are two possible choices for the Boltzmann factor:
\begin{equation}
\begin{aligned}
\frac{\Omega_B(1)}{\Omega_B(0)} &= \frac{n_0}{N-n_0+1},\\
\frac{\Omega_B(2)}{\Omega_B(1)} &= \frac{n_0-1}{N-n_0+2},
\end{aligned}
\end{equation}
where we consider an $N=14$-qubit bath with the two fermionic sites occupied with $k$ particles. These two ratios disagree because $N$ and $n_0$ are finite; they both converge to $\rho/(1-\rho)$ in the $N,n_0\to\infty$ limit at fixed filling fraction $\rho=n_0/N$.

In our experiment with finite $N$ and $n_0$, we resolve this ambiguity by treating each fermionic site separately. Taking one of the fermionic sites as the subsystem and the remaining $N=15$ qubits as the bath, we obtain the Boltzmann factor
\begin{equation}
x=\frac{\Omega_B(1)}{\Omega_B(0)}=\frac{n_0}{N-n_0+1}.
\end{equation}
The average population of each fermionic site is then
\begin{equation}
\braket{n} = \frac{x}{1+x} = \frac{n_0}{N+1} = \frac{n_0}{16},
\end{equation}
which is shown in Fig.~\ref{fig: thermalization}d as the black straight line.

\clearpage

\section{Extended data on the thermalization experiment}

\begin{figure*}[ht]
\centering
\includegraphics[width=\textwidth]{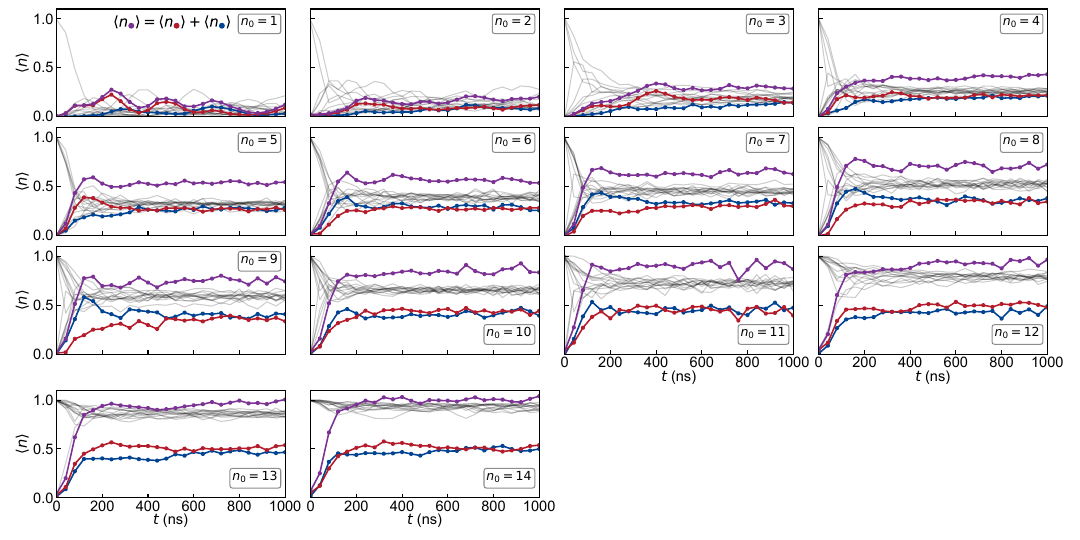}
\caption{Time dynamics in the thermalization experiment with a paraparticle site. The paraparticle site population is shown as a purple curve, which is the sum of the populations of the red and blue qubits. Each panel shows an instance with a given number of initial excitations $n_0$. The paraparticle site, with three possible microstates ($\ket{0}$, $\ket{r}$, and $\ket{b}$), thermalizes to a higher average population than the bath qubits, each of which has two possible microstates ($\ket{0}$ and $\ket{1}$). Individual populations of the red and blue qubits are clearly suppressed relative to the bath qubits due to the strong $ZZ$ interaction between them. The data is post-selected on shots with total $n_0$ excitations across all 16 qubits.}
\label{fig: supp. paraparticle trace}
\end{figure*}

\clearpage

\begin{figure*}[ht]
\centering
\includegraphics[width=\textwidth]{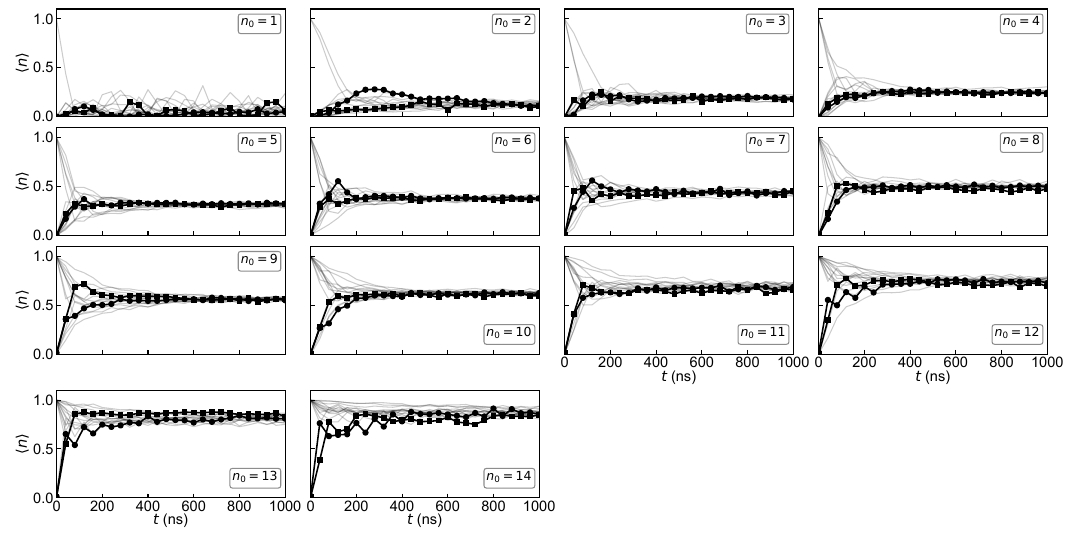}
\caption{Time dynamics in the thermalization experiment with fermionic sites. As discussed in Section~\ref{sec: thermalization details}, the same pair of qubits designated as the paraparticle site (QB7 and QB11 in Fig.~\ref{fig: supp. thermalization frequency layout}b) are shifted in frequency to \textit{suppress} the $ZZ$ interaction. These qubits now become two fermionic sites, for which simultaneous occupation is not forbidden. The circle and square markers denote each qubit in the pair. Their average population at $t=\qty{1}{\micro\s}$ is taken as the fermionic thermal population shown in Fig.~\ref{fig: thermalization} of the main text. The data is post-selected on shots with total $n_0$ excitations across all 16 qubits.}
\label{fig: supp. fermion trace}
\end{figure*}

\clearpage
\putbib[ref]
\clearpage

\end{bibunit}

\end{bibunit}


\begin{thebibliography}{48}
\makeatletter
\providecommand \@ifxundefined [1]{\@ifx{#1\undefined} }
\providecommand \@ifnum [1]{\ifnum #1\expandafter \@firstoftwo \else \expandafter \@secondoftwo \fi }
\providecommand \@ifx [1]{\ifx #1\expandafter \@firstoftwo \else \expandafter \@secondoftwo \fi }
\providecommand \natexlab [1]{#1}
\providecommand \enquote  [1]{``#1''}
\providecommand \bibnamefont  [1]{#1}
\providecommand \bibfnamefont [1]{#1}
\providecommand \citenamefont [1]{#1}
\providecommand \href@noop [0]{\@secondoftwo}
\providecommand \href [0]{\begingroup \@sanitize@url \@href}
\providecommand \@href[1]{\@@startlink{#1}\@@href}
\providecommand \@@href[1]{\endgroup#1\@@endlink}
\providecommand \@sanitize@url [0]{\catcode `\\12\catcode `\$12\catcode `\&12\catcode `\#12\catcode `\^12\catcode `\_12\catcode `\%12\relax}
\providecommand \@@startlink[1]{}
\providecommand \@@endlink[0]{}
\providecommand \url  [0]{\begingroup\@sanitize@url \@url }
\providecommand \@url [1]{\endgroup\@href {#1}{\urlprefix }}
\providecommand \urlprefix  [0]{URL }
\providecommand \Eprint [0]{\href }
\providecommand \doibase [0]{https://doi.org/}
\providecommand \selectlanguage [0]{\@gobble}
\providecommand \bibinfo  [0]{\@secondoftwo}
\providecommand \bibfield  [0]{\@secondoftwo}
\providecommand \translation [1]{[#1]}
\providecommand \BibitemOpen [0]{}
\providecommand \bibitemStop [0]{}
\providecommand \bibitemNoStop [0]{.\EOS\space}
\providecommand \EOS [0]{\spacefactor3000\relax}
\providecommand \BibitemShut  [1]{\csname bibitem#1\endcsname}
\let\auto@bib@innerbib\@empty
\bibitem [{\citenamefont {Wang}\ and\ \citenamefont {Hazzard}(2025)}]{wangParticleExchangeStatistics2025}\BibitemOpen
  \bibfield  {author} {\bibinfo {author} {\bibfnamefont {Z.}~\bibnamefont {Wang}}\ and\ \bibinfo {author} {\bibfnamefont {K.~R.~A.}\ \bibnamefont {Hazzard}},\ }\href {https://doi.org/10.1038/s41586-024-08262-7} {\bibfield  {journal} {\bibinfo  {journal} {Nature}\ }\textbf {\bibinfo {volume} {637}},\ \bibinfo {pages} {314} (\bibinfo {year} {2025})}\BibitemShut {NoStop}\relax
\bibitem [{\citenamefont {Pathria}\ and\ \citenamefont {Beale}(2011)}]{Pathria2011}\BibitemOpen
  \bibfield  {author} {\bibinfo {author} {\bibfnamefont {R.~K.}\ \bibnamefont {Pathria}}\ and\ \bibinfo {author} {\bibfnamefont {P.~D.}\ \bibnamefont {Beale}},\ }\href@noop {} {\emph {\bibinfo {title} {Statistical Mechanics}}},\ \bibinfo {edition} {3rd}\ ed.\ (\bibinfo  {publisher} {Academic Press},\ \bibinfo {year} {2011})\BibitemShut {NoStop}\relax
\bibitem [{\citenamefont {Chen}\ \emph {et~al.}(2024)\citenamefont {Chen}, \citenamefont {Wang}, \citenamefont {Boyack}, \citenamefont {Yang},\ and\ \citenamefont {Levin}}]{chenWhenSuperconductivityCrosses2024}\BibitemOpen
  \bibfield  {author} {\bibinfo {author} {\bibfnamefont {Q.}~\bibnamefont {Chen}}, \bibinfo {author} {\bibfnamefont {Z.}~\bibnamefont {Wang}}, \bibinfo {author} {\bibfnamefont {R.}~\bibnamefont {Boyack}}, \bibinfo {author} {\bibfnamefont {S.}~\bibnamefont {Yang}},\ and\ \bibinfo {author} {\bibfnamefont {K.}~\bibnamefont {Levin}},\ }\href {https://doi.org/10.1103/RevModPhys.96.025002} {\bibfield  {journal} {\bibinfo  {journal} {Reviews of Modern Physics}\ }\textbf {\bibinfo {volume} {96}},\ \bibinfo {pages} {025002} (\bibinfo {year} {2024})}\BibitemShut {NoStop}\relax
\bibitem [{\citenamefont {Greiter}\ and\ \citenamefont {Wilczek}(2024)}]{greiterFractionalStatistics2024}\BibitemOpen
  \bibfield  {author} {\bibinfo {author} {\bibfnamefont {M.}~\bibnamefont {Greiter}}\ and\ \bibinfo {author} {\bibfnamefont {F.}~\bibnamefont {Wilczek}},\ }\href {https://doi.org/10.1146/annurev-conmatphys-040423-014045} {\bibfield  {journal} {\bibinfo  {journal} {Annual Review of Condensed Matter Physics}\ }\textbf {\bibinfo {volume} {15}},\ \bibinfo {pages} {131} (\bibinfo {year} {2024})}\BibitemShut {NoStop}\relax
\bibitem [{\citenamefont {Green}(1953)}]{Green1953Parastatistics}\BibitemOpen
  \bibfield  {author} {\bibinfo {author} {\bibfnamefont {H.~S.}\ \bibnamefont {Green}},\ }\href {https://doi.org/10.1103/PhysRev.90.270} {\bibfield  {journal} {\bibinfo  {journal} {Physical Review}\ }\textbf {\bibinfo {volume} {90}},\ \bibinfo {pages} {270} (\bibinfo {year} {1953})}\BibitemShut {NoStop}\relax
\bibitem [{\citenamefont {Leinaas}\ and\ \citenamefont {Myrheim}(1977)}]{LeinaasMyrheim1977Identical}\BibitemOpen
  \bibfield  {author} {\bibinfo {author} {\bibfnamefont {J.~M.}\ \bibnamefont {Leinaas}}\ and\ \bibinfo {author} {\bibfnamefont {J.}~\bibnamefont {Myrheim}},\ }\href {https://doi.org/10.1007/BF02727953} {\bibfield  {journal} {\bibinfo  {journal} {Il Nuovo Cimento B}\ }\textbf {\bibinfo {volume} {37}},\ \bibinfo {pages} {1} (\bibinfo {year} {1977})}\BibitemShut {NoStop}\relax
\bibitem [{\citenamefont {Goldin}\ \emph {et~al.}(1981)\citenamefont {Goldin}, \citenamefont {Menikoff},\ and\ \citenamefont {Sharp}}]{goldinRepresentationsLocalCurrent1981}\BibitemOpen
  \bibfield  {author} {\bibinfo {author} {\bibfnamefont {G.~A.}\ \bibnamefont {Goldin}}, \bibinfo {author} {\bibfnamefont {R.}~\bibnamefont {Menikoff}},\ and\ \bibinfo {author} {\bibfnamefont {D.~H.}\ \bibnamefont {Sharp}},\ }\href {https://doi.org/10.1063/1.525110} {\bibfield  {journal} {\bibinfo  {journal} {Journal of Mathematical Physics}\ }\textbf {\bibinfo {volume} {22}},\ \bibinfo {pages} {1664} (\bibinfo {year} {1981})}\BibitemShut {NoStop}\relax
\bibitem [{\citenamefont {Wilczek}(1982)}]{Wilczek1982FractionalSpin}\BibitemOpen
  \bibfield  {author} {\bibinfo {author} {\bibfnamefont {F.}~\bibnamefont {Wilczek}},\ }\href {https://doi.org/10.1103/PhysRevLett.49.957} {\bibfield  {journal} {\bibinfo  {journal} {Physical Review Letters}\ }\textbf {\bibinfo {volume} {49}},\ \bibinfo {pages} {957} (\bibinfo {year} {1982})}\BibitemShut {NoStop}\relax
\bibitem [{\citenamefont {Wu}(1984)}]{wuGeneralTheoryQuantum1984}\BibitemOpen
  \bibfield  {author} {\bibinfo {author} {\bibfnamefont {Y.-S.}\ \bibnamefont {Wu}},\ }\href {https://doi.org/10.1103/PhysRevLett.52.2103} {\bibfield  {journal} {\bibinfo  {journal} {Physical Review Letters}\ }\textbf {\bibinfo {volume} {52}},\ \bibinfo {pages} {2103} (\bibinfo {year} {1984})}\BibitemShut {NoStop}\relax
\bibitem [{\citenamefont {Arovas}\ \emph {et~al.}(1984)\citenamefont {Arovas}, \citenamefont {Schrieffer},\ and\ \citenamefont {Wilczek}}]{arovasFractionalStatisticsQuantum1984}\BibitemOpen
  \bibfield  {author} {\bibinfo {author} {\bibfnamefont {D.}~\bibnamefont {Arovas}}, \bibinfo {author} {\bibfnamefont {J.~R.}\ \bibnamefont {Schrieffer}},\ and\ \bibinfo {author} {\bibfnamefont {F.}~\bibnamefont {Wilczek}},\ }\href {https://doi.org/10.1103/PhysRevLett.53.722} {\bibfield  {journal} {\bibinfo  {journal} {Physical Review Letters}\ }\textbf {\bibinfo {volume} {53}},\ \bibinfo {pages} {722} (\bibinfo {year} {1984})}\BibitemShut {NoStop}\relax
\bibitem [{\citenamefont {Greenberg}(1990)}]{Greenberg1990InfiniteStats}\BibitemOpen
  \bibfield  {author} {\bibinfo {author} {\bibfnamefont {O.~W.}\ \bibnamefont {Greenberg}},\ }\href {https://doi.org/10.1103/PhysRevLett.64.705} {\bibfield  {journal} {\bibinfo  {journal} {Physical Review Letters}\ }\textbf {\bibinfo {volume} {64}},\ \bibinfo {pages} {705} (\bibinfo {year} {1990})}\BibitemShut {NoStop}\relax
\bibitem [{\citenamefont {Greenberg}(1991)}]{greenbergParticlesSmallViolations1991}\BibitemOpen
  \bibfield  {author} {\bibinfo {author} {\bibfnamefont {O.~W.}\ \bibnamefont {Greenberg}},\ }\href {https://doi.org/10.1103/PhysRevD.43.4111} {\bibfield  {journal} {\bibinfo  {journal} {Physical Review D}\ }\textbf {\bibinfo {volume} {43}},\ \bibinfo {pages} {4111} (\bibinfo {year} {1991})}\BibitemShut {NoStop}\relax
\bibitem [{\citenamefont {Haldane}(1991)}]{haldaneFractionalStatisticsArbitrary1991a}\BibitemOpen
  \bibfield  {author} {\bibinfo {author} {\bibfnamefont {F.~D.~M.}\ \bibnamefont {Haldane}},\ }\href {https://doi.org/10.1103/PhysRevLett.67.937} {\bibfield  {journal} {\bibinfo  {journal} {Physical Review Letters}\ }\textbf {\bibinfo {volume} {67}},\ \bibinfo {pages} {937} (\bibinfo {year} {1991})}\BibitemShut {NoStop}\relax
\bibitem [{\citenamefont {Wu}(1994)}]{wuStatisticalDistributionGeneralized1994}\BibitemOpen
  \bibfield  {author} {\bibinfo {author} {\bibfnamefont {Y.-S.}\ \bibnamefont {Wu}},\ }\href {https://doi.org/10.1103/PhysRevLett.73.922} {\bibfield  {journal} {\bibinfo  {journal} {Physical Review Letters}\ }\textbf {\bibinfo {volume} {73}},\ \bibinfo {pages} {922} (\bibinfo {year} {1994})}\BibitemShut {NoStop}\relax
\bibitem [{\citenamefont {Bartolomei}\ \emph {et~al.}(2020)\citenamefont {Bartolomei}, \citenamefont {Kumar}, \citenamefont {Bisognin}, \citenamefont {Marguerite}, \citenamefont {Berroir}, \citenamefont {Bocquillon}, \citenamefont {Pla{\c c}ais}, \citenamefont {Cavanna}, \citenamefont {Dong}, \citenamefont {Gennser}, \citenamefont {Jin},\ and\ \citenamefont {F{\`e}ve}}]{Bartolomei2020AnyonCollisions}\BibitemOpen
  \bibfield  {author} {\bibinfo {author} {\bibfnamefont {H.}~\bibnamefont {Bartolomei}}, \bibinfo {author} {\bibfnamefont {M.}~\bibnamefont {Kumar}}, \bibinfo {author} {\bibfnamefont {R.}~\bibnamefont {Bisognin}}, \bibinfo {author} {\bibfnamefont {A.}~\bibnamefont {Marguerite}}, \bibinfo {author} {\bibfnamefont {J.-M.}\ \bibnamefont {Berroir}}, \bibinfo {author} {\bibfnamefont {E.}~\bibnamefont {Bocquillon}}, \bibinfo {author} {\bibfnamefont {B.}~\bibnamefont {Pla{\c c}ais}}, \bibinfo {author} {\bibfnamefont {A.}~\bibnamefont {Cavanna}}, \bibinfo {author} {\bibfnamefont {Q.}~\bibnamefont {Dong}}, \bibinfo {author} {\bibfnamefont {U.}~\bibnamefont {Gennser}}, \bibinfo {author} {\bibfnamefont {Y.}~\bibnamefont {Jin}},\ and\ \bibinfo {author} {\bibfnamefont {G.}~\bibnamefont {F{\`e}ve}},\ }\href {https://doi.org/10.1126/science.aaz5601} {\bibfield  {journal} {\bibinfo  {journal} {Science}\ }\textbf {\bibinfo {volume} {368}},\ \bibinfo {pages} {173} (\bibinfo {year} {2020})}\BibitemShut {NoStop}\relax
\bibitem [{\citenamefont {Nakamura}\ \emph {et~al.}(2020)\citenamefont {Nakamura}, \citenamefont {Liang}, \citenamefont {Gardner},\ and\ \citenamefont {Manfra}}]{nakamuraDirectObservationAnyonic2020}\BibitemOpen
  \bibfield  {author} {\bibinfo {author} {\bibfnamefont {J.}~\bibnamefont {Nakamura}}, \bibinfo {author} {\bibfnamefont {S.}~\bibnamefont {Liang}}, \bibinfo {author} {\bibfnamefont {G.~C.}\ \bibnamefont {Gardner}},\ and\ \bibinfo {author} {\bibfnamefont {M.~J.}\ \bibnamefont {Manfra}},\ }\href {https://doi.org/10.1038/s41567-020-1019-1} {\bibfield  {journal} {\bibinfo  {journal} {Nature Physics}\ }\textbf {\bibinfo {volume} {16}},\ \bibinfo {pages} {931} (\bibinfo {year} {2020})}\BibitemShut {NoStop}\relax
\bibitem [{\citenamefont {Greenberg}\ and\ \citenamefont {Messiah}(1965)}]{greenbergSelectionRulesParafields1965}\BibitemOpen
  \bibfield  {author} {\bibinfo {author} {\bibfnamefont {O.~W.}\ \bibnamefont {Greenberg}}\ and\ \bibinfo {author} {\bibfnamefont {A.~M.~L.}\ \bibnamefont {Messiah}},\ }\href {https://doi.org/10.1103/PhysRev.138.B1155} {\bibfield  {journal} {\bibinfo  {journal} {Physical Review}\ }\textbf {\bibinfo {volume} {138}},\ \bibinfo {pages} {B1155} (\bibinfo {year} {1965})}\BibitemShut {NoStop}\relax
\bibitem [{\citenamefont {Dr{\"u}hl}\ \emph {et~al.}(1970)\citenamefont {Dr{\"u}hl}, \citenamefont {Haag},\ and\ \citenamefont {Roberts}}]{druhlParastatistics1970}\BibitemOpen
  \bibfield  {author} {\bibinfo {author} {\bibfnamefont {K.}~\bibnamefont {Dr{\"u}hl}}, \bibinfo {author} {\bibfnamefont {R.}~\bibnamefont {Haag}},\ and\ \bibinfo {author} {\bibfnamefont {J.~E.}\ \bibnamefont {Roberts}},\ }\href {https://doi.org/10.1007/BF01649433} {\bibfield  {journal} {\bibinfo  {journal} {Communications in Mathematical Physics}\ }\textbf {\bibinfo {volume} {18}},\ \bibinfo {pages} {204} (\bibinfo {year} {1970})}\BibitemShut {NoStop}\relax
\bibitem [{\citenamefont {Doplicher}\ \emph {et~al.}(1971)\citenamefont {Doplicher}, \citenamefont {Haag},\ and\ \citenamefont {Roberts}}]{DHR1971LocalObsStatsI}\BibitemOpen
  \bibfield  {author} {\bibinfo {author} {\bibfnamefont {S.}~\bibnamefont {Doplicher}}, \bibinfo {author} {\bibfnamefont {R.}~\bibnamefont {Haag}},\ and\ \bibinfo {author} {\bibfnamefont {J.~E.}\ \bibnamefont {Roberts}},\ }\href {https://doi.org/10.1007/BF01877742} {\bibfield  {journal} {\bibinfo  {journal} {Communications in Mathematical Physics}\ }\textbf {\bibinfo {volume} {23}},\ \bibinfo {pages} {199} (\bibinfo {year} {1971})}\BibitemShut {NoStop}\relax
\bibitem [{\citenamefont {Doplicher}\ \emph {et~al.}(1974)\citenamefont {Doplicher}, \citenamefont {Haag},\ and\ \citenamefont {Roberts}}]{DHR1974LocalObsStatsII}\BibitemOpen
  \bibfield  {author} {\bibinfo {author} {\bibfnamefont {S.}~\bibnamefont {Doplicher}}, \bibinfo {author} {\bibfnamefont {R.}~\bibnamefont {Haag}},\ and\ \bibinfo {author} {\bibfnamefont {J.~E.}\ \bibnamefont {Roberts}},\ }\href {https://doi.org/10.1007/BF01646454} {\bibfield  {journal} {\bibinfo  {journal} {Communications in Mathematical Physics}\ }\textbf {\bibinfo {volume} {35}},\ \bibinfo {pages} {49} (\bibinfo {year} {1974})}\BibitemShut {NoStop}\relax
\bibitem [{\citenamefont {Doplicher}\ and\ \citenamefont {Roberts}(1990)}]{doplicherWhyThereField1990}\BibitemOpen
  \bibfield  {author} {\bibinfo {author} {\bibfnamefont {S.}~\bibnamefont {Doplicher}}\ and\ \bibinfo {author} {\bibfnamefont {J.~E.}\ \bibnamefont {Roberts}},\ }\href {https://doi.org/10.1007/BF02097680} {\bibfield  {journal} {\bibinfo  {journal} {Communications in Mathematical Physics}\ }\textbf {\bibinfo {volume} {131}},\ \bibinfo {pages} {51} (\bibinfo {year} {1990})}\BibitemShut {NoStop}\relax
\bibitem [{\citenamefont {Wang}\ and\ \citenamefont {Hazzard}(2026)}]{wangRparastatisticsFoundation2026}\BibitemOpen
  \bibfield  {author} {\bibinfo {author} {\bibfnamefont {Z.}~\bibnamefont {Wang}}\ and\ \bibinfo {author} {\bibfnamefont {K.~R.~A.}\ \bibnamefont {Hazzard}},\ }\href {https://arxiv.org/abs/2607.26351} {\bibinfo {title} {On {$R$}-parastatistics {I}: Foundation}} (\bibinfo {year} {2026}),\ \Eprint {https://arxiv.org/abs/2607.26351} {arXiv:2607.26351 [quant-ph]} \BibitemShut {NoStop}\relax
\bibitem [{\citenamefont {Salinel}\ and\ \citenamefont {Villegas}(2026)}]{salinelBosonization$R$paraparticleLuttinger2026}\BibitemOpen
  \bibfield  {author} {\bibinfo {author} {\bibfnamefont {D.~F.}\ \bibnamefont {Salinel}}\ and\ \bibinfo {author} {\bibfnamefont {K.~H.~A.}\ \bibnamefont {Villegas}},\ }\href {https://doi.org/10.1103/51qz-bmh5} {\bibfield  {journal} {\bibinfo  {journal} {Physical Review B}\ }\textbf {\bibinfo {volume} {113}},\ \bibinfo {pages} {245425} (\bibinfo {year} {2026})}\BibitemShut {NoStop}\relax
\bibitem [{\citenamefont {Schuricht}\ and\ \citenamefont {Sirker}(2026)}]{schurichtParastatisticsInteractingPeriodic2026}\BibitemOpen
  \bibfield  {author} {\bibinfo {author} {\bibfnamefont {D.}~\bibnamefont {Schuricht}}\ and\ \bibinfo {author} {\bibfnamefont {J.}~\bibnamefont {Sirker}},\ }\href {https://doi.org/10.1103/3spr-jfq4} {\bibfield  {journal} {\bibinfo  {journal} {Physical Review Letters}\ }\textbf {\bibinfo {volume} {137}},\ \bibinfo {pages} {066503} (\bibinfo {year} {2026})}\BibitemShut {NoStop}\relax
\bibitem [{\citenamefont {Thakur}\ and\ \citenamefont {Singh}(2025)}]{thakurStatisticalMechanicsParaparticles2025}\BibitemOpen
  \bibfield  {author} {\bibinfo {author} {\bibfnamefont {N.}~\bibnamefont {Thakur}}\ and\ \bibinfo {author} {\bibfnamefont {N.}~\bibnamefont {Singh}},\ }\href {https://doi.org/10.48550/arXiv.2509.00112} {\bibinfo {title} {Statistical {{Mechanics}} of {{Paraparticles}}}} (\bibinfo {year} {2025}),\ \Eprint {https://arxiv.org/abs/2509.00112} {arXiv:2509.00112 [cond-mat.stat-mech]} \BibitemShut {NoStop}\relax
\bibitem [{\citenamefont {Liu}(2026)}]{liuParaparticlesIntrinsicallyExhibit2026}\BibitemOpen
  \bibfield  {author} {\bibinfo {author} {\bibfnamefont {K.}~\bibnamefont {Liu}},\ }\href {https://doi.org/10.48550/arXiv.2607.11867} {\bibinfo {title} {Paraparticles intrinsically exhibit {{Hardy-space}} breakdown}} (\bibinfo {year} {2026}),\ \Eprint {https://arxiv.org/abs/2607.11867} {arXiv:2607.11867 [quant-ph]} \BibitemShut {NoStop}\relax
\bibitem [{\citenamefont {Mishra}\ and\ \citenamefont {Rajasekaran}(1991)}]{mishraAlgebraFermionsNew1991}\BibitemOpen
  \bibfield  {author} {\bibinfo {author} {\bibfnamefont {A.~K.}\ \bibnamefont {Mishra}}\ and\ \bibinfo {author} {\bibfnamefont {G.}~\bibnamefont {Rajasekaran}},\ }\href {https://doi.org/10.1007/BF02894926} {\bibfield  {journal} {\bibinfo  {journal} {Pramana}\ }\textbf {\bibinfo {volume} {36}},\ \bibinfo {pages} {537} (\bibinfo {year} {1991})}\BibitemShut {NoStop}\relax
\bibitem [{\citenamefont {Mishra}\ and\ \citenamefont {Rajasekaran}(2000)}]{mishraGeneralizedFockSpaces2000}\BibitemOpen
  \bibfield  {author} {\bibinfo {author} {\bibfnamefont {A.~K.}\ \bibnamefont {Mishra}}\ and\ \bibinfo {author} {\bibfnamefont {G.}~\bibnamefont {Rajasekaran}},\ }\href {https://doi.org/10.1063/1.1337725} {\bibfield  {journal} {\bibinfo  {journal} {AIP Conference Proceedings}\ }\textbf {\bibinfo {volume} {545}},\ \bibinfo {pages} {162} (\bibinfo {year} {2000})}\BibitemShut {NoStop}\relax
\bibitem [{\citenamefont {Jordan}\ and\ \citenamefont {Wigner}(1928)}]{JordanWigner1928}\BibitemOpen
  \bibfield  {author} {\bibinfo {author} {\bibfnamefont {P.}~\bibnamefont {Jordan}}\ and\ \bibinfo {author} {\bibfnamefont {E.}~\bibnamefont {Wigner}},\ }\href {https://doi.org/10.1007/BF01331938} {\bibfield  {journal} {\bibinfo  {journal} {Zeitschrift f\"ur Physik}\ }\textbf {\bibinfo {volume} {47}},\ \bibinfo {pages} {631} (\bibinfo {year} {1928})}\BibitemShut {NoStop}\relax
\bibitem [{\citenamefont {Fors}\ \emph {et~al.}(2024)\citenamefont {Fors}, \citenamefont {Fern{\'a}ndez-Pend{\'a}s},\ and\ \citenamefont {Kockum}}]{Fors2024ZZComprehensive}\BibitemOpen
  \bibfield  {author} {\bibinfo {author} {\bibfnamefont {S.~P.}\ \bibnamefont {Fors}}, \bibinfo {author} {\bibfnamefont {J.}~\bibnamefont {Fern{\'a}ndez-Pend{\'a}s}},\ and\ \bibinfo {author} {\bibfnamefont {A.~F.}\ \bibnamefont {Kockum}},\ }\href {https://arxiv.org/abs/2408.15402} {\bibinfo {title} {Comprehensive explanation of {{$ZZ$}} coupling in superconducting qubits}} (\bibinfo {year} {2024}),\ \Eprint {https://arxiv.org/abs/2408.15402} {arXiv:2408.15402 [quant-ph]} \BibitemShut {NoStop}\relax
\bibitem [{\citenamefont {Zhao}\ \emph {et~al.}(2021)\citenamefont {Zhao}, \citenamefont {Lan}, \citenamefont {Xu}, \citenamefont {Xue}, \citenamefont {Blank}, \citenamefont {Tan}, \citenamefont {Yu},\ and\ \citenamefont {Yu}}]{Zhao2021SuppressStaticZZ}\BibitemOpen
  \bibfield  {author} {\bibinfo {author} {\bibfnamefont {P.}~\bibnamefont {Zhao}}, \bibinfo {author} {\bibfnamefont {D.}~\bibnamefont {Lan}}, \bibinfo {author} {\bibfnamefont {P.}~\bibnamefont {Xu}}, \bibinfo {author} {\bibfnamefont {G.}~\bibnamefont {Xue}}, \bibinfo {author} {\bibfnamefont {M.}~\bibnamefont {Blank}}, \bibinfo {author} {\bibfnamefont {X.}~\bibnamefont {Tan}}, \bibinfo {author} {\bibfnamefont {H.}~\bibnamefont {Yu}},\ and\ \bibinfo {author} {\bibfnamefont {Y.}~\bibnamefont {Yu}},\ }\href {https://doi.org/10.1103/PhysRevApplied.16.024037} {\bibfield  {journal} {\bibinfo  {journal} {Physical Review Applied}\ }\textbf {\bibinfo {volume} {16}},\ \bibinfo {pages} {024037} (\bibinfo {year} {2021})}\BibitemShut {NoStop}\relax
\bibitem [{\citenamefont {Zhao}\ \emph {et~al.}(2020)\citenamefont {Zhao}, \citenamefont {Xu}, \citenamefont {Lan}, \citenamefont {Chu}, \citenamefont {Tan}, \citenamefont {Yu},\ and\ \citenamefont {Yu}}]{zhao2020highcontrast}\BibitemOpen
  \bibfield  {author} {\bibinfo {author} {\bibfnamefont {P.}~\bibnamefont {Zhao}}, \bibinfo {author} {\bibfnamefont {P.}~\bibnamefont {Xu}}, \bibinfo {author} {\bibfnamefont {D.}~\bibnamefont {Lan}}, \bibinfo {author} {\bibfnamefont {J.}~\bibnamefont {Chu}}, \bibinfo {author} {\bibfnamefont {X.}~\bibnamefont {Tan}}, \bibinfo {author} {\bibfnamefont {H.}~\bibnamefont {Yu}},\ and\ \bibinfo {author} {\bibfnamefont {Y.}~\bibnamefont {Yu}},\ }\href {https://doi.org/10.1103/PhysRevLett.125.200503} {\bibfield  {journal} {\bibinfo  {journal} {Physical Review Letters}\ }\textbf {\bibinfo {volume} {125}},\ \bibinfo {pages} {200503} (\bibinfo {year} {2020})}\BibitemShut {NoStop}\relax
\bibitem [{\citenamefont {Strand}\ \emph {et~al.}(2013)\citenamefont {Strand}, \citenamefont {Ware}, \citenamefont {Beaudoin}, \citenamefont {Ohki}, \citenamefont {Johnson}, \citenamefont {Blais},\ and\ \citenamefont {Plourde}}]{strandFirstorderSidebandTransitions2013a}\BibitemOpen
  \bibfield  {author} {\bibinfo {author} {\bibfnamefont {J.~D.}\ \bibnamefont {Strand}}, \bibinfo {author} {\bibfnamefont {M.}~\bibnamefont {Ware}}, \bibinfo {author} {\bibfnamefont {F.}~\bibnamefont {Beaudoin}}, \bibinfo {author} {\bibfnamefont {T.~A.}\ \bibnamefont {Ohki}}, \bibinfo {author} {\bibfnamefont {B.~R.}\ \bibnamefont {Johnson}}, \bibinfo {author} {\bibfnamefont {A.}~\bibnamefont {Blais}},\ and\ \bibinfo {author} {\bibfnamefont {B.~L.~T.}\ \bibnamefont {Plourde}},\ }\href {https://doi.org/10.1103/PhysRevB.87.220505} {\bibfield  {journal} {\bibinfo  {journal} {Physical Review B}\ }\textbf {\bibinfo {volume} {87}},\ \bibinfo {pages} {220505} (\bibinfo {year} {2013})}\BibitemShut {NoStop}\relax
\bibitem [{\citenamefont {Rosen}\ \emph {et~al.}(2024)\citenamefont {Rosen}, \citenamefont {Muschinske}, \citenamefont {Barrett}, \citenamefont {Chatterjee}, \citenamefont {Hays}, \citenamefont {DeMarco}, \citenamefont {Karamlou}, \citenamefont {Rower}, \citenamefont {Das}, \citenamefont {Kim}, \citenamefont {Niedzielski}, \citenamefont {Schuldt}, \citenamefont {Serniak}, \citenamefont {Schwartz}, \citenamefont {Yoder}, \citenamefont {Grover},\ and\ \citenamefont {Oliver}}]{rosenSyntheticMagneticVector2024}\BibitemOpen
  \bibfield  {author} {\bibinfo {author} {\bibfnamefont {I.~T.}\ \bibnamefont {Rosen}}, \bibinfo {author} {\bibfnamefont {S.}~\bibnamefont {Muschinske}}, \bibinfo {author} {\bibfnamefont {C.~N.}\ \bibnamefont {Barrett}}, \bibinfo {author} {\bibfnamefont {A.}~\bibnamefont {Chatterjee}}, \bibinfo {author} {\bibfnamefont {M.}~\bibnamefont {Hays}}, \bibinfo {author} {\bibfnamefont {M.~A.}\ \bibnamefont {DeMarco}}, \bibinfo {author} {\bibfnamefont {A.~H.}\ \bibnamefont {Karamlou}}, \bibinfo {author} {\bibfnamefont {D.~A.}\ \bibnamefont {Rower}}, \bibinfo {author} {\bibfnamefont {R.}~\bibnamefont {Das}}, \bibinfo {author} {\bibfnamefont {D.~K.}\ \bibnamefont {Kim}}, \bibinfo {author} {\bibfnamefont {B.~M.}\ \bibnamefont {Niedzielski}}, \bibinfo {author} {\bibfnamefont {M.}~\bibnamefont {Schuldt}}, \bibinfo {author} {\bibfnamefont {K.}~\bibnamefont {Serniak}}, \bibinfo {author} {\bibfnamefont {M.~E.}\ \bibnamefont {Schwartz}}, \bibinfo {author} {\bibfnamefont {J.~L.}\ \bibnamefont {Yoder}}, \bibinfo {author} {\bibfnamefont {J.~A.}\ \bibnamefont {Grover}},\ and\ \bibinfo {author} {\bibfnamefont {W.~D.}\ \bibnamefont {Oliver}},\ }\href {https://doi.org/10.1038/s41567-024-02661-3} {\bibfield  {journal} {\bibinfo  {journal} {Nature Physics}\ }\textbf {\bibinfo {volume} {20}},\ \bibinfo {pages} {1881} (\bibinfo {year} {2024})}\BibitemShut {NoStop}\relax
\bibitem [{\citenamefont {Aharonov}\ and\ \citenamefont {Anandan}(1987)}]{aharonovPhaseChangeCyclic1987}\BibitemOpen
  \bibfield  {author} {\bibinfo {author} {\bibfnamefont {Y.}~\bibnamefont {Aharonov}}\ and\ \bibinfo {author} {\bibfnamefont {J.}~\bibnamefont {Anandan}},\ }\href {https://doi.org/10.1103/PhysRevLett.58.1593} {\bibfield  {journal} {\bibinfo  {journal} {Physical Review Letters}\ }\textbf {\bibinfo {volume} {58}},\ \bibinfo {pages} {1593} (\bibinfo {year} {1987})}\BibitemShut {NoStop}\relax
\bibitem [{\citenamefont {Loudon}(1998)}]{loudon1998fermion}\BibitemOpen
  \bibfield  {author} {\bibinfo {author} {\bibfnamefont {R.}~\bibnamefont {Loudon}},\ }\href {https://doi.org/10.1103/PhysRevA.58.4904} {\bibfield  {journal} {\bibinfo  {journal} {Physical Review A}\ }\textbf {\bibinfo {volume} {58}},\ \bibinfo {pages} {4904} (\bibinfo {year} {1998})}\BibitemShut {NoStop}\relax
\bibitem [{\citenamefont {Hance}(2025)}]{hance2025fermion}\BibitemOpen
  \bibfield  {author} {\bibinfo {author} {\bibfnamefont {J.~R.}\ \bibnamefont {Hance}},\ }\href {https://doi.org/10.1088/2515-7647/ad8d96} {\bibfield  {journal} {\bibinfo  {journal} {Journal of Physics: Photonics}\ }\textbf {\bibinfo {volume} {7}},\ \bibinfo {pages} {012001} (\bibinfo {year} {2025})}\BibitemShut {NoStop}\relax
\bibitem [{\citenamefont {Hong}\ \emph {et~al.}(1987)\citenamefont {Hong}, \citenamefont {Ou},\ and\ \citenamefont {Mandel}}]{Hong1987HOM}\BibitemOpen
  \bibfield  {author} {\bibinfo {author} {\bibfnamefont {C.~K.}\ \bibnamefont {Hong}}, \bibinfo {author} {\bibfnamefont {Z.~Y.}\ \bibnamefont {Ou}},\ and\ \bibinfo {author} {\bibfnamefont {L.}~\bibnamefont {Mandel}},\ }\href {https://doi.org/10.1103/PhysRevLett.59.2044} {\bibfield  {journal} {\bibinfo  {journal} {Physical Review Letters}\ }\textbf {\bibinfo {volume} {59}},\ \bibinfo {pages} {2044} (\bibinfo {year} {1987})}\BibitemShut {NoStop}\relax
\bibitem [{\citenamefont {Lopes}\ \emph {et~al.}(2015)\citenamefont {Lopes}, \citenamefont {Imanaliev}, \citenamefont {Aspect}, \citenamefont {Cheneau}, \citenamefont {Boiron},\ and\ \citenamefont {Westbrook}}]{Lopes2015AtomicHOM}\BibitemOpen
  \bibfield  {author} {\bibinfo {author} {\bibfnamefont {R.}~\bibnamefont {Lopes}}, \bibinfo {author} {\bibfnamefont {A.}~\bibnamefont {Imanaliev}}, \bibinfo {author} {\bibfnamefont {A.}~\bibnamefont {Aspect}}, \bibinfo {author} {\bibfnamefont {M.}~\bibnamefont {Cheneau}}, \bibinfo {author} {\bibfnamefont {D.}~\bibnamefont {Boiron}},\ and\ \bibinfo {author} {\bibfnamefont {C.~I.}\ \bibnamefont {Westbrook}},\ }\href {https://doi.org/10.1038/nature14331} {\bibfield  {journal} {\bibinfo  {journal} {Nature}\ }\textbf {\bibinfo {volume} {520}},\ \bibinfo {pages} {66} (\bibinfo {year} {2015})}\BibitemShut {NoStop}\relax
\bibitem [{\citenamefont {Liu}\ \emph {et~al.}(1998)\citenamefont {Liu}, \citenamefont {Odom}, \citenamefont {Yamamoto},\ and\ \citenamefont {Tarucha}}]{liu1998QuantumInterferenceElectron}\BibitemOpen
  \bibfield  {author} {\bibinfo {author} {\bibfnamefont {R.~C.}\ \bibnamefont {Liu}}, \bibinfo {author} {\bibfnamefont {B.}~\bibnamefont {Odom}}, \bibinfo {author} {\bibfnamefont {Y.}~\bibnamefont {Yamamoto}},\ and\ \bibinfo {author} {\bibfnamefont {S.}~\bibnamefont {Tarucha}},\ }\href {https://doi.org/10.1038/34611} {\bibfield  {journal} {\bibinfo  {journal} {Nature}\ }\textbf {\bibinfo {volume} {391}},\ \bibinfo {pages} {263} (\bibinfo {year} {1998})}\BibitemShut {NoStop}\relax
\bibitem [{\citenamefont {Bocquillon}\ \emph {et~al.}(2013)\citenamefont {Bocquillon}, \citenamefont {Freulon}, \citenamefont {Berroir}, \citenamefont {Degiovanni}, \citenamefont {Pla{\c{c}}ais}, \citenamefont {Cavanna}, \citenamefont {Jin},\ and\ \citenamefont {F{\`e}ve}}]{Bocquillon2013ElectronicHOM}\BibitemOpen
  \bibfield  {author} {\bibinfo {author} {\bibfnamefont {E.}~\bibnamefont {Bocquillon}}, \bibinfo {author} {\bibfnamefont {V.}~\bibnamefont {Freulon}}, \bibinfo {author} {\bibfnamefont {J.-M.}\ \bibnamefont {Berroir}}, \bibinfo {author} {\bibfnamefont {P.}~\bibnamefont {Degiovanni}}, \bibinfo {author} {\bibfnamefont {B.}~\bibnamefont {Pla{\c{c}}ais}}, \bibinfo {author} {\bibfnamefont {A.}~\bibnamefont {Cavanna}}, \bibinfo {author} {\bibfnamefont {Y.}~\bibnamefont {Jin}},\ and\ \bibinfo {author} {\bibfnamefont {G.}~\bibnamefont {F{\`e}ve}},\ }\href {https://doi.org/10.1126/science.1232572} {\bibfield  {journal} {\bibinfo  {journal} {Science}\ }\textbf {\bibinfo {volume} {339}},\ \bibinfo {pages} {1054} (\bibinfo {year} {2013})}\BibitemShut {NoStop}\relax
\bibitem [{\citenamefont {Zotos}\ \emph {et~al.}(1997)\citenamefont {Zotos}, \citenamefont {Naef},\ and\ \citenamefont {Prelovsek}}]{zotos1997TransportConservationLaws}\BibitemOpen
  \bibfield  {author} {\bibinfo {author} {\bibfnamefont {X.}~\bibnamefont {Zotos}}, \bibinfo {author} {\bibfnamefont {F.}~\bibnamefont {Naef}},\ and\ \bibinfo {author} {\bibfnamefont {P.}~\bibnamefont {Prelovsek}},\ }\href {https://doi.org/10.1103/PhysRevB.55.11029} {\bibfield  {journal} {\bibinfo  {journal} {Physical Review B}\ }\textbf {\bibinfo {volume} {55}},\ \bibinfo {pages} {11029} (\bibinfo {year} {1997})}\BibitemShut {NoStop}\relax
\bibitem [{\citenamefont {Rigol}\ \emph {et~al.}(2008)\citenamefont {Rigol}, \citenamefont {Dunjko},\ and\ \citenamefont {Olshanii}}]{rigol2008ThermalizationItsMechanism}\BibitemOpen
  \bibfield  {author} {\bibinfo {author} {\bibfnamefont {M.}~\bibnamefont {Rigol}}, \bibinfo {author} {\bibfnamefont {V.}~\bibnamefont {Dunjko}},\ and\ \bibinfo {author} {\bibfnamefont {M.}~\bibnamefont {Olshanii}},\ }\href {https://doi.org/10.1038/nature06838} {\bibfield  {journal} {\bibinfo  {journal} {Nature}\ }\textbf {\bibinfo {volume} {452}},\ \bibinfo {pages} {854} (\bibinfo {year} {2008})}\BibitemShut {NoStop}\relax
\bibitem [{\citenamefont {Kaufman}\ \emph {et~al.}(2016)\citenamefont {Kaufman}, \citenamefont {Tai}, \citenamefont {Lukin}, \citenamefont {Rispoli}, \citenamefont {Schittko}, \citenamefont {Preiss},\ and\ \citenamefont {Greiner}}]{kaufmanETH2016}\BibitemOpen
  \bibfield  {author} {\bibinfo {author} {\bibfnamefont {A.~M.}\ \bibnamefont {Kaufman}}, \bibinfo {author} {\bibfnamefont {M.~E.}\ \bibnamefont {Tai}}, \bibinfo {author} {\bibfnamefont {A.}~\bibnamefont {Lukin}}, \bibinfo {author} {\bibfnamefont {M.}~\bibnamefont {Rispoli}}, \bibinfo {author} {\bibfnamefont {R.}~\bibnamefont {Schittko}}, \bibinfo {author} {\bibfnamefont {P.~M.}\ \bibnamefont {Preiss}},\ and\ \bibinfo {author} {\bibfnamefont {M.}~\bibnamefont {Greiner}},\ }\href {https://doi.org/10.1126/science.aaf6725} {\bibfield  {journal} {\bibinfo  {journal} {Science}\ }\textbf {\bibinfo {volume} {353}},\ \bibinfo {pages} {794} (\bibinfo {year} {2016})}\BibitemShut {NoStop}\relax
\bibitem [{\citenamefont {Deutsch}(2018)}]{deutsch2018EigenstateThermalizationHypothesis}\BibitemOpen
  \bibfield  {author} {\bibinfo {author} {\bibfnamefont {J.~M.}\ \bibnamefont {Deutsch}},\ }\href {https://doi.org/10.1088/1361-6633/aac9f1} {\bibfield  {journal} {\bibinfo  {journal} {Reports on Progress in Physics}\ }\textbf {\bibinfo {volume} {81}},\ \bibinfo {pages} {082001} (\bibinfo {year} {2018})}\BibitemShut {NoStop}\relax
\bibitem [{\citenamefont {Luttinger}(1963)}]{luttingerExactlySolubleModel1963}\BibitemOpen
  \bibfield  {author} {\bibinfo {author} {\bibfnamefont {J.~M.}\ \bibnamefont {Luttinger}},\ }\href {https://doi.org/10.1063/1.1704046} {\bibfield  {journal} {\bibinfo  {journal} {Journal of Mathematical Physics}\ }\textbf {\bibinfo {volume} {4}},\ \bibinfo {pages} {1154} (\bibinfo {year} {1963})}\BibitemShut {NoStop}\relax
\bibitem [{\citenamefont {Haldane}(1981)}]{haldaneLuttingerLiquidTheory1981}\BibitemOpen
  \bibfield  {author} {\bibinfo {author} {\bibfnamefont {F.~D.~M.}\ \bibnamefont {Haldane}},\ }\href {https://doi.org/10.1088/0022-3719/14/19/010} {\bibfield  {journal} {\bibinfo  {journal} {Journal of Physics C: Solid State Physics}\ }\textbf {\bibinfo {volume} {14}},\ \bibinfo {pages} {2585} (\bibinfo {year} {1981})}\BibitemShut {NoStop}\relax
\bibitem [{\citenamefont {Terhal}(2015)}]{terhalQuantumErrorCorrection2015}\BibitemOpen
  \bibfield  {author} {\bibinfo {author} {\bibfnamefont {B.~M.}\ \bibnamefont {Terhal}},\ }\href {https://doi.org/10.1103/RevModPhys.87.307} {\bibfield  {journal} {\bibinfo  {journal} {Reviews of Modern Physics}\ }\textbf {\bibinfo {volume} {87}},\ \bibinfo {pages} {307} (\bibinfo {year} {2015})}\BibitemShut {NoStop}\relax
\end{thebibliography}


\begin{thebibliography}{16}
\makeatletter
\providecommand \@ifxundefined [1]{\@ifx{#1\undefined} }
\providecommand \@ifnum [1]{\ifnum #1\expandafter \@firstoftwo \else \expandafter \@secondoftwo \fi }
\providecommand \@ifx [1]{\ifx #1\expandafter \@firstoftwo \else \expandafter \@secondoftwo \fi }
\providecommand \natexlab [1]{#1}
\providecommand \enquote  [1]{``#1''}
\providecommand \bibnamefont  [1]{#1}
\providecommand \bibfnamefont [1]{#1}
\providecommand \citenamefont [1]{#1}
\providecommand \href@noop [0]{\@secondoftwo}
\providecommand \href [0]{\begingroup \@sanitize@url \@href}
\providecommand \@href[1]{\@@startlink{#1}\@@href}
\providecommand \@@href[1]{\endgroup#1\@@endlink}
\providecommand \@sanitize@url [0]{\catcode `\\12\catcode `\$12\catcode `\&12\catcode `\#12\catcode `\^12\catcode `\_12\catcode `\%12\relax}
\providecommand \@@startlink[1]{}
\providecommand \@@endlink[0]{}
\providecommand \url  [0]{\begingroup\@sanitize@url \@url }
\providecommand \@url [1]{\endgroup\@href {#1}{\urlprefix }}
\providecommand \urlprefix  [0]{URL }
\providecommand \Eprint [0]{\href }
\providecommand \doibase [0]{https://doi.org/}
\providecommand \selectlanguage [0]{\@gobble}
\providecommand \bibinfo  [0]{\@secondoftwo}
\providecommand \bibfield  [0]{\@secondoftwo}
\providecommand \translation [1]{[#1]}
\providecommand \BibitemOpen [0]{}
\providecommand \bibitemStop [0]{}
\providecommand \bibitemNoStop [0]{.\EOS\space}
\providecommand \EOS [0]{\spacefactor3000\relax}
\providecommand \BibitemShut  [1]{\csname bibitem#1\endcsname}
\let\auto@bib@innerbib\@empty
\bibitem [{\citenamefont {Koch}\ \emph {et~al.}(2007)\citenamefont {Koch}, \citenamefont {Yu}, \citenamefont {Gambetta}, \citenamefont {Houck}, \citenamefont {Schuster}, \citenamefont {Majer}, \citenamefont {Blais}, \citenamefont {Devoret}, \citenamefont {Girvin},\ and\ \citenamefont {Schoelkopf}}]{kochChargeinsensitiveQubitDesign2007}\BibitemOpen
  \bibfield  {author} {\bibinfo {author} {\bibfnamefont {J.}~\bibnamefont {Koch}}, \bibinfo {author} {\bibfnamefont {T.~M.}\ \bibnamefont {Yu}}, \bibinfo {author} {\bibfnamefont {J.}~\bibnamefont {Gambetta}}, \bibinfo {author} {\bibfnamefont {A.~A.}\ \bibnamefont {Houck}}, \bibinfo {author} {\bibfnamefont {D.~I.}\ \bibnamefont {Schuster}}, \bibinfo {author} {\bibfnamefont {J.}~\bibnamefont {Majer}}, \bibinfo {author} {\bibfnamefont {A.}~\bibnamefont {Blais}}, \bibinfo {author} {\bibfnamefont {M.~H.}\ \bibnamefont {Devoret}}, \bibinfo {author} {\bibfnamefont {S.~M.}\ \bibnamefont {Girvin}},\ and\ \bibinfo {author} {\bibfnamefont {R.~J.}\ \bibnamefont {Schoelkopf}},\ }\href {https://doi.org/10.1103/PhysRevA.76.042319} {\bibfield  {journal} {\bibinfo  {journal} {Physical Review A}\ }\textbf {\bibinfo {volume} {76}},\ \bibinfo {pages} {042319} (\bibinfo {year} {2007})}\BibitemShut {NoStop}\relax
\bibitem [{\citenamefont {Krantz}\ \emph {et~al.}(2019)\citenamefont {Krantz}, \citenamefont {Kjaergaard}, \citenamefont {Yan}, \citenamefont {Orlando}, \citenamefont {Gustavsson},\ and\ \citenamefont {Oliver}}]{krantzQuantumEngineersGuide2019}\BibitemOpen
  \bibfield  {author} {\bibinfo {author} {\bibfnamefont {P.}~\bibnamefont {Krantz}}, \bibinfo {author} {\bibfnamefont {M.}~\bibnamefont {Kjaergaard}}, \bibinfo {author} {\bibfnamefont {F.}~\bibnamefont {Yan}}, \bibinfo {author} {\bibfnamefont {T.~P.}\ \bibnamefont {Orlando}}, \bibinfo {author} {\bibfnamefont {S.}~\bibnamefont {Gustavsson}},\ and\ \bibinfo {author} {\bibfnamefont {W.~D.}\ \bibnamefont {Oliver}},\ }\href {https://doi.org/10.1063/1.5089550} {\bibfield  {journal} {\bibinfo  {journal} {Applied Physics Reviews}\ }\textbf {\bibinfo {volume} {6}},\ \bibinfo {pages} {021318} (\bibinfo {year} {2019})}\BibitemShut {NoStop}\relax
\bibitem [{\citenamefont {Naghiloo}(2019)}]{naghilooIntroductionExperimentalQuantum2019}\BibitemOpen
  \bibfield  {author} {\bibinfo {author} {\bibfnamefont {M.}~\bibnamefont {Naghiloo}},\ }\href@noop {} {\bibinfo {title} {Introduction to experimental quantum measurement with superconducting qubits}} (\bibinfo {year} {2019}),\ \Eprint {https://arxiv.org/abs/1904.09291} {arXiv:1904.09291 [quant-ph]} \BibitemShut {NoStop}\relax
\bibitem [{\citenamefont {Karamlou}\ \emph {et~al.}(2024)\citenamefont {Karamlou}, \citenamefont {Rosen}, \citenamefont {Muschinske}, \citenamefont {Barrett}, \citenamefont {Di~Paolo}, \citenamefont {Ding}, \citenamefont {Harrington}, \citenamefont {Hays}, \citenamefont {Das}, \citenamefont {Kim}, \citenamefont {Niedzielski}, \citenamefont {Schuldt}, \citenamefont {Serniak}, \citenamefont {Schwartz}, \citenamefont {Yoder}, \citenamefont {Gustavsson}, \citenamefont {Yanay}, \citenamefont {Grover},\ and\ \citenamefont {Oliver}}]{Karamlou2024EntanglementHCBH}\BibitemOpen
  \bibfield  {author} {\bibinfo {author} {\bibfnamefont {A.~H.}\ \bibnamefont {Karamlou}}, \bibinfo {author} {\bibfnamefont {I.~T.}\ \bibnamefont {Rosen}}, \bibinfo {author} {\bibfnamefont {S.~E.}\ \bibnamefont {Muschinske}}, \bibinfo {author} {\bibfnamefont {C.~N.}\ \bibnamefont {Barrett}}, \bibinfo {author} {\bibfnamefont {A.}~\bibnamefont {Di~Paolo}}, \bibinfo {author} {\bibfnamefont {L.}~\bibnamefont {Ding}}, \bibinfo {author} {\bibfnamefont {P.~M.}\ \bibnamefont {Harrington}}, \bibinfo {author} {\bibfnamefont {M.}~\bibnamefont {Hays}}, \bibinfo {author} {\bibfnamefont {R.}~\bibnamefont {Das}}, \bibinfo {author} {\bibfnamefont {D.~K.}\ \bibnamefont {Kim}}, \bibinfo {author} {\bibfnamefont {B.~M.}\ \bibnamefont {Niedzielski}}, \bibinfo {author} {\bibfnamefont {M.}~\bibnamefont {Schuldt}}, \bibinfo {author} {\bibfnamefont {K.}~\bibnamefont {Serniak}}, \bibinfo {author} {\bibfnamefont {M.~E.}\ \bibnamefont {Schwartz}}, \bibinfo {author} {\bibfnamefont {J.~L.}\ \bibnamefont {Yoder}}, \bibinfo {author} {\bibfnamefont {S.}~\bibnamefont {Gustavsson}}, \bibinfo {author} {\bibfnamefont {Y.}~\bibnamefont {Yanay}}, \bibinfo {author} {\bibfnamefont {J.~A.}\ \bibnamefont {Grover}},\ and\ \bibinfo {author} {\bibfnamefont {W.~D.}\ \bibnamefont {Oliver}},\ }\href {https://doi.org/10.1038/s41586-024-07325-z} {\bibfield  {journal} {\bibinfo  {journal} {Nature}\ }\textbf {\bibinfo {volume} {629}},\ \bibinfo {pages} {561} (\bibinfo {year} {2024})}\BibitemShut {NoStop}\relax
\bibitem [{\citenamefont {Barrett}\ \emph {et~al.}(2023)\citenamefont {Barrett}, \citenamefont {Karamlou}, \citenamefont {Muschinske}, \citenamefont {Rosen}, \citenamefont {Braum{\"u}ller}, \citenamefont {Das}, \citenamefont {Kim}, \citenamefont {Niedzielski}, \citenamefont {Schuldt}, \citenamefont {Serniak}, \citenamefont {Schwartz}, \citenamefont {Yoder}, \citenamefont {Orlando}, \citenamefont {Gustavsson}, \citenamefont {Grover},\ and\ \citenamefont {Oliver}}]{barrett2023LearningBasedCalibrationFlux}\BibitemOpen
  \bibfield  {author} {\bibinfo {author} {\bibfnamefont {C.~N.}\ \bibnamefont {Barrett}}, \bibinfo {author} {\bibfnamefont {A.~H.}\ \bibnamefont {Karamlou}}, \bibinfo {author} {\bibfnamefont {S.~E.}\ \bibnamefont {Muschinske}}, \bibinfo {author} {\bibfnamefont {I.~T.}\ \bibnamefont {Rosen}}, \bibinfo {author} {\bibfnamefont {J.}~\bibnamefont {Braum{\"u}ller}}, \bibinfo {author} {\bibfnamefont {R.}~\bibnamefont {Das}}, \bibinfo {author} {\bibfnamefont {D.~K.}\ \bibnamefont {Kim}}, \bibinfo {author} {\bibfnamefont {B.~M.}\ \bibnamefont {Niedzielski}}, \bibinfo {author} {\bibfnamefont {M.}~\bibnamefont {Schuldt}}, \bibinfo {author} {\bibfnamefont {K.}~\bibnamefont {Serniak}}, \bibinfo {author} {\bibfnamefont {M.~E.}\ \bibnamefont {Schwartz}}, \bibinfo {author} {\bibfnamefont {J.~L.}\ \bibnamefont {Yoder}}, \bibinfo {author} {\bibfnamefont {T.~P.}\ \bibnamefont {Orlando}}, \bibinfo {author} {\bibfnamefont {S.}~\bibnamefont {Gustavsson}}, \bibinfo {author} {\bibfnamefont {J.~A.}\ \bibnamefont {Grover}},\ and\ \bibinfo {author} {\bibfnamefont {W.~D.}\ \bibnamefont {Oliver}},\ }\href {https://doi.org/10.1103/PhysRevApplied.20.024070} {\bibfield  {journal} {\bibinfo  {journal} {Physical Review Applied}\ }\textbf {\bibinfo {volume} {20}},\ \bibinfo {pages} {024070} (\bibinfo {year} {2023})}\BibitemShut {NoStop}\relax
\bibitem [{\citenamefont {Rol}\ \emph {et~al.}(2020)\citenamefont {Rol}, \citenamefont {Ciorciaro}, \citenamefont {Malinowski}, \citenamefont {Tarasinski}, \citenamefont {Sagastizabal}, \citenamefont {Bultink}, \citenamefont {Salathe}, \citenamefont {Haandbaek}, \citenamefont {Sedivy},\ and\ \citenamefont {DiCarlo}}]{rolTimedomainCharacterizationCorrection2020a}\BibitemOpen
  \bibfield  {author} {\bibinfo {author} {\bibfnamefont {M.~A.}\ \bibnamefont {Rol}}, \bibinfo {author} {\bibfnamefont {L.}~\bibnamefont {Ciorciaro}}, \bibinfo {author} {\bibfnamefont {F.~K.}\ \bibnamefont {Malinowski}}, \bibinfo {author} {\bibfnamefont {B.~M.}\ \bibnamefont {Tarasinski}}, \bibinfo {author} {\bibfnamefont {R.~E.}\ \bibnamefont {Sagastizabal}}, \bibinfo {author} {\bibfnamefont {C.~C.}\ \bibnamefont {Bultink}}, \bibinfo {author} {\bibfnamefont {Y.}~\bibnamefont {Salathe}}, \bibinfo {author} {\bibfnamefont {N.}~\bibnamefont {Haandbaek}}, \bibinfo {author} {\bibfnamefont {J.}~\bibnamefont {Sedivy}},\ and\ \bibinfo {author} {\bibfnamefont {L.}~\bibnamefont {DiCarlo}},\ }\href {https://doi.org/10.1063/1.5133894} {\bibfield  {journal} {\bibinfo  {journal} {Applied Physics Letters}\ }\textbf {\bibinfo {volume} {116}},\ \bibinfo {pages} {054001} (\bibinfo {year} {2020})}\BibitemShut {NoStop}\relax
\bibitem [{\citenamefont {Hellings}\ \emph {et~al.}(2025)\citenamefont {Hellings}, \citenamefont {Lacroix}, \citenamefont {Remm}, \citenamefont {Boell}, \citenamefont {Herrmann}, \citenamefont {Laz{\u a}r}, \citenamefont {Krinner}, \citenamefont {Swiadek}, \citenamefont {Andersen}, \citenamefont {Eichler},\ and\ \citenamefont {Wallraff}}]{hellingsCalibratingMagneticFlux2025a}\BibitemOpen
  \bibfield  {author} {\bibinfo {author} {\bibfnamefont {C.}~\bibnamefont {Hellings}}, \bibinfo {author} {\bibfnamefont {N.}~\bibnamefont {Lacroix}}, \bibinfo {author} {\bibfnamefont {A.}~\bibnamefont {Remm}}, \bibinfo {author} {\bibfnamefont {R.}~\bibnamefont {Boell}}, \bibinfo {author} {\bibfnamefont {J.}~\bibnamefont {Herrmann}}, \bibinfo {author} {\bibfnamefont {S.}~\bibnamefont {Laz{\u a}r}}, \bibinfo {author} {\bibfnamefont {S.}~\bibnamefont {Krinner}}, \bibinfo {author} {\bibfnamefont {F.}~\bibnamefont {Swiadek}}, \bibinfo {author} {\bibfnamefont {C.~K.}\ \bibnamefont {Andersen}}, \bibinfo {author} {\bibfnamefont {C.}~\bibnamefont {Eichler}},\ and\ \bibinfo {author} {\bibfnamefont {A.}~\bibnamefont {Wallraff}},\ }\href {https://doi.org/10.1103/1qhb-r4fb} {\bibfield  {journal} {\bibinfo  {journal} {Physical Review Research}\ }\textbf {\bibinfo {volume} {7}},\ \bibinfo {pages} {043142} (\bibinfo {year} {2025})}\BibitemShut {NoStop}\relax
\bibitem [{\citenamefont {Hofheinz}\ \emph {et~al.}(2009)\citenamefont {Hofheinz}, \citenamefont {Wang}, \citenamefont {Ansmann}, \citenamefont {Bialczak}, \citenamefont {Lucero}, \citenamefont {Neeley}, \citenamefont {O'Connell}, \citenamefont {Sank}, \citenamefont {Wenner}, \citenamefont {Martinis},\ and\ \citenamefont {Cleland}}]{hofheinzSynthesizingArbitraryQuantum2009}\BibitemOpen
  \bibfield  {author} {\bibinfo {author} {\bibfnamefont {M.}~\bibnamefont {Hofheinz}}, \bibinfo {author} {\bibfnamefont {H.}~\bibnamefont {Wang}}, \bibinfo {author} {\bibfnamefont {M.}~\bibnamefont {Ansmann}}, \bibinfo {author} {\bibfnamefont {R.~C.}\ \bibnamefont {Bialczak}}, \bibinfo {author} {\bibfnamefont {E.}~\bibnamefont {Lucero}}, \bibinfo {author} {\bibfnamefont {M.}~\bibnamefont {Neeley}}, \bibinfo {author} {\bibfnamefont {A.~D.}\ \bibnamefont {O'Connell}}, \bibinfo {author} {\bibfnamefont {D.}~\bibnamefont {Sank}}, \bibinfo {author} {\bibfnamefont {J.}~\bibnamefont {Wenner}}, \bibinfo {author} {\bibfnamefont {J.~M.}\ \bibnamefont {Martinis}},\ and\ \bibinfo {author} {\bibfnamefont {A.~N.}\ \bibnamefont {Cleland}},\ }\href {https://doi.org/10.1038/nature08005} {\bibfield  {journal} {\bibinfo  {journal} {Nature}\ }\textbf {\bibinfo {volume} {459}},\ \bibinfo {pages} {546} (\bibinfo {year} {2009})}\BibitemShut {NoStop}\relax
\bibitem [{\citenamefont {Guo}\ \emph {et~al.}(2024)\citenamefont {Guo}, \citenamefont {Duan}, \citenamefont {Zhang}, \citenamefont {Yang}, \citenamefont {Zhang}, \citenamefont {Du}, \citenamefont {Zhang}, \citenamefont {Tao}, \citenamefont {Wang}, \citenamefont {Jia}, \citenamefont {Chen},\ and\ \citenamefont {Guo}}]{guoUniversalScalableCharacterization2024}\BibitemOpen
  \bibfield  {author} {\bibinfo {author} {\bibfnamefont {L.-L.}\ \bibnamefont {Guo}}, \bibinfo {author} {\bibfnamefont {P.}~\bibnamefont {Duan}}, \bibinfo {author} {\bibfnamefont {S.}~\bibnamefont {Zhang}}, \bibinfo {author} {\bibfnamefont {X.-X.}\ \bibnamefont {Yang}}, \bibinfo {author} {\bibfnamefont {C.}~\bibnamefont {Zhang}}, \bibinfo {author} {\bibfnamefont {L.}~\bibnamefont {Du}}, \bibinfo {author} {\bibfnamefont {H.-F.}\ \bibnamefont {Zhang}}, \bibinfo {author} {\bibfnamefont {H.-R.}\ \bibnamefont {Tao}}, \bibinfo {author} {\bibfnamefont {T.-L.}\ \bibnamefont {Wang}}, \bibinfo {author} {\bibfnamefont {Z.-L.}\ \bibnamefont {Jia}}, \bibinfo {author} {\bibfnamefont {Z.-Y.}\ \bibnamefont {Chen}},\ and\ \bibinfo {author} {\bibfnamefont {G.-P.}\ \bibnamefont {Guo}},\ }\href {https://doi.org/10.1103/PhysRevApplied.21.064060} {\bibfield  {journal} {\bibinfo  {journal} {Physical Review Applied}\ }\textbf {\bibinfo {volume} {21}},\ \bibinfo {pages} {064060} (\bibinfo {year} {2024})}\BibitemShut {NoStop}\relax
\bibitem [{\citenamefont {An}\ \emph {et~al.}(2026)\citenamefont {An}, \citenamefont {Zhang}, \citenamefont {Ding}, \citenamefont {Ding}, \citenamefont {Sung}, \citenamefont {Winik}, \citenamefont {Kim}, \citenamefont {Rosen}, \citenamefont {Azar}, \citenamefont {Pi{\~n}ero}, \citenamefont {Gertler}, \citenamefont {Gingras}, \citenamefont {Niedzielski}, \citenamefont {Stickler}, \citenamefont {Schwartz}, \citenamefont {Wang}, \citenamefont {Orlando}, \citenamefont {Gustavsson}, \citenamefont {Hays}, \citenamefont {Grover}, \citenamefont {Serniak},\ and\ \citenamefont {Oliver}}]{an$ZZ$freeTwotransmon$CZ$2026}\BibitemOpen
  \bibfield  {author} {\bibinfo {author} {\bibfnamefont {J.}~\bibnamefont {An}}, \bibinfo {author} {\bibfnamefont {H.}~\bibnamefont {Zhang}}, \bibinfo {author} {\bibfnamefont {Q.}~\bibnamefont {Ding}}, \bibinfo {author} {\bibfnamefont {L.}~\bibnamefont {Ding}}, \bibinfo {author} {\bibfnamefont {Y.}~\bibnamefont {Sung}}, \bibinfo {author} {\bibfnamefont {R.}~\bibnamefont {Winik}}, \bibinfo {author} {\bibfnamefont {J.}~\bibnamefont {Kim}}, \bibinfo {author} {\bibfnamefont {I.~T.}\ \bibnamefont {Rosen}}, \bibinfo {author} {\bibfnamefont {K.}~\bibnamefont {Azar}}, \bibinfo {author} {\bibfnamefont {R.~D.}\ \bibnamefont {Pi{\~n}ero}}, \bibinfo {author} {\bibfnamefont {J.~M.}\ \bibnamefont {Gertler}}, \bibinfo {author} {\bibfnamefont {M.}~\bibnamefont {Gingras}}, \bibinfo {author} {\bibfnamefont {B.~M.}\ \bibnamefont {Niedzielski}}, \bibinfo {author} {\bibfnamefont {H.}~\bibnamefont {Stickler}}, \bibinfo {author} {\bibfnamefont {M.~E.}\ \bibnamefont {Schwartz}}, \bibinfo {author} {\bibfnamefont {J.~{\^I-j}.}\ \bibnamefont {Wang}}, \bibinfo {author} {\bibfnamefont {T.~P.}\ \bibnamefont {Orlando}}, \bibinfo {author} {\bibfnamefont {S.}~\bibnamefont {Gustavsson}}, \bibinfo {author} {\bibfnamefont {M.}~\bibnamefont {Hays}}, \bibinfo {author} {\bibfnamefont {J.~A.}\ \bibnamefont {Grover}}, \bibinfo {author} {\bibfnamefont {K.}~\bibnamefont {Serniak}},\ and\ \bibinfo {author} {\bibfnamefont {W.~D.}\ \bibnamefont {Oliver}},\ }\href {https://doi.org/10.1103/9k19-c99g} {\bibfield  {journal} {\bibinfo  {journal} {Physical Review Applied}\ }\textbf {\bibinfo {volume} {26}},\ \bibinfo {pages} {014076} (\bibinfo {year} {2026})}\BibitemShut {NoStop}\relax
\bibitem [{\citenamefont {Strand}\ \emph {et~al.}(2013)\citenamefont {Strand}, \citenamefont {Ware}, \citenamefont {Beaudoin}, \citenamefont {Ohki}, \citenamefont {Johnson}, \citenamefont {Blais},\ and\ \citenamefont {Plourde}}]{strandFirstorderSidebandTransitions2013a}\BibitemOpen
  \bibfield  {author} {\bibinfo {author} {\bibfnamefont {J.~D.}\ \bibnamefont {Strand}}, \bibinfo {author} {\bibfnamefont {M.}~\bibnamefont {Ware}}, \bibinfo {author} {\bibfnamefont {F.}~\bibnamefont {Beaudoin}}, \bibinfo {author} {\bibfnamefont {T.~A.}\ \bibnamefont {Ohki}}, \bibinfo {author} {\bibfnamefont {B.~R.}\ \bibnamefont {Johnson}}, \bibinfo {author} {\bibfnamefont {A.}~\bibnamefont {Blais}},\ and\ \bibinfo {author} {\bibfnamefont {B.~L.~T.}\ \bibnamefont {Plourde}},\ }\href {https://doi.org/10.1103/PhysRevB.87.220505} {\bibfield  {journal} {\bibinfo  {journal} {Physical Review B}\ }\textbf {\bibinfo {volume} {87}},\ \bibinfo {pages} {220505} (\bibinfo {year} {2013})}\BibitemShut {NoStop}\relax
\bibitem [{\citenamefont {Rosen}\ \emph {et~al.}(2024)\citenamefont {Rosen}, \citenamefont {Muschinske}, \citenamefont {Barrett}, \citenamefont {Chatterjee}, \citenamefont {Hays}, \citenamefont {DeMarco}, \citenamefont {Karamlou}, \citenamefont {Rower}, \citenamefont {Das}, \citenamefont {Kim}, \citenamefont {Niedzielski}, \citenamefont {Schuldt}, \citenamefont {Serniak}, \citenamefont {Schwartz}, \citenamefont {Yoder}, \citenamefont {Grover},\ and\ \citenamefont {Oliver}}]{rosenSyntheticMagneticVector2024}\BibitemOpen
  \bibfield  {author} {\bibinfo {author} {\bibfnamefont {I.~T.}\ \bibnamefont {Rosen}}, \bibinfo {author} {\bibfnamefont {S.}~\bibnamefont {Muschinske}}, \bibinfo {author} {\bibfnamefont {C.~N.}\ \bibnamefont {Barrett}}, \bibinfo {author} {\bibfnamefont {A.}~\bibnamefont {Chatterjee}}, \bibinfo {author} {\bibfnamefont {M.}~\bibnamefont {Hays}}, \bibinfo {author} {\bibfnamefont {M.~A.}\ \bibnamefont {DeMarco}}, \bibinfo {author} {\bibfnamefont {A.~H.}\ \bibnamefont {Karamlou}}, \bibinfo {author} {\bibfnamefont {D.~A.}\ \bibnamefont {Rower}}, \bibinfo {author} {\bibfnamefont {R.}~\bibnamefont {Das}}, \bibinfo {author} {\bibfnamefont {D.~K.}\ \bibnamefont {Kim}}, \bibinfo {author} {\bibfnamefont {B.~M.}\ \bibnamefont {Niedzielski}}, \bibinfo {author} {\bibfnamefont {M.}~\bibnamefont {Schuldt}}, \bibinfo {author} {\bibfnamefont {K.}~\bibnamefont {Serniak}}, \bibinfo {author} {\bibfnamefont {M.~E.}\ \bibnamefont {Schwartz}}, \bibinfo {author} {\bibfnamefont {J.~L.}\ \bibnamefont {Yoder}}, \bibinfo {author} {\bibfnamefont {J.~A.}\ \bibnamefont {Grover}},\ and\ \bibinfo {author} {\bibfnamefont {W.~D.}\ \bibnamefont {Oliver}},\ }\href {https://doi.org/10.1038/s41567-024-02661-3} {\bibfield  {journal} {\bibinfo  {journal} {Nature Physics}\ }\textbf {\bibinfo {volume} {20}},\ \bibinfo {pages} {1881} (\bibinfo {year} {2024})}\BibitemShut {NoStop}\relax
\bibitem [{\citenamefont {Vidal}\ \emph {et~al.}(2021)\citenamefont {Vidal}, \citenamefont {Bera}, \citenamefont {Riera}, \citenamefont {Lewenstein},\ and\ \citenamefont {Bera}}]{vidalQuantumOperationsInformation2021a}\BibitemOpen
  \bibfield  {author} {\bibinfo {author} {\bibfnamefont {N.~T.}\ \bibnamefont {Vidal}}, \bibinfo {author} {\bibfnamefont {M.~L.}\ \bibnamefont {Bera}}, \bibinfo {author} {\bibfnamefont {A.}~\bibnamefont {Riera}}, \bibinfo {author} {\bibfnamefont {M.}~\bibnamefont {Lewenstein}},\ and\ \bibinfo {author} {\bibfnamefont {M.~N.}\ \bibnamefont {Bera}},\ }\href {https://doi.org/10.1103/PhysRevA.104.032411} {\bibfield  {journal} {\bibinfo  {journal} {Physical Review A}\ }\textbf {\bibinfo {volume} {104}},\ \bibinfo {pages} {032411} (\bibinfo {year} {2021})}\BibitemShut {NoStop}\relax
\bibitem [{\citenamefont {Wang}\ and\ \citenamefont {Hazzard}(2025)}]{wangParticleExchangeStatistics2025}\BibitemOpen
  \bibfield  {author} {\bibinfo {author} {\bibfnamefont {Z.}~\bibnamefont {Wang}}\ and\ \bibinfo {author} {\bibfnamefont {K.~R.~A.}\ \bibnamefont {Hazzard}},\ }\href {https://doi.org/10.1038/s41586-024-08262-7} {\bibfield  {journal} {\bibinfo  {journal} {Nature}\ }\textbf {\bibinfo {volume} {637}},\ \bibinfo {pages} {314} (\bibinfo {year} {2025})}\BibitemShut {NoStop}\relax
\bibitem [{\citenamefont {Wang}\ and\ \citenamefont {Hazzard}(2026)}]{wangRparastatisticsFoundation2026}\BibitemOpen
  \bibfield  {author} {\bibinfo {author} {\bibfnamefont {Z.}~\bibnamefont {Wang}}\ and\ \bibinfo {author} {\bibfnamefont {K.~R.~A.}\ \bibnamefont {Hazzard}},\ }\href {https://arxiv.org/abs/2607.26351} {\bibinfo {title} {On {$R$}-parastatistics {I}: Foundation}} (\bibinfo {year} {2026}),\ \Eprint {https://arxiv.org/abs/2607.26351} {arXiv:2607.26351 [quant-ph]} \BibitemShut {NoStop}\relax
\bibitem [{\citenamefont {Nielsen}\ and\ \citenamefont {Chuang}(2010)}]{NielsenChuangQCQI}\BibitemOpen
  \bibfield  {author} {\bibinfo {author} {\bibfnamefont {M.~A.}\ \bibnamefont {Nielsen}}\ and\ \bibinfo {author} {\bibfnamefont {I.~L.}\ \bibnamefont {Chuang}},\ }\href {https://doi.org/10.1017/CBO9780511976667} {\emph {\bibinfo {title} {Quantum Computation and Quantum Information}}}\ (\bibinfo  {publisher} {Cambridge University Press},\ \bibinfo {year} {2010})\BibitemShut {NoStop}\relax
\end{thebibliography}
\end{document}